\documentclass[twocolumn]{svjour3}

\usepackage[latin1]{inputenc}
\usepackage{times}

\usepackage{psfrag, amsmath, amssymb, amscd, amsfonts, latexsym, epsf, graphicx}
\usepackage{natbib}
\usepackage[switch]{lineno}
\usepackage{framed}

\def\bbbr{{\mathbb R}} 
\def\bbbz{{\mathbb Z}}
\def\bbbs{{\mathbb S}}

\def\diag{{\operatorname{diag}}}
\def\norm{\scriptsize\mbox{norm}}
\def\affnorm{\scriptsize\mbox{affnorm}}
\def\normtiny{\tiny\mbox{norm}}
\def\affnormtiny{\tiny\mbox{affnorm}}

\newcommand{\orth}{\bot}

\journalname{arXiv preprint}

\begin{document}

\title{\bf On sources to variabilities of simple cells in the primary visual
  cortex: A principled theory for the interaction between geometric
  image transformations and receptive field responses%
\thanks{The support from the Swedish Research Council 
              (contract 2022-02969) is gratefully acknowledged. }}

\titlerunning{On sources to variabilities of simple cells in the primary visual
  cortex as implied by the influence of geometric image
  transformations}

\author{Tony Lindeberg}

\institute{Tony Lindeberg \at
                Computational Brain Science Lab,
              Division of Computational Science and Technology,
              KTH Royal Institute of Technology,
              SE-100 44 Stockholm, Sweden.
              \email{tony@kth.se}}

\date{Received: date / Accepted: date}

\maketitle

\begin{abstract}
  When a visual observer views objects and spatio-temporal events in
  the environment, the image data may undergo substantial
  variabilities, as caused by variations in the viewing distance,
  the viewing direction and the relative motion between the object or
  event and the observer. 
  A non-trivial question concerns how the vision system should be able
  to establish an identity between the receptive field responses
  obtained from different observations of the same object or event
  under such variations in the viewing conditions, since the responses of the
  receptive fields in the first layers of the visual hierarchy will be
  strongly influenced by the underlying geometric image transformations.
  
  This paper presents a systematic methodology for addressing this problem,
  by giving an overview of a principled theory for modelling the
  interaction between geometric image transformations and receptive
  field responses, for a visual observer that views objects and
  spatio-temporal events in the environment. Specifically, the paper
  presents fundamental results regarding the influence on the receptive field
  responses due to 4 main classes of geometric image transformations,
  which represent the main sources of variabilities in the geometry between the object
  or the event and the observer, based on locally linear approximations of
  the perspective or projective image projection models between
  different views.

  By postulating that the family of receptive fields should be
  well-behaved (covariant) under these classes of geometric image transformations, it
  follows that the receptive field shapes should be expanded over the
  degrees of freedom of the corresponding image transformations, to
  enable a formal matching between the receptive field responses
  computed under different viewing conditions for the same scene or
  for a structurally similar spatio-temporal event.

  We develop this theory for the idealized generalized Gaussian
  derivative model of visual receptive fields, which has been
  previously demonstrated to rather well model the qualitative
  shape of simple cells for higher mammals.
  Formal transformation properties of the receptive fields are stated
  for the 4 main fundamental classes of
  primitive geometric image transformations, and it is shown that a
  visual system, based on such computational primitives, will 
  have the ability to match the spatio-temporal receptive
  responses computed from dynamic scenes under the variabilities
  caused by composed variations in the viewing conditions.

  We conclude the treatment by discussing and providing partial support for a
  working hypothesis that the receptive fields of simple cells in the
  primary visual cortex ought to have the shapes of their
  receptive fields expanded over the degrees of freedom of the
  corresponding geometric image transformations.
  We also outline directions for further research, to seek additional
  characterization and neurophysiological or psychophysical support for the stated
  predictions, regarding variabilities in the shapes of the spatial or
  spatio-temporal receptive fields of simple cells in the primary
  visual cortex.

\end{abstract}

\section{Introduction}

When seeking to understand the function of the brain, one may ask if there
could be overarching principles that have guided how the neural circuits have
evolved to process the information from the sensory organs.
Specifically, regarding the visual and auditory senses, which provide
ways for an observer to infer cues about distant objects in the environment
without touching the objects physically, one may ask if there
could be special structures in the incoming flows of light and sound
waves, that have shaped how the visual and the auditory senses have
evolved over the ages of evolution.

Concerning vision, where one of the tasks is to recognize objects from
the light that is reflected from them, it is clear that the identity of an object remains,
even though we may view the object at different positions in the world,
from different distances, different viewing directions and
different relative motions between the object and the
observer. Concerning the related task of recognizing spatio-temporal
events, we may also regard different instantiations of spatio-temporal
events, that occur either faster or slower relative to each other,
as similar types of spatio-temporal events, under variations of the
temporal scales in the time-varying phenomena.

\begin{figure*}[hbt]
  \begin{center}
    {\em\small Uniform spatial scaling transformations caused by varying the distance between the object and the observer\/}

    \smallskip
    
    \begin{tabular}{ccc}
      \includegraphics[width=0.31\textwidth]{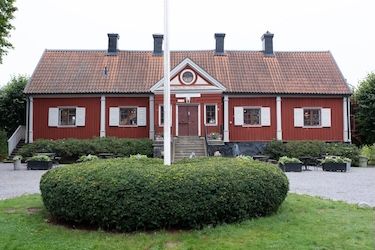}
      & \includegraphics[width=0.31\textwidth]{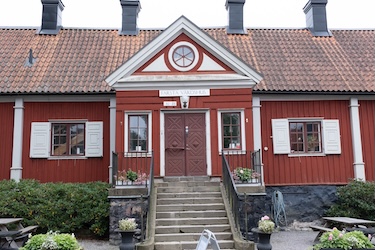}
      & \includegraphics[width=0.31\textwidth]{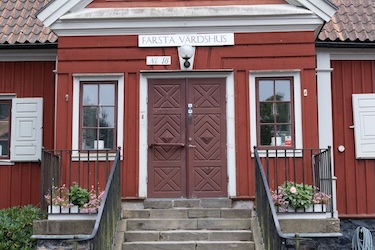}        
    \end{tabular}

    \medskip

     {\em\small Non-isotropic spatial affine transformations caused by varying the viewing direction relative to the object\/}

    \smallskip
    
    \begin{tabular}{ccc}
      \includegraphics[width=0.31\textwidth]{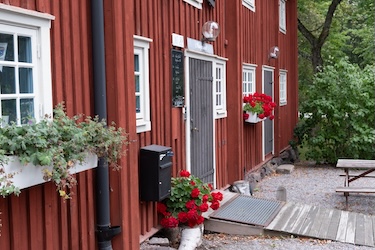}
      & \includegraphics[width=0.31\textwidth]{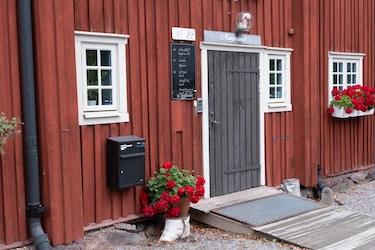}
      & \includegraphics[width=0.31\textwidth]{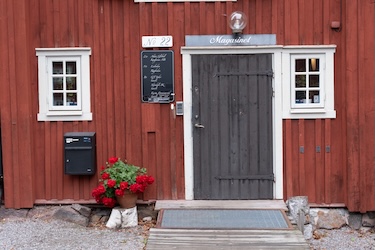}        
    \end{tabular}

    \medskip

     {\em\small Galilean transformations caused by relative motions
       between objects in the environment and the viewing direction\/}

     \smallskip
     
     \begin{tabular}{ccc}
      \includegraphics[width=0.31\textwidth]{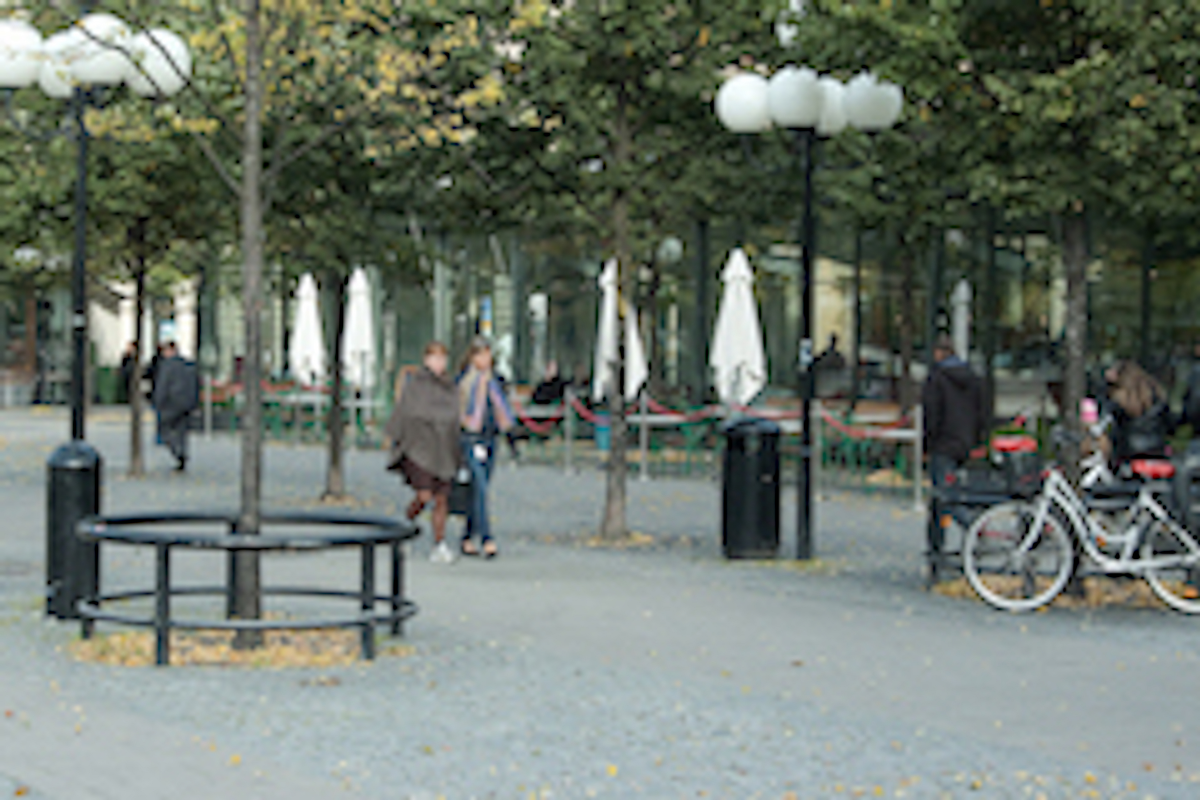}
      & \includegraphics[width=0.31\textwidth]{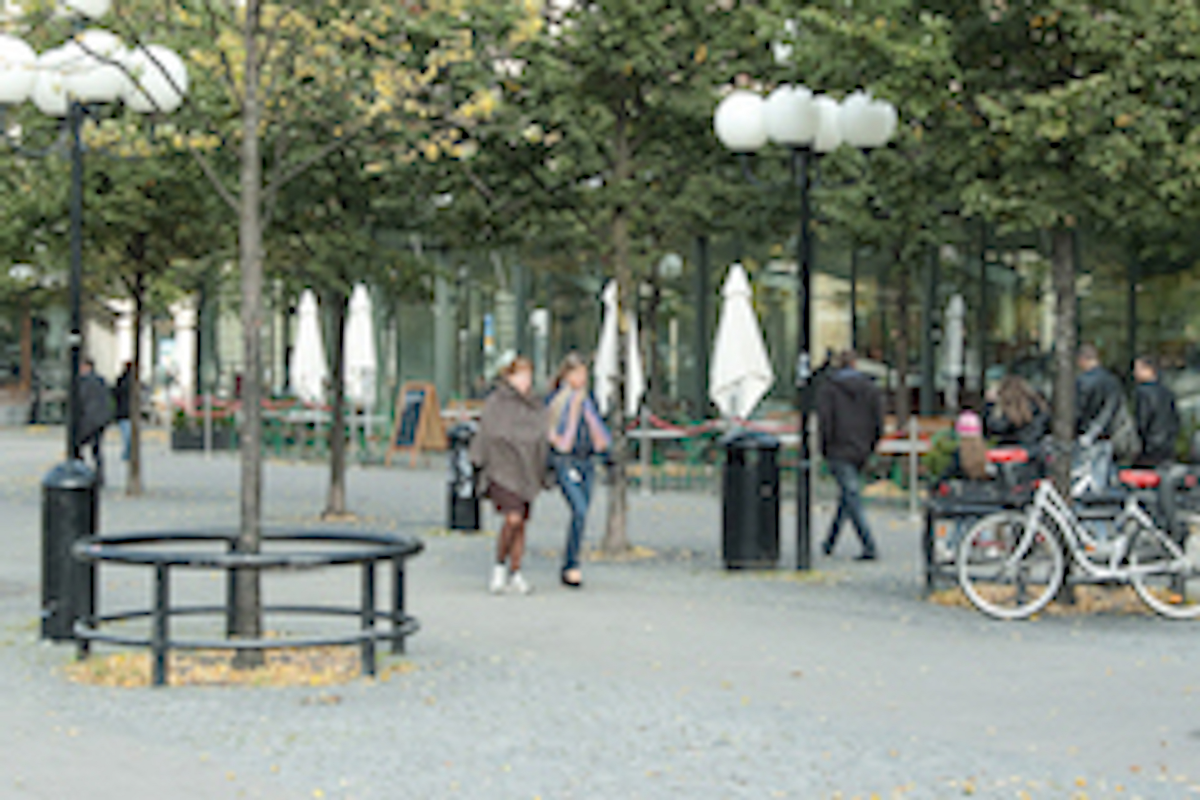}
      & \includegraphics[width=0.31\textwidth]{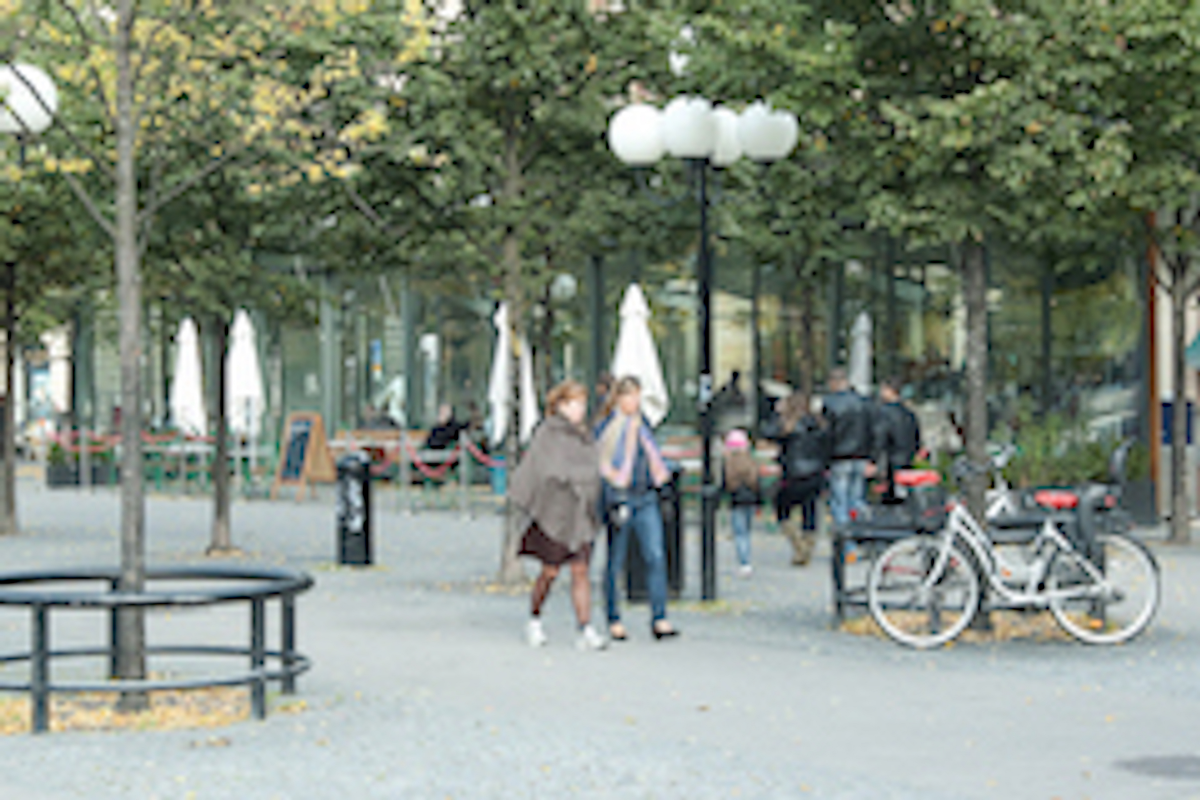}        
    \end{tabular}

  \end{center}
  \caption{Illustrations of variabilities in spatial and spatio-temporal image
    structures as caused by natural geometric image transformations.
    (top row) When the distance between the object and the observer is
  varied, this will lead to perspective image transformations, that to
  first order of approximation can be modelled as local uniform spatial
  scaling transformations.
  (middle row) When the viewing direction is varied relative to the
  object, this will lead to projective transformations between the two
  views, that to first order of approximation can be modelled as local
  spatial
  affine transformations.
  (bottom row) When there is relative motion between the object and
  the observer, the corresponding spatio-temporal image
  transformations can to first order of approximation be modelled as
  local Galilean transformations.
  (These figures have all been taken by a moving physical camera in either an
  approximately static environment (the top and middle rows) or a dynamic
  environment (the bottom row).
  The figures in the bottom row are reproduced
  from Lindeberg (\citeyear{Lin23-FrontCompNeuroSci}) with permission (Open Access).)}
  \label{fig-natural-img-transf}
\end{figure*}

The overall theme of this paper is to address this general problem,
regarding vision, from the
viewpoint of the interactions between geometric image transformations 
and receptive field responses, specifically manifested in terms of a certain
geometric structure on the families of receptive fields corresponding
to the simple cells in the primary visual cortex. Formally, we will
express these relations in terms of provable covariance%
\footnote{Structurally, the notion of covariance is related to the
  notion of invariance. The notion of covariance is, however, a less restrictive
  and more general property, which means that covariant visual processes
  may carry more complementary information. Invariant visual
  processes can, in turn, be constructed on top of covariant visual
  processes, by adding complementary selection mechanisms over the
  parameters of covariant processes, for example,
  the shape parameters of the here considered families of receptive fields.}
properties
of the receptive fields under a specific set of geometric image
transformations, to be described further below, based on the
following normative arguments, which for a general purpose vision system under
sufficient evolutionary pressure could determine how the neural
representations would be adapted to structural properties of the
stimuli that originate from the environment:

When a visual observer views objects in the environment, the resulting
image data on the retina or the image plane in the camera can exhibit a substantial
variability, as caused by the geometric image transformations induced
by variabilities in the viewing conditions. Specifically, by the variabilities caused
by varying (i)~the distance, (ii)~the viewing direction and (iii)~the relative motion
between the object and the observer, this will result in geometric image transformations,
that to first order of approximation can be modelled in terms of
(i)~spatial scaling transformations, (ii)~spatial affine deformations and
(iii)~Galilean transformations, see
Figure~\ref{fig-natural-img-transf} for illustrations.
By additionally (iv)~viewing different instances of a similar
spatio-temporal event that occurs either faster or slower relative to
a previously observed reference view, a visual observer can also experience variabilities
in terms of (iv)~temporal scaling transformations.

When the visual system operates on the resulting spatial or
spatio-temporal image data by either purely spatial or joint
spatio-temporal receptive fields in the earliest layers of the visual
hierarchy, the effects of these types of geometric image
transformations will have a very strong influence on the receptive
field responses. Despite such huge variabilities occurring regularly
when observing natural scenes under generic viewing conditions,
we as visual observers are nevertheless able to maintain the identity
of objects and to function robustly under such variabilities in the
image data.

Given these facts, one may ask if the biological vision systems
for different species have evolved to handle the influence of geometric
image transformations on the image data, to be able to maintain an
identity of the responses from the spatial and spatio-temporal
receptive fields under the huge variabilities in image data that can
be generated from similar scenes as depending on variabilities in
the viewing conditions.

The subject of this paper is to give both (i)~an overview,
(ii)~a set of conceptual extensions, and (iii)~a set of biological
implications and predictions from a theory that has
been developed to address this topic in a recent series of papers in
Lindeberg (\citeyear{Lin21-Heliyon}, \citeyear{Lin23-FrontCompNeuroSci},
\citeyear{Lin25-JMIV}, \citeyear{Lin25-BICY},
\citeyear{Lin25-JCompNeurSci-spanelong},
\citeyear{Lin25-JCompNeurSci-orisel},
\citeyear{Lin25-PONE}), and which constitutes a
substantially extended version of an earlier prototype to this theory
in Lindeberg (\citeyear{Lin10-JMIV}, \citeyear{Lin13-BICY}).
Compared to the original papers, the presentation in this review will
be simplified, with less focus on the mathematical details and more
emphasis on the main ideas and concepts, thus aimed at making
overall results from the proposed theory more easily
accessible for a wider audience.

Compared to the previous publications
on this topic, this presentation will be significantly more integrated, where the
different types of variabilities, that were addressed individually 
in Lindeberg (\citeyear{Lin25-JCompNeurSci-spanelong,Lin25-BICY}),
are here addressed in a unified
manner, thus reframing the individually addressed variabilities
in Lindeberg (\citeyear{Lin25-JCompNeurSci-spanelong,Lin25-BICY})
into a wider coherent context. In relation to the early approach to such
variabilities in Lindeberg (\citeyear{Lin23-FrontCompNeuroSci}),
the resulting biological hypotheses will also be formulated based on
significantly further theoretical depth,%
\footnote{While this paper is primarily written as a review paper,
  presenting a joint summarizing treatment of contributions that have
  been addressed in a set of other papers, beyond the first contributions in Lindeberg
  (\citeyear{Lin25-JCompNeurSci-spanelong,Lin25-BICY}), the theoretical
  background to the biological hypotheses in this paper will be based on the
  set of affine covariant derivative-based models for visual receptive fields
  to be described in Section~\ref{sec-aff-norm-ders},
  that were proposed in Lindeberg (\citeyear{Lin25-JMIV})
  after the publications of the earlier more
  biologically oriented contributions, and which will be built upon
  in the formulation of the more explicit affine-covariant models
  of visual receptive fields in
  Section~\ref{sec-joint-cov-props-spat-spattemo-rfs-comp-transf}
  and~\ref{sec-expl-ex-cov-props-rfs}.

  This material will then
  constitute a more solid foundation for the biological
  predictions to be presented in Section~\ref{sec-span-vars},
  with specifically substantial extensions regarding the predictions
  about variabilities with
  respect to Galilean transformations and temporal scaling 
  in Section~\ref{sec-variabil-temp},
  and the formulation of overall
  neuroscientific questions in Section~\ref{sec-outlines-neurophys}.

  For this purpose we will also include main cores of the theory for
  direction-selective visual receptive fields from the unpublished
  arXiv paper (Lindeberg \citeyear{Lin25-arXiv-dirsel}), however, without
  reproducing any figures from that paper.
  In addition, we provide an alternative model for handling
  variabilities with respect to uniform spatial scaling
  transformations and temporal scaling transformations
  in Section~\ref{sec-span-vars}, by only
  representing the finest spatial and temporal scales in the
  sensorium, as enabled by a formal theory for cascade smoothing
  properties, and as developed in the recent companion paper
  on hybrid Lie semi-group and cascade theory for visual receptive
  fields (Lindeberg \citeyear{Lin26-JMIV}).}
as enabled by the generalized mathematical
theory presented in Lindeberg
(\citeyear{Lin25-JMIV,Lin26-JMIV})
and the complementary analysis of the directional selectivity
properties of the idealized models of the spatio-temporal receptive
fields of simple cells recently performed in Lindeberg (\citeyear{Lin25-arXiv-dirsel}).

We will additionally describe a set of significant extensions%
\footnote{The explicit statements in
  Sections~\ref{sec-joint-cov-props-spat-spattemo-rfs-comp-transf},
  \ref{sec-expl-ex-cov-props-rfs} and~\ref{sec-span-vars} go
  significantly further compared to the related results in
  the previous publications. 
  The outlines to further research outlined in the
  beginning of Section~\ref{sec-outlines-neurophys} are specifically also new.}
relative to the
previously presented theoretical results,
as enabled by describing the overall
theory in a unified manner relative to the previous more specialized
technical contributions.
Specifically, we will use results from this theory to address the
topic of variabilities of simple cells%
\footnote{According to the taxonomy of neurons in the primary visual
cortex by Hubel and Wiesel
(\citeyear{HubWie59-Phys,HubWie62-Phys,HubWie68-JPhys,HubWie05-book}),
a visual neuron is said to be simple if: (i)~it has distinct
excitatory and inhibitory subregions, (ii)~it obeys roughly linear
summation properties and (iii) the excitatory
and inhibitory regions balance each other in diffuse lighting.}
in the primary visual cortex,
based on a hypothesis that the shapes of the spatial or
spatio-temporal receptive fields ought to span the degrees of the
geometric image transformations that are involved in the image
formation process, as well as provide further depth to the alternative
working hypothesis that a vision system could choose to only implement
the lowest levels of spatial and temporal scales in the sensorium
corresponding to the simple cells.
This will be done by making use of cascade smoothing
properties over spatial and temporal scales, as enabled by the
recently developed hybrid Lie semi-group theory over the filter shape
parameters in the
multi-parameter generalized Gaussian derivative model for visual
receptive fields, presented in Lindeberg (\citeyear{Lin26-JMIV}).

Based on these theoretical results, we will then propose an extended
set of theoretical predictions in Section~\ref{sec-span-vars}.
In the most idealized version of the theory, these predictions
state that in order to cover the degrees
of freedom of the four classes of geometric image transformations,
the shapes of the receptive fields ought to be expanded over the
following seven parameters:
(i)~the spatial size (spatial scale),
(ii)~the preferred orientation in image space,
(iii)~the degree of elongation,
(iv)~a fourth spatial degree of freedom,
(v)~the preferred direction in joint space-time,
(vi)~the preferred speed in joint space-time and
(vii)~the temporal duration (temporal scale).
Specifically, we will relate these predictions
to the neurophysiological results that we currently find as
available, and also propose new neurophysiological and psychophysical
experiments to address questions posed from this theoretical
study, and which previous neurophysiological and psychophysical
experiments may not have been designed to address.

An overall theme of the presented theoretical framework
is to postulate an identity between
the receptive field responses computed from different observations of
the same scene or the same spatio-temporal event, by requiring the family of
receptive fields to be covariant under local linearizations of the
considered classes of geometric image transformations.
Covariance, also referred to as equivariance in some literature,
essentially means that the families of receptive fields are to be
well-behaved under the corresponding classes of geometric image
transformations, in such a way that the result of computing a
receptive field response from geometrically transformed image data
should correspond to applying the same type of geometric
transformation to the receptive field response of the original image
data before the image transformation.

In this way, the notion of covariance makes it possible to establish a
notion of identity between the receptive field responses computed
from sets of image data that have been
transformed by the geometric image transformations, as induced by varying
the viewing conditions in terms of the distance, the viewing
direction, the relative motion between objects in the scene and
observer, and also the image transformations induced by viewing
similar types of spatio-temporal events that may occur either faster
or slower relative to a previously observed reference view.

Specifically, the notion of covariance makes it possible to, up to
first order of approximation, perfectly
match the receptive field responses between different views of the
same scene or spatio-temporal event, in such a way that the receptive
field responses are either exactly equal or very similar for local linearizations of the
geometric image transformations, for a particular way of formulating
the receptive fields corresponding to idealized models of simple cells
according to the considered generalized Gaussian derivative model for
visual receptive fields.

A more general underlying biological motivation to this study is that, while it is
well-known that the neurons in the brain tend to
exhibit different types of variabilities, it may on the other hand be usually less known what
are the sources to those variabilities.
See also M{\l}ynarski (\citeyear{Mly25-VisRes}) for a recent perspective of
this, regarding the retina and of explaining the functional variability in the retina over the
dimensions of sensory coding, stimulus and noise
statistics and collective coding over neuronal populations,
with special emphasis on the need for normative theories.
In this treatment, we present a systematic and theoretically principled
study, that predicts a set of variabilities regarding the shapes of the
receptive fields of simple cells in the primary cortex,
based on a normative theory of visual receptive fields, and with a
theoretically well-founded explanation in terms of covariance
properties under geometric image transformations
of the receptive fields in the lower layers of the visual
hierarchy, to in turn enable the computation of invariant visual
representations with regard to the influence of geometric image
transformations at higher layers in the visual pathway.

A condensed summary of the main core of messages to be presented
in this paper is given in Figure~\ref{fig-main-messages}.

\begin{figure}[hbt]
  \begin{framed}
  {\bf Main messages in this paper:}
  \begin{itemize}
  \item
    The overall prediction put forward in this paper is that the simple
    cells in the primary visual cortex of higher mammals ought to have the
    shapes of their receptive fields {\em expanded\/} over the degrees of
    freedom of the main types of geometric image transformations in
    terms of {\em joint\/} (i)~spatial scaling transformations, (ii)~spatial affine
    transformations, (iii)~Galilean transformations and (iv)~temporal
    scaling transformations.
  \item
    In the most idealized version of the theory, this prediction
    comprises {\em variabilities in the shapes of the receptive fields over
    the following 7~parameters:}
    (i)~the spatial size (spatial scale),
    (ii)~the preferred orientation in image space,
    (iii)~the degree of elongation,
    (iv)~a fourth spatial degree of freedom,
    (v)~the preferred direction in joint space-time,
    (vi)~the preferred speed in joint space-time and
    (vii)~the temporal duration (temporal scale).
  \item
    This prediction is motivated by the desired ability to be able to
    {\em match\/} the
    receptive field responses computed under different viewing
    conditions of the same object or the same spatio-temporal event,
    as based on {\em covariance properties\/} of the underlying visual receptive fields
    under geometric image transformations.
  \item
    A variation of that approach is also presented, in that the
    visual receptive fields in the primary visual cortex could
    alternatively explicitly represent only {\em the finest spatial
    and temporal scales\/} for each point in the visual field and for each
    time moment. Then, coarser-scale representations for
    the purpose of processing at higher levels in the visual hierarchy
    could be computed by either direct or indirect cascade smoothing
    properties from finer to coarser scales.
  \item
    In addition, a set of suggestions for further
    {\em neurophysiological measurements\/} is outlined in
    Section~\ref{sec-outlines-neurophys}, concerning
    mapping the more fine-grained structure of the
    visual receptive fields in the primary visual
    cortex, based on predictions from the presented theory.
  \item
    As theoretical basis for the analysis, as well as further
    modelling and theoretical analysis of visual receptive fields,
    a {\em formal theory\/} is put forward concerning the
    {\em interaction between receptive field
    responses and geometric image transformations\/}.
    \end{itemize}
    \end{framed}
   \caption{Condensed summary of main messages in this paper. A
     primary common theme of these contributions is that the influence of geometric
     image transformations on the incoming image data is considered as a
     primary factor for the development of the computational
     mechanisms in the receptive fields in the primary visual cortex.}
   \label{fig-main-messages}
\end{figure}

\subsection{Structure of this presentation}

In this paper, we will summarize the main components of this theory,
infer biological interpretations as well as state biological
predictions of the theory in the following way: 
After a brief overview of related work in Section~\ref{sec-rel-work},
Section~\ref{sec-geom-im-transf} starts by first defining the main classes of
geometric image transformations that we consider.

Then, Section~\ref{sec-ideal-rfs}
defines the axiomatically determined idealized models for visual
receptive fields, that we base this work upon,
complemented in Section~\ref{sec-cov-props} by describing the
covariance properties of the idealized receptive fields under
the considered classes of
geometric image transformations. Section~\ref{sec-span-vars}
then addresses whether we can regard the spatial and the spatio-temporal
shapes of simple cells in the primary visual cortex of higher
mammals to span the variabilities of geometric image transformations,
to support explicitly covariant families of visual receptive fields.
Section~\ref{sec-span-vars} also describes ideas to future
neurophysiological and psychophysical experiments to investigate
this topic in more detail.

Section~\ref{sec-model-framework} complements with a retrospective
description about properties of the theoretical framework that we 
use for modelling the relations between the geometric image
transformations and the receptive field responses, including the types
of conceptual simplifications that we have made for the purpose of performing
a closed-form theoretical analysis, as opposed to resorting to
numerical simulations, while at the same time describing
why those assumptions do not in any critical way limit the
generality of the overall approach. 
Finally, Section~\ref{sec-summ-disc} concludes with a summary and
discussion.

As a guide to the reader, this presentation is aimed at both
researchers interested in theoretical and computational modelling of
visual receptive fields and researchers interested in characterizing
the properties of the visual neurons in the visual pathway, 
including neurophysiological and psychophysical experimentalists.
For readers without a strong mathematical background, a
few of the sections in the main Section~\ref{sec-cov-props}
in this paper may by necessity be somewhat technical, to make
it possible to reproduce the main ideas, concepts and implications from the
underlying mathematical theory. For a reader more interested in the biological
interpretations and implications,
a shorter path should be possible,
by first reading the introductory
Sections~\ref{sec-geom-im-transf} and~\ref{sec-ideal-rfs} concerning
the image geometry and the receptive field models and then proceeding
directly to the treatment in Section~\ref{sec-span-vars},
about whether the shapes of the receptive
fields of simple cells in the primary visual cortex can be regarded as
spanning the degrees of freedom of the set of primitive geometric
image transformations. For filling in possible missing complementary
details, a good start could then be to backtrack from the references to
specific sections and equations from the summary and discussion
in Section~\ref{sec-summ-disc}.

\section{Relations to previous work}
\label{sec-rel-work}

Concerning variabilities of image data under spatial scaling
transformations, there are several sources of evidence that
demonstrate scale-invariant processing in the primate visual
cortex; see Biederman and Cooper (\citeyear{BieCoo92-ExpPhys}),
Logothetis {\em et al.\/} (\citeyear{LogPauPog95-CurrBiol}),
Ito {\em et al.\/} (\citeyear{ItoTamFujTan95-JNeuroPhys}),
Furmanski and Engel (\citeyear{FurEng00-VisRes}),
Hung {\em et al.\/} (\citeyear{HunKrePogDiC05-Science}) and
Isik {\em et al.\/} (\citeyear{IsiMeyLeiPog13-JNPhys}).

Given that scale-covariant image operations in the lower layers
constitute a powerful precursor to scale-invariant image
operations in higher layers in the visual hierarchy
(see Lindeberg (\citeyear{Lin21-Heliyon}) Appendix~I), one may hence ask
if the earliest layers of the visual system of higher mammals can be
regarded as able to process the image data in a way that can be
modelled in terms of scale covariance.
In a corresponding manner, one may ask if such a covariance property
would also extend to other types of geometric image transformations, in
relation to viewing objects, scenes and spatio-temporal events under
different types of viewing conditions.
Notably, the correspondence of invariance principles has been
postulated for biological vision systems by
Rolls (\citeyear{Rol94-BehavProc}),
DiCarlo and Maunsell (\citeyear{DiCMau00-Nature}),
Grimes and Rao (\citeyear{GriRao05-NeurComp}),
Quiroga {\em et al.\/}\ (\citeyear{QuiRedKreKocFri05-Nature}),
DiCarlo and Cox (\citeyear{DiCCox07-TICS}),
and Goris and Beek (\citeyear{GorBee09-FCNS}) and
Lindeberg (\citeyear{Lin13-PONE}).

Our knowledge about the functional properties of the receptive fields
of simple cells in the primary visual cortex originates from the
pioneering work by Hubel and Wiesel
(\citeyear{HubWie59-Phys,HubWie62-Phys,HubWie68-JPhys,HubWie05-book})
followed by more detailed characterizations by
DeAngelis {\em et al.\/}\ (\citeyear{DeAngOhzFre95-TINS,deAngAnz04-VisNeuroSci}),
Ringach (\citeyear{Rin01-JNeuroPhys,Rin04-JPhys}),
Conway and Livingstone (\citeyear{ConLiv06-JNeurSci}),
Johnson {\em et al.\/}\ (\citeyear{JohHawSha08-JNeuroSci}),
Walker {\em et al.\/}
(\citeyear{WalSinCobMuhFroFahEckReiPitTol19-NatNeurSci}) and
De and Horwitz (\citeyear{DeHor21-JNPhys}).

Computational models of simple cells have specifically
been expressed in terms of Gabor filters 
by Marcelja (\citeyear{Mar80-JOSA}),
Jones and Palmer (\citeyear{JonPal87a,JonPal87b}),
Porat and Zeevi (\citeyear{PorZee88-PAMI}),
Ringach (\citeyear{Rin01-JNeuroPhys,Rin04-JPhys}),
Citti and Sarti {\em et al.\/} (\citeyear{CitSar06-JMIV}),
Serre {\em et al.\/} (\citeyear{SerWolBilRiePog07-PAMI}),
Sarti {\em et al.\/} (\citeyear{SarCitPet08-BICY}),
Cocci {\em et al.\/} (\citeyear{CocBarSar11-JOSA}),
Barbieri {\em et al.\/} (\citeyear{BarCitCocSar14-JMIV}),
Baspinar {\em et al.\/} (\citeyear{BasCitSar18-JMIV,BasSarCit20-MathNeuroSci})
and
De and Horwitz (\citeyear{DeHor21-JNPhys}),
and in terms of Gaussian derivatives by
Koenderink and van Doorn (\citeyear{Koe84,KoeDoo87-BC,KoeDoo92-PAMI}),
Young (\citeyear{You87-SV}),
Young {\em et al.\/}\ (\citeyear{YouLesMey01-SV,YouLes01-SV}) and
Lindeberg (\citeyear{Lin10-JMIV,Lin13-BICY,Lin21-Heliyon}).
Theoretical models of early visual processes have also been
formulated based on Gaussian derivatives by
Lowe (\citeyear{Low00-BIO}),
May and Georgeson (\citeyear{MayGeo05-VisRes}),
Hesse and Georgeson (\citeyear{HesGeo05-VisRes}),
Georgeson  {\em et al.\/}\ (\citeyear{GeoMayFreHes07-JVis}),
Hansen and Neumann (\citeyear{HanNeu09-JVis}),
Wallis and Georgeson (\citeyear{WalGeo09-VisRes}),  
Wang and Spratling (\citeyear{WanSpra16-CognComp}),
Pei {\em et al.\/}\ (\citeyear{PeiGaoHaoQiaAi16-NeurRegen}),
Ghodrati {\em et al.\/}\ (\citeyear{GhoKhaLeh17-ProNeurobiol}),
Kristensen and Sandberg (\citeyear{KriSan21-SciRep}),
Abballe and Asari (\citeyear{AbbAsa22-PONE}),
Ruslim {\em et al.\/}\ (\citeyear{RusBurLia23-bioRxiv}) and
Wendt and Faul (\citeyear{WenFay24-JVis}).

Learning-based schemes to model visual receptive fields from training
data have also been proposed by
Rao and Ballard (\citeyear{RaoBal98-CompNeurSyst}),
Olshausen and Field (\citeyear{OlsFie96-Nature,OlsFie97-VR}),
Simoncelli and Olshausen (\citeyear{SimOls01-AnnRevNeurSci}),
Geisler (\citeyear{Wil08-AnnRevPsychol}),
Hyv{\"a}rinen {\em et al.\/} (\citeyear{HyvHurHoy09-NatImgStat}),
L{\"o}rincz {\em et al.\/} (\citeyear{LoePalSzi12-PLOS-CB}) and
Singer {\em et al.\/} (\citeyear{SinTerWilSchKinHar18-Elife}).
Poggio and Anselmi (\citeyear{PogAns16-book}) did on the other hand propose to model
learning of invariant receptive fields based on group theory.
More recently, deep learning approaches have been applied
for modelling visual receptive fields
(Keshishian {\em et al.\/} \citeyear{KesAkbKhaHerMehMes20-Elife}),
although one may also raise issues concerning the applicability of such approaches,
see Bae {\em et al.\/} (\citeyear{BaeKimKim21-FrontSystNeuroSci}),
Bowers {\em et al.\/}\
 (\citeyear{BowMalDujMonTsvBisPueAdoHumHeaEvaMitBly22-BehBrainSci}),
Heinke {\em et al.\/}\ (\citeyear{HeiLeoLee22-VisRes}),
Wichmann and Geirhos (\citeyear{WichGei23-AnnRevVisSci}) and the references therein.

The main subject of this paper is to model the receptive fields of
simple cells based on the normative theory for visual receptive fields
proposed in Lindeberg (\citeyear{Lin21-Heliyon}), in terms of the
generalized Gaussian derivative model, and then consider the
influence on the resulting receptive field responses caused by
variabilities in geometric image transformations.

This approach does specifically have structural similarities to the
recently developed area of geometric deep learning
(Bronstein {\em et al.\/} \citeyear{BroBruCohVel21-arXiv},
Gerken {\em et al.\/} \citeyear{GerAroCarLinOhlPetPer23-AIRev}),
where deep networks are formulated from the constraint that they
should be well-behaved under the influence of geometric image
transformations.
For examples of deep networks that are covariant (equivariant) under spatial scaling
transformations, see Worrall and Welling (\citeyear{WorWel19-NeuroIPS}),
Bekkers (\citeyear{Bek20-ICLR}),
Sosnovik {\em et al.\/} (\citeyear{SosSzmSme20-ICLR}, \citeyear{SosMosSme21-BMVC}),
Zhu {\em et al.\/} (\citeyear{ZhuQiuCalSapChe22-JMLR}),
Jansson and Lindeberg (\citeyear{JanLin22-JMIV}),
Lindeberg (\citeyear{Lin22-JMIV}),
Zhan {\em et al.\/} (\citeyear{ZhaSunLi22-ICCRE}),
Wimmer {\em et al.\/} (\citeyear{WimGolDa23-arXiv}) and
Perzanowski and Lindeberg (\citeyear{PerLin25-JMIV}).
For deep networks that are covariant or equivariant under spatial affine
transformations, see Li {\em et al.\/}
(\citeyear{LiQiuCheHeLin24-CVPR}),
regarding Galilean transformations, see Keller
(\citeyear{Kel25-NeurIPS}),
and regarding temporal scaling transformations,
see Jacques {\em et al.\/} (\citeyear{JacTigSarHowSed22-ICML}).

While the above cited works on covariant/equivariant networks have
set out to study the influence of one {\em single\/} type of geometric image
transformations in isolation, the subject of the treatment in this
paper is, on the other hand, to consider the
influence of all these types of geometric image transformations
{\em jointly\/}.
Additionally, in relation to such covariant/equivariant networks, the
proposed idealized models of simple cells considered in this paper
can be seen as and constitute a first linear layer that is
simultaneously covariant with respect to {\em all\/} the four types
of basic geometric image transformations in terms of (i)~spatial scaling
transformations, (ii)~spatial affine transformations, (iii)~Galilean
transformations and (iv)~temporal scaling transformations.

\section{Main classes of locally linearized geometric image transformations}
\label{sec-geom-im-transf}

For a monocular observer, that views the objects in a 3-D scene by a
planar 2-D image sensor, the projection is described by a non-linear
perspective projection model. For a binocular observer or multiple monocular
observers, that view the same 3-D scene from multiple observation
points and multiple viewing directions, the transformations between the
different views of the same scene are described by non-linear
projective transformations. To substantially simplify these non-linear
projection models, we will linearize them {\em locally\/}%
\footnote{This means that a new model for local linearization is
  defined for each point in space and each moment in time, thereby
  implying a substantially better accuracy compared to a global
  linearization. As will be described in more detail in
  Section~\ref{sec-loc-lin-approx}, a main assumption is that the form
  of local linearization should vary {\em slowly\/} within the support region
  of each receptive field. If that would not be the case, then
  receptive field responses at finer spatial and temporal scales
  should be considered. Specifically, by this model, we consider the
  projections of smooth local surface patches and disregard the
  effects of (i)~non-linear components of the perspective or
  projective mappings, (ii)~the curvature of the local surface
  patches, (iii)~strong non-rigid deformations and (iv)~rapid
  accelerations within the temporal duration of the receptive fields.
  See Section~\ref{sec-model-framework} for a more extensive treatment
  regarding the underlying assumptions behind the model and why these do
  not in any essential way affect to ability to infer the predictions
  about the 7~types of variabilities in the shapes of the receptive
  fields of simple cells as presented in this study.}
around each point
in terms of local first-order derivatives, which will then result in the
following classes of linear projection models applied to the image
coordinates of the form $x = (x_1, x_2)^T \in \bbbr^2$
and the temporal variable $t \in \bbbr$:
\begin{description}
\item[\em Uniform spatial scaling transformations:]
  \begin{equation}
    \label{eq-spat-sc-transf}
    f'(x') = f(x)  \quad\quad\mbox{for}\quad\quad  x' = S_x \, x,
  \end{equation}
  where $S_x \in \bbbr_+$ is a spatial scaling factor.
  \medskip
  
\item[\em Spatial affine transformations:]
  \begin{equation}
    \label{eq-spat-aff-transf}
    f'(x') = f(x) \quad\quad\mbox{for}\quad\quad x' = A \, x,
  \end{equation}
  where $A$ is an orientation-preserving non-singular
  $2 \times 2$ matrix with $\det A > 0$.
  \medskip
  
\item[\em Galilean transformations:]
    \begin{equation}
    \label{eq-gal-transf}
    f'(x', t') = f(x, t) \quad\quad\mbox{for}\quad\quad x' = x + u \, t, \; t' = t,
  \end{equation}
  where $u = (u_1, u_2)^T \in \bbbr^2$ is a 2-D velocity vector.
  \medskip
  
\item[\em Temporal scaling transformations:]
    \begin{equation}
      \label{eq-temp-sc-transf}
      f'(x', t') = f(x, t)  \quad\quad\mbox{for}\quad\quad  t' = S_t \, t, \; x' = x,
  \end{equation}
  where $S_t \in \bbbr_+$ is a temporal scaling factor.
\end{description}

\medskip

\noindent
Of particular interest is to compose these geometric transformations
in the following way when observing dynamic scenes with either
monocular or binocular locally linearized camera models
(Lindeberg \citeyear{Lin25-JMIV} Equations~(222)--(223)):%
\footnote{In the special case when the image velocity $u = 0$,
  the transformation (\ref{eq-x-transf}) reduces to a pure affine
  transformation. In the special case when the affine transformation
  matrix $A$ is equal to the identity matrix $A = I$,
  the transformation (\ref{eq-x-transf}) reduces to
  a scaled Galilean transformation.}
\begin{align}
  \begin{split}
     \label{eq-x-transf}
     x' = S_x \, (A \,  x + u \, t),
   \end{split}\\
  \begin{split}
     \label{eq-t-transf}
     t' = S_t \, t.
   \end{split}
\end{align}

\begin{figure}[btp]
  \vspace{-40mm}
  \begin{center}
     \includegraphics[width=0.50\textwidth]{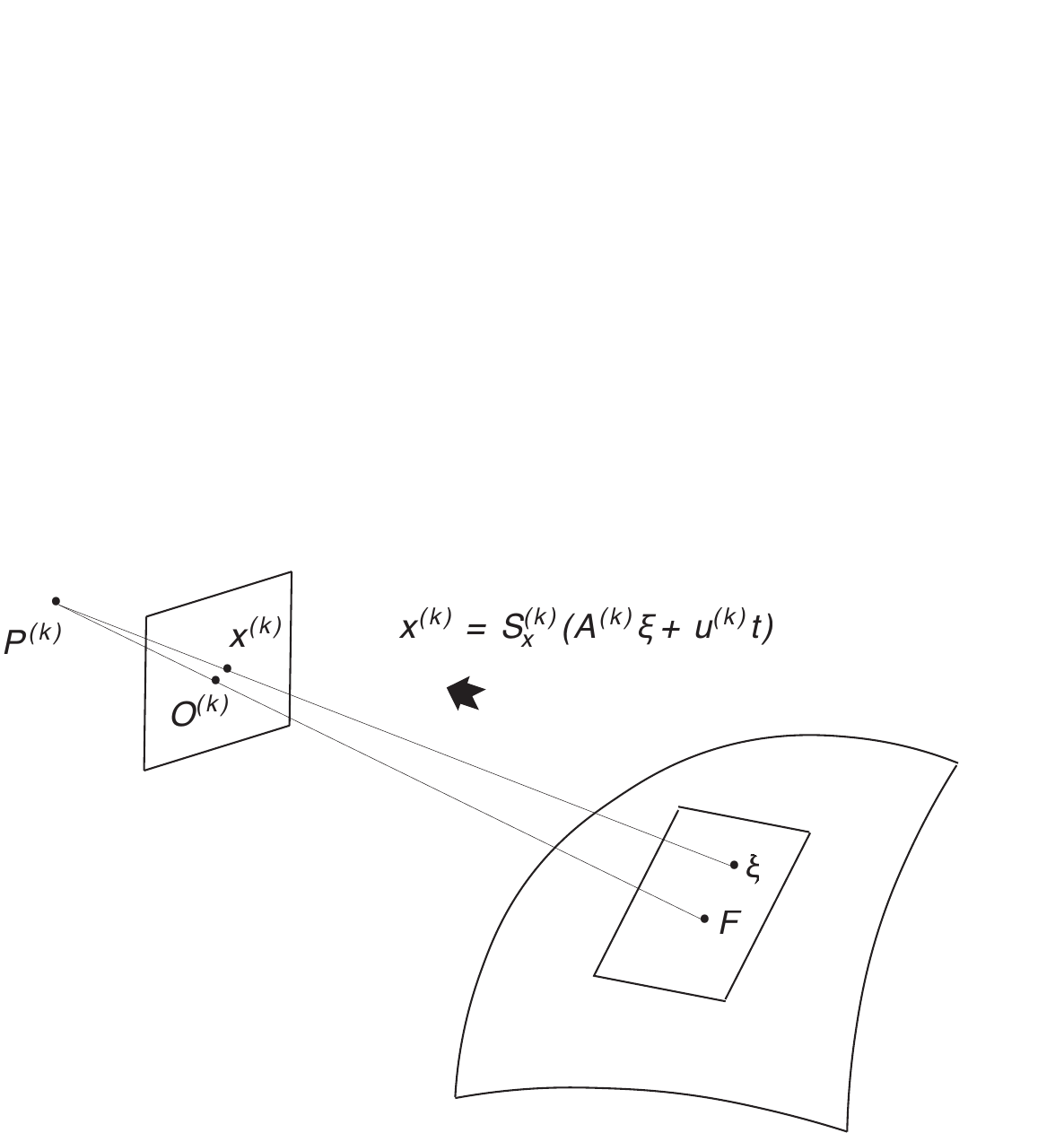} 
   \end{center}
   \caption{Illustration of the geometry underlying the composed
     locally linearized projection model in Equations~(\ref{eq-x-transf}) and
     (\ref{eq-t-transf}) for a single monocular view.
     Here, a local, possibly moving, surface patch is projected to an
     arbitrary view indexed by $k$ in a multi-view locally linearized
     projection model, with
     the fixation point $F$ on the surface mapped to the origin
     $O^{(k)} = 0$ in the image plane for the observer with the
     optic center $P^{(k)}$. Then, any point in the
     tangent plane to the surface at the fixation point, as
     parameterized by the local coordinates $\xi$ in a coordinate
     frame attached to the tangent plane of the surface with $\xi = 0$
     at the fixation point $F$, is by the
     local linearization mapped to the image point $x^{(k)}$.
     (Figure reproduced from Lindeberg (\citeyear{Lin25-JMIV}) with
     permission (OpenAccess).)}
   \label{fig-singlegeom}

   \bigskip
   
   \vspace{-25mm}
   \begin{center}
     \includegraphics[width=0.50\textwidth]{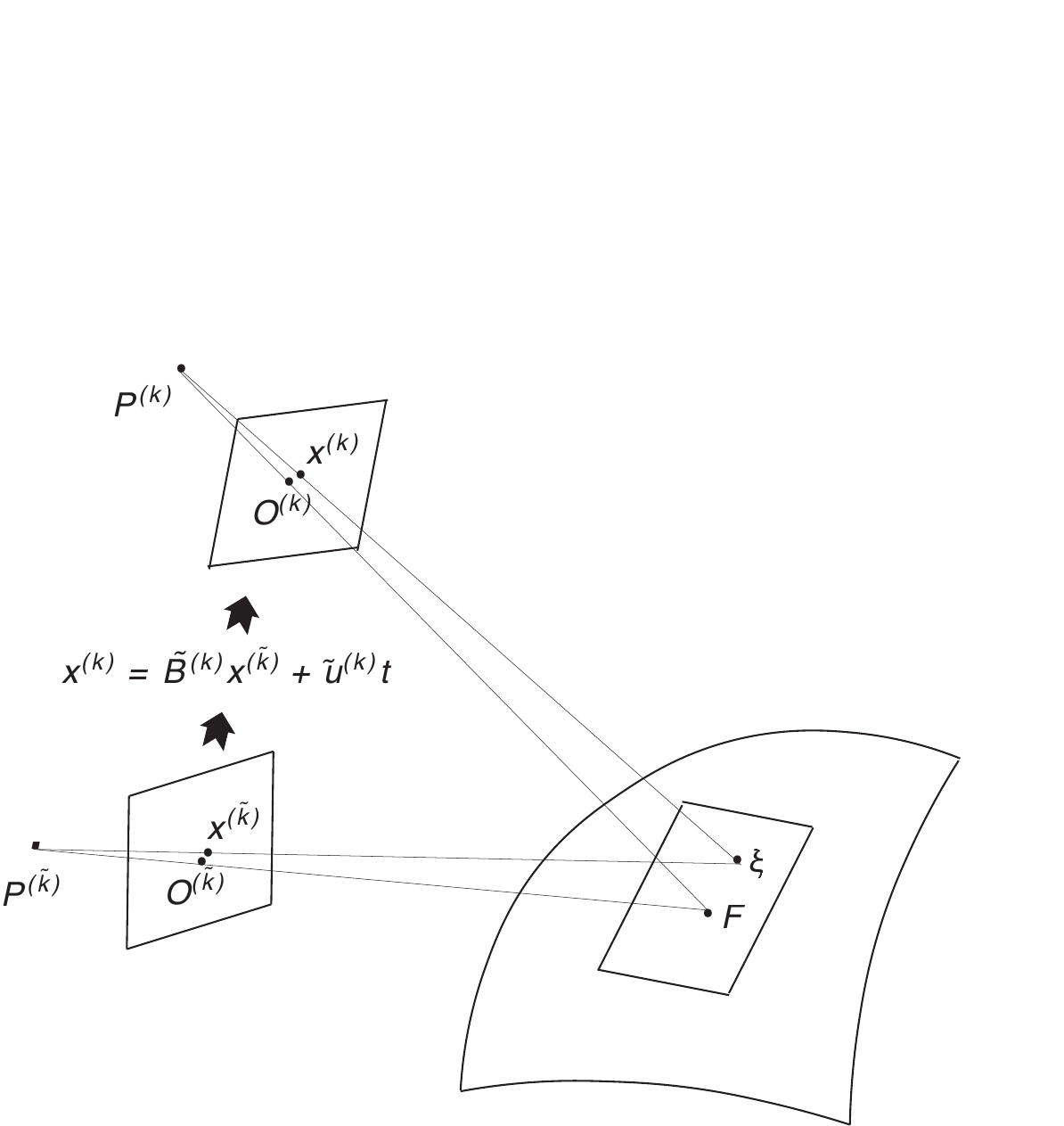} 
   \end{center}
   \caption{Illustration of the underlying geometric situtation for
     the locally linearized transformations between pairwise views of
     the same, possibly moving, local surface patch
     according to Equation~(\ref{eq-sc-aff-vel-transf-alt-obs-model}).
     Here, the view
     indexed by $\tilde{k}$ constitutes the reference view and the
     view indexed by $k$ constitutes an arbitrary view.
     By a connection of the point $\xi$ being the same in two instances
     of Figure~\ref{fig-singlegeom}, we can from the parameters $S_t$,
     $A$ and $u$ of the monocular mappings for optical centers based on
     the indices $k$ and $\tilde{k}$ establish a relationship between
     the matching image points $x^{(k)}$ and $x^{(\tilde{k})}$ in these
     two views. (Figure reproduced from Lindeberg (\citeyear{Lin25-JMIV}) with
     permission (OpenAccess).)}
   \label{fig-pairgeom}   
 \end{figure}

\noindent
Then, specifically 
\begin{itemize}
\item
  the $2 \times 2$ affine transformation matrix $A$ models the
  orthographic%
\footnote{Restricting the affine transformation matrix $A$ to
  correspond to an orthographic projection model corresponds to
  normalizing this matrix to the maximum eigenvalue $\lambda_{\max} = 1$.}
  projection of
  surface patterns from the tangent plane of a local surface patch
  to a plane, parallel with the image plane of the observer,
\item
  the velocity vector $u = (u_1, u_2)^T \in \bbbr^2$ models
  the projection of the 3-D motion vector
  $U = (U_1, U_2, U_3)^T$ of local surface patterns onto a plane,
  parallel to the image plane, by local orthographic projection,
\item
  the spatial scaling factor $S_x \in \bbbr_+$ models the perspective
  scaling factor proportional to the inverse depth $Z$, which will
  then affect both the projection of a spatial surface pattern and the
  magnitude of the perceived motion in the image plane, and
\item
  the temporal scaling factor $S_t \in \bbbr_+$ models the variability of
  similar spatio-temporal events, that may occur either faster or
  slower, when observing different instances of a similar event at
  different occasions.
\end{itemize}
Thereby, the composed image transformation model captures the variabilities
of the scaled orthographic projection model, complemented with a
variability over projections of 3-D motions between an observed object
and the observer, including spatio-temporal
events that may occur faster or slower relative to a previously
observed reference view,
see Figure~\ref{fig-singlegeom} for an illustration.

By further considering a pair of such projection equations for the
indices $k$ and $\tilde{k}$ of the observation points, and introducing
the alternative parameterizations of the parameters
according to Lindeberg (\citeyear{Lin25-JMIV}) Equations~(308)--(309)
\begin{equation}
  \tilde{B}^{(k)}
  = \frac{S_x^{(k)}}{S_x^{(\tilde{k})}} \, A^{(k)} (A^{(\tilde{k})})^{-1}
\end{equation}
and
\begin{equation}
  \tilde{u}^{(k)} = S_x^{(k)} \, A^{(k)} \left( u^{(k)} - u^{(\tilde{k})} \right),
\end{equation}
we have that corresponding image points $x^{(k)}$
and $x^{(\tilde{k})}$ between these views can be
expressed as
(Lindeberg \citeyear{Lin25-JMIV} Equation~(299))%
\footnote{In these relations, it should be noted that we have here abstracted
  away the epipolar constraints, which for a given geometric
  configuration determine with linear transformation matrices $B$ are
  compatible with the geometry. The important point with respect to
  the presented analysis is instead that the general form of the
  locally linearized geometric transformation
  (\ref{eq-sc-aff-vel-transf-alt-obs-model}) completely covers the
  space of all possible geometric transformations that are, up to first
  order of approximation, possible for all non-degenerate
  geometric configurations.}
\begin{equation}
  \label{eq-sc-aff-vel-transf-alt-obs-model}
  x^{(k)} = \tilde{B}^{(k)} \, x^{(\tilde{k})} + \tilde{u}^{(k)} \, t,
\end{equation}
where
\begin{itemize}
\item
  $x^{(k)} \in \bbbr^2$ is the locally linearized projection of the physical point
  on the surface pattern in the view from the observer with index $k$
  at time $t$,
\item
  $x^{(\tilde{k})} \in \bbbr^2$ is the locally linearized projection of the physical point
  on the surface pattern in the view from the observer with index
  $\tilde{k}$ at time $t$,
\item
  $\tilde{B}^{(k)}$ is a non-singular $2 \times 2$ affine projection matrix
  for the observer with index $k$ in relation to an observation from 
  a reference view with index $\tilde{k}$, and
\item
  $\tilde{u}^{(k)} \in \bbbr^2$ is a corresponding
  2-D relative motion vector for the observer with index $k$ in
  relation to an observation from a reference view with index $\tilde{k}$,
\end{itemize}
see Figure~\ref{fig-pairgeom} for an illustration.

In these ways, we can based on the four primitive geometric image
transformations according to
Equations~(\ref{eq-spat-sc-transf})--(\ref{eq-temp-sc-transf}) model both
locally linearized monocular perspective projections and locally
linearized binocular projective projections for a possibly moving
observer that observes possibly moving objects in multiple views of
dynamic scenes, as based on joint
compositions of these primitives according to
Equations~(\ref{eq-x-transf}), (\ref{eq-t-transf}) and~(\ref{eq-sc-aff-vel-transf-alt-obs-model}).

\begin{figure*}[hbtp]
  \begin{center}
    \begin{tabular}{ccccc}
      $\sigma_x = 1$ & $\sigma_x = \sqrt{2}$ & $\sigma_x = 2$ & $\sigma_x = 2\sqrt{2}$ & $\sigma_x = 4$ \\
      \includegraphics[width=0.17\textwidth]{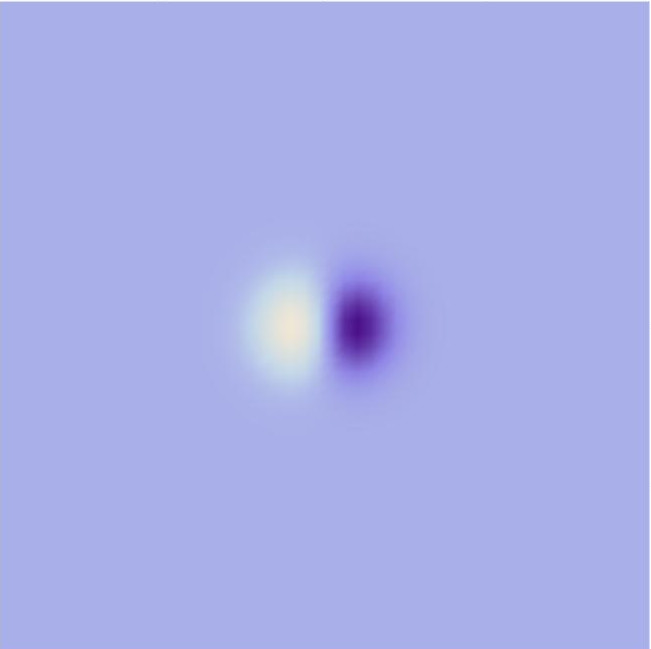}
      & \includegraphics[width=0.17\textwidth]{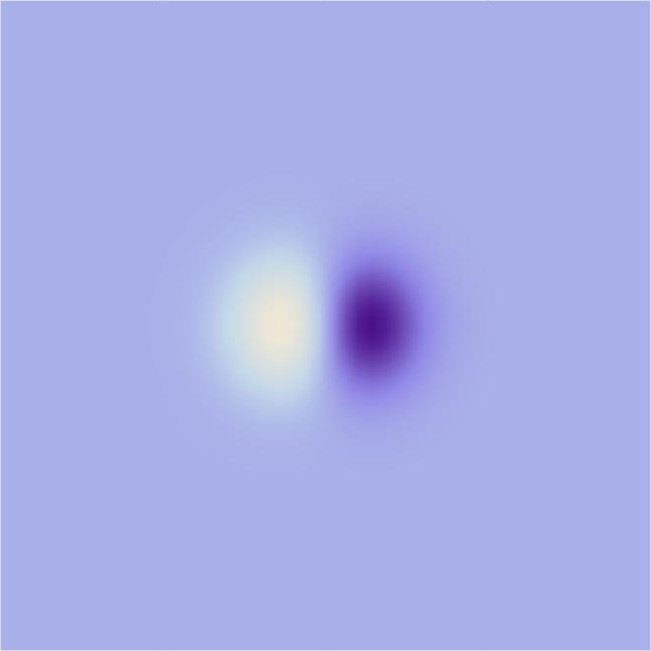}
      & \includegraphics[width=0.17\textwidth]{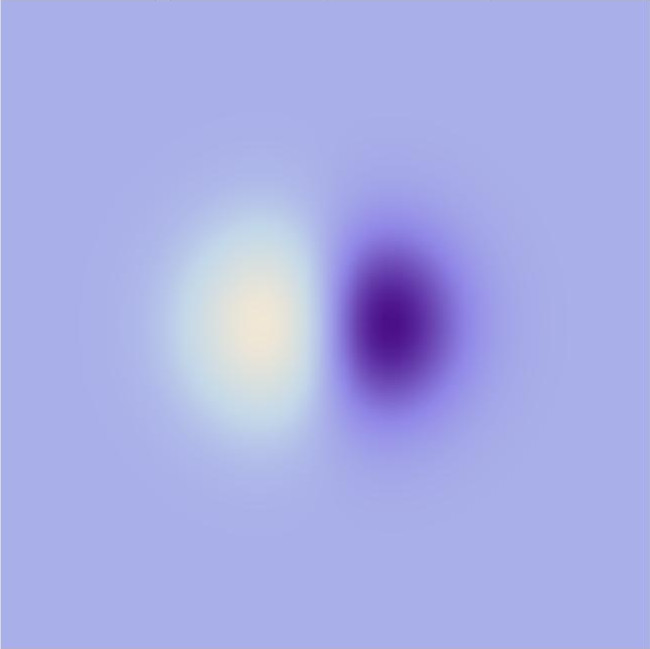}
      & \includegraphics[width=0.17\textwidth]{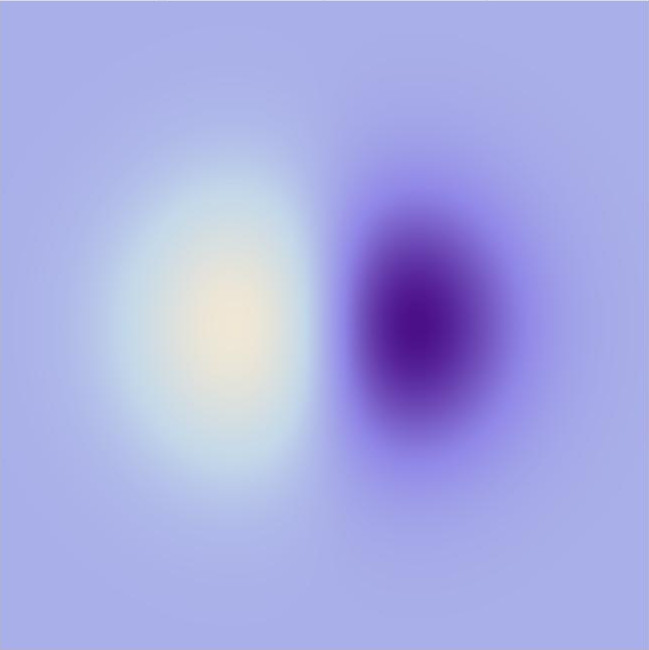}
      & \includegraphics[width=0.17\textwidth]{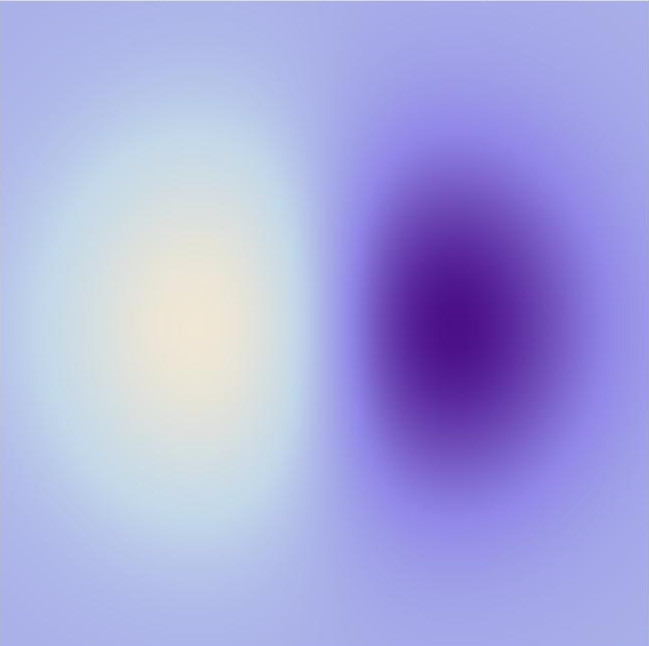}
     \end{tabular}
   \end{center}
   \caption{Illustration of the variability of spatial receptive fields under
     uniform spatial scaling transformations.
     Here, the first-order directional  derivative of the Gaussian kernel
     $T_{\varphi}(x;\; s, \Sigma) = \partial_{\varphi} (g(x;\; s, \Sigma))$
     in the horizontal direction $\varphi = 0$ is shown for
     different values of the spatial scale parameter
     $\sigma_x = \sqrt{s}$ for the special case of using an isotropic
     spatial covariant matrix with $\Sigma = I$.
     The variability of this spatial scale parameter makes it possible
     to handle objects of different size in the world as well as
     objects at different distances to the camera.
     (Horizontal axes: spatial coordinate $x_1 \in [-10, 10]$.
     Vertical axes: spatial coordinate $x_2 \in [-10, 10]$.)}
   \label{fig-1spatders-scale-var}
 \end{figure*}

\begin{figure}[hbtp]
  \begin{center}
     \includegraphics[width=0.45\textwidth]{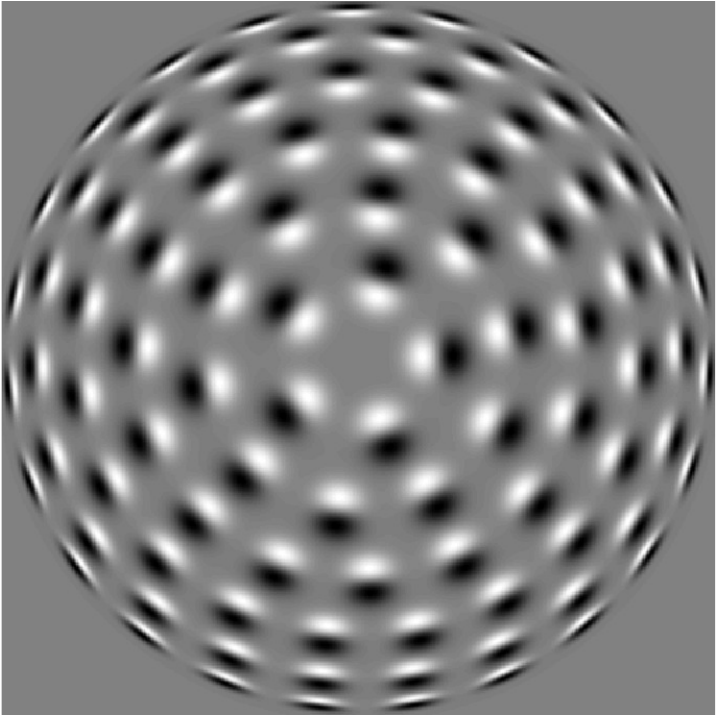} 
   \end{center}
   \caption{Illustration of the variability of spatial receptive fields under
     non-isotropic spatial affine transformations.
     Here, first-order directional spatial derivatives of
     Gaussian kernels
     $T_{\varphi}(x;\; s, \Sigma) = \partial_{\varphi} (g(x;\; s, \Sigma))$
     are shown under variations of the preferred image orientation and
     the degree of elongation of receptive fields of the spatial
     covariance matrices $\Sigma$, corresponding to a
     uniform distribution on a hemisphere.
     The variability of the spatial covariance matrix $\Sigma$
     corresponds to varying the slant and the tilt angles of the surface
     patch over all the angles on the visible hemisphere.
     (In this figure, the receptive fields are illustrated on a
     hemisphere, with each affine receptive field constituting an orthographic
     projection of a corresponding isotropic receptive field in each
     tangent plane of the hemisphere, corresponding to the first-order
     derivative of a rotationally symmetric Gaussian kernel in the
     tilt direction of the tangent plane. Specifically, the spatial
     scale parameters of the receptive
     fields have been normalized, such that the maximum eigenvalue
     of the spatial covariance matrix $\Sigma$ is the same for all
     the receptive fields, also corresponding to what will be the
     geometric situation with a local orthographic projection model.)
     (Horizontal and vertical axes: the spatial coordinates $x_1$ and $x_2$,
     for multiple spatial receptive fields shown within the same frame.)}
   \label{fig-1dir-gaussder}
\end{figure}

\begin{figure*}[hbtp]
  \begin{center}
    \begin{tabular}{ccccc}
      $v = -1$ & $v = -1/2$ & $v = 0$ & $v = 1/2$ & $v = 1$ \\
      \includegraphics[width=0.17\textwidth]{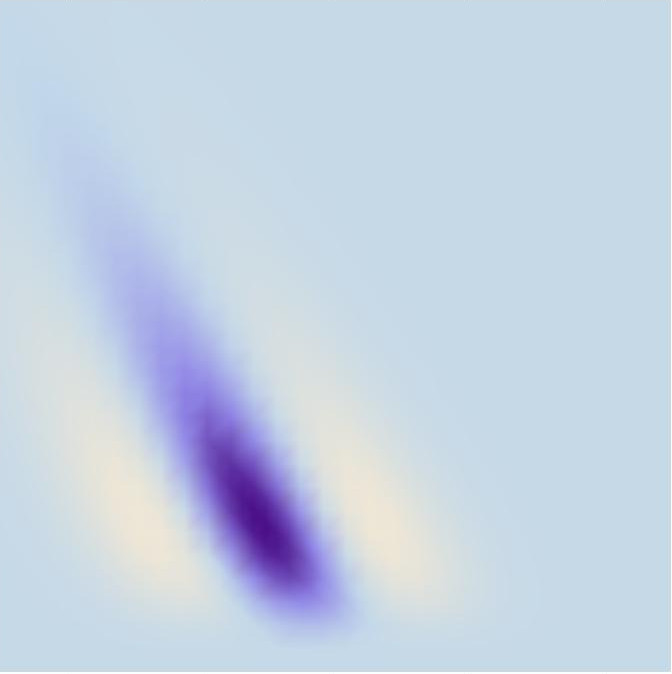}
      & \includegraphics[width=0.17\textwidth]{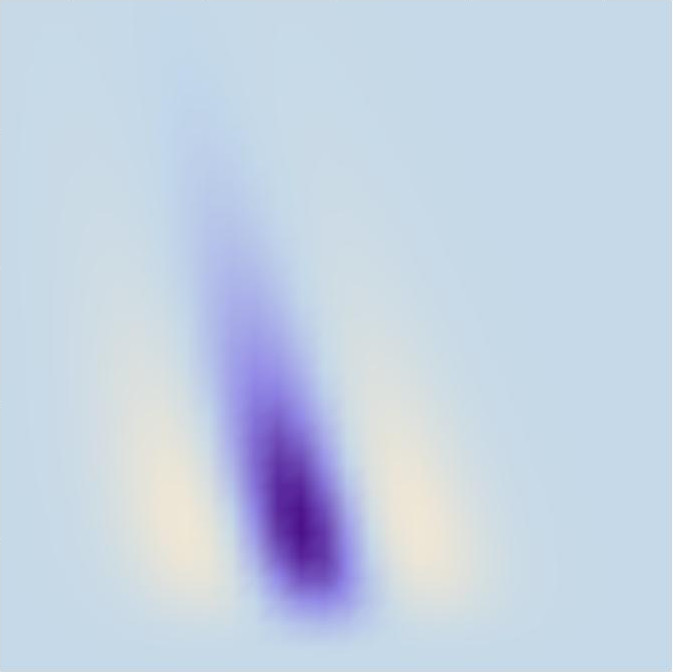}
      & \includegraphics[width=0.17\textwidth]{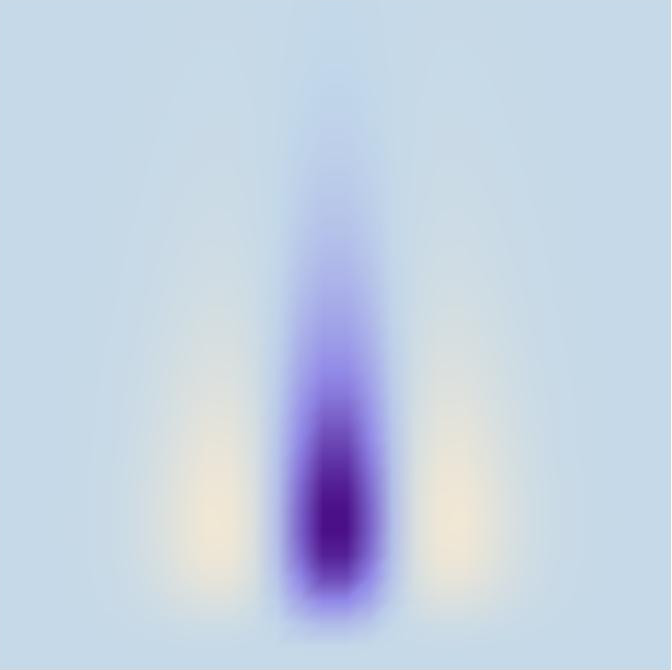}
      & \includegraphics[width=0.17\textwidth]{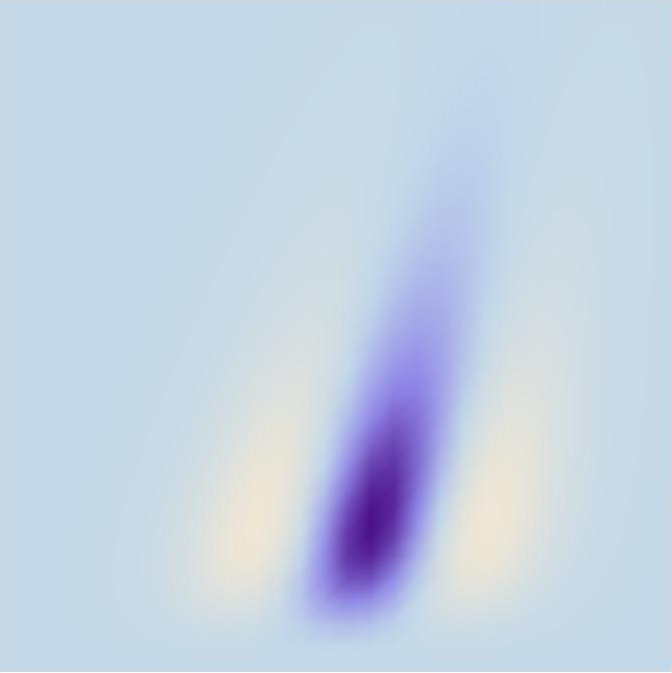}
      & \includegraphics[width=0.17\textwidth]{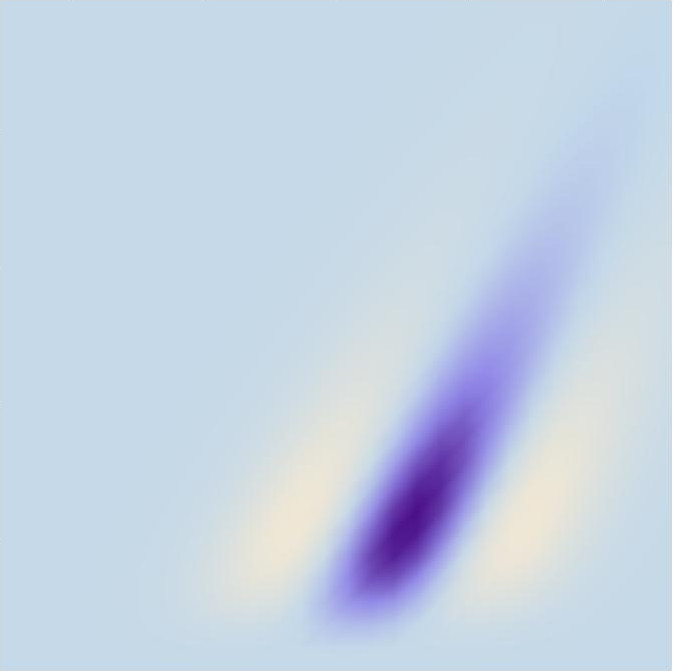}
     \end{tabular}
   \end{center}
   \caption{Illustration of the variability of spatio-temporal
     receptive fields under Galilean transformations.
     Here, the second-order spatial derivative $T_{xx}(x, t;\; s, \tau, v)
   = \partial_x \, \partial_{\bar t} \left( g(x - v \, t;\; s) \,
     \Psi(t;\; \tau, c) \right)$
   over a 1+1-D spatio-temporal domain is shown for different values
   of the velocity parameter $v$, based on using a first-order
   Gaussian derivative over the spatial domain and
   the zero-order derivative of the time-causal limit kernel over the temporal domain,  
   for $s = \sigma_x^2$, $\tau = \sigma_t^2$ and $c =2$
   with $\sigma_x = 1$ and $\sigma_t = 1$.
   This variability corresponds to varying the relative motion between
   the viewing direction and a moving local surface patch.
   (Horizontal axes: spatial coordinate $x \in [-5, 5]$.
   Vertical axes: time $t \in [0, 5]$.)}
   \label{fig-1spat1tempdir-timecaus-spattempscsp}
\end{figure*}

\begin{figure*}[hbtp]
  \begin{center}
    \begin{tabular}{ccc}
      $\sigma_t = 1$ & $\sigma_t = 2$ & $\sigma_t = 4$ \\
      \includegraphics[width=0.30\textwidth]{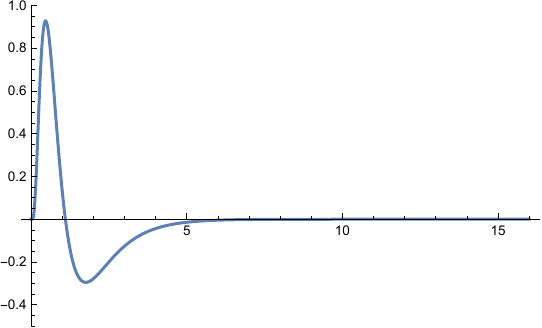}
      & \includegraphics[width=0.30\textwidth]{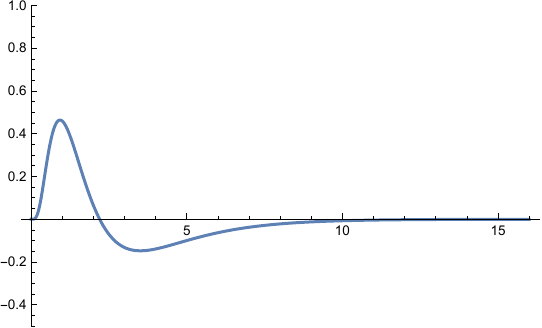}
      & \includegraphics[width=0.30\textwidth]{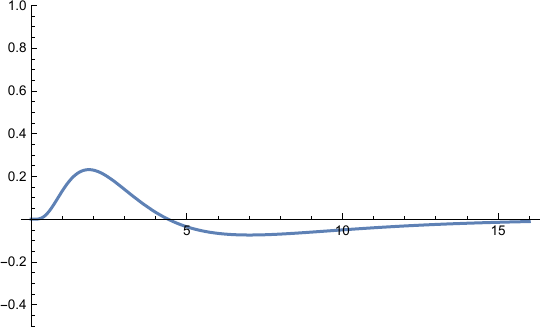}
     \end{tabular}
   \end{center}
   \caption{Illustration of the variability of purely temporal
     receptive fields under temporal scaling transformations.
     Here, the first-order scale-normalized temporal
     derivative of the time-causal limit kernel
     $T_{t}(t;\; \tau) = \sqrt{\tau} \, \partial_t \left( \Psi(t;\; \tau, c) \right)$
     for $c = 2$ is shown for different values of
     the temporal scale parameter $\sigma_t = \sqrt{\tau}$.
     This variability corresponds to observing temporal structures that
     occur either faster or slower relative to a previously observed
     reference view.
     (Horizontal axes: time $t \in [0, 16]$.
      Vertical axes: magnitude of the scale-normalized derivative $\in [-0.5, 1]$.)}
   \label{fig-timecaus-1tempders}
\end{figure*}

\section{Idealized receptive fields according to the generalized
  Gaussian derivative model for visual receptive fields}
\label{sec-ideal-rfs}

In this section, we will review main components in the generalized
Gaussian derivative theory for visual receptive fields, as arising
from the normative theory for visual receptive fields in
Lindeberg (\citeyear{Lin21-Heliyon}).

\subsection{Receptive field models in terms of linear spatial or spatio-temporal convolution operations}

Given spatial image data $f \colon \bbbr^2 \rightarrow \bbbr$ expressed on
the form $f(x) = f(x_1, x_2)$ for the image coordinates $x = (x_1, x_2)^T \in \bbbr^2$
or spatio-temporal image data
$f \colon \bbbr^2 \times \bbbr \rightarrow \bbbr$
expressed on the form $f(x, t) = f(x_1, x_2, t)$ with an additional dependency on the
temporal variable $t \in \bbbr$, a (linear) spatial receptive field
$T \colon \bbbr^2 \rightarrow \bbbr$ 
or a (linear) spatio-temporal receptive field
$T \colon \bbbr^2 \times \bbbr \rightarrow \bbbr$
can be seen as a spatial or a spatio-temporal convolution kernel,
that is to be applied to the image data $f$ according to
\begin{equation}
  (T * f)(x) = \int_{\xi \in \bbbr^2} T(\xi) \, f(x - \xi) \, d\xi
\end{equation}
in the case of a purely spatial image domain, or according to
\begin{equation}
  (T * f)(x, t)
  = \int_{\xi \in \bbbr^2} \int_{\eta \in \bbbr}
         T(\xi, \eta) \, f(x - \xi, t - \eta) \, d\xi \, d\eta
\end{equation}
in the case of a joint spatio-temporal image domain.

\subsection{Covariance properties of spatial and spatio-temporal
  receptive responses under spatial translations in the image plane
  and temporal shifts}

Because of this
convolution structure, the receptive field responses are covariant
under spatial translations in the image plane according to
\begin{equation}
  f'(x') = f(x) \quad\quad\mbox{or}\quad\quad f'(x', t') = f(x, t)
\end{equation}
for
\begin{equation}
  x' = x + \Delta x \quad\quad\mbox{where}\quad\quad \Delta x \in \bbbr^2
\end{equation}
and $t' = t$, in the sense that the corresponding spatial or
spatio-temporal receptive field responses $L = T * f$ and
$L' = T* f'$ then satisfy
\begin{equation}
  L'(x') = L(x) \quad\quad\mbox{or}\quad\quad L'(x', t') = L(x, t).
\end{equation}
Similarly, under a temporal shift of spatio-temporal image data of the form
\begin{equation}
  f'(x', t') = f(x, t)
\end{equation}
for
\begin{equation}
  t' = t + \Delta t \quad\quad\mbox{where}\quad\quad \Delta t \in \bbbr
\end{equation}
and $x' = x$, the corresponding spatio-temporal receptive field
responses $L = T * f$ and $L' = T* f'$ satisfy
\begin{equation}
  L'(x', t') = L(x, t).
\end{equation}
Because of these covariance properties, a vision system based on receptive field
responses, that can be modelled in terms of convolution operations, will
handle objects at different positions%
\footnote{In this treatment, we disregard the effects of a spatially
  varying sampling density of the receptive fields on a foveated
  sensor, such as the primate retina. For a principled treatment of
  such spatial sampling effects with respect to the
  receptive field responses, see
  Lindeberg and Florack (\citeyear{CVAP166}), with a condensed summary
  of some of the main results in Lindeberg (\citeyear{Lin13-BICY}) Section~7.}
in the image plane as well as
temporal events, that occur at different time moments, in a similar
manner.

A main subject of this paper is to present a set of theoretical
extensions to this linear convolution structure, to make it possible for
receptive fields with structurally similar properties as the simple
cells in the primary visual cortex, to handle more developed sets of
geometric image transformations applied to the image data used as
input to a vision system.

\subsection{Idealized spatial or spatio-temporal models for simple
  cells in the primary visual cortex}

To handle the additional influence on the receptive fields, due to  
the in Section~\ref{sec-geom-im-transf} described classes of
geometric image transformations, we will consider
idealized models for simple cells, based on the generalized Gaussian derivative
model for visual receptive fields, as initiated in the early work in
Lindeberg (\citeyear{Lin10-JMIV, Lin13-BICY}) and then further refined
regarding the temporal domain in Lindeberg (\citeyear{Lin16-JMIV,Lin21-Heliyon}).

The receptive fields according to this model have been obtained based
on axiomatic derivations, based on first principles
that reflect symmetry properties of the
environment in combination with internal consistency requirements to
guarantee theoretically well-founded treatment of image structures over
different spatial and temporal scales. In this respect, the families
of receptive fields have been formulated in a theoretically
well-founded manner.

According to the underlying normative theory for visual receptive
fields, the shapes of the receptive fields are
parameterized by a set of filter parameters,
with linear models of purely spatial receptive fields
corresponding to simple cells formulated in terms of affine Gaussian
derivatives of the form
\begin{multline}
  \label{eq-spat-RF-model}
  T_{\text{simple}}(x_1, x_2;\; \sigma_{\varphi}, \varphi, \Sigma_{\varphi}, m) = \\
  = T_{\varphi^m,\norm}(x_1, x_2;\; \sigma_{\varphi}, \Sigma_{\varphi})
  = \sigma_{\varphi}^{m} \, \partial_{\varphi}^{m} \left( g(x_1, x_2;\; \Sigma_{\varphi}) \right),
\end{multline}
where
\begin{itemize}
\item
   $\varphi \in [-\pi, \pi[$ is the preferred orientation of the receptive
   field,
\item
  $\sigma_{\varphi} \in \bbbr_+$ is the amount of spatial smoothing
  in units of the spatial standard deviation, and which may be
  defined as
  $\sigma_{\varphi} = \sqrt{e_{\varphi}^T \, \Sigma_{\varphi} \, e_{\varphi}}$
  where $e_{\varphi}$ is the unit vector in the direction of $\varphi$,
\item
  $\partial_{\varphi}^m =
  (\cos \varphi \, \partial_{x_1} + \sin  \varphi \, \partial_{x_2})^m$
  is an $m$th-order directional derivative operator,
  which in a reduced submanifold of receptive field shapes may
  be restricted to be in the direction $\varphi$, but more generally
  needs to be decoupled from the direction of $\varphi$ in order to
  allow for full covariance properties,
 \item
   $\Sigma_{\varphi}$ is a $2 \times 2$ symmetric positive definite
   covariance matrix, where in a reduced submanifold of the receptive field
   shapes, one may consider one of the eigenvectors of $\Sigma_{\varphi}$ as being in
   the direction of $\varphi$; to obtain full ability of the receptive
   fields to span all the degrees of freedom of the geometric image
   transformations, such a criterion about alignment does, however,
   need to be relaxed,
 \item
   $g(x;\; \Sigma_{\varphi})$ is a 2-D affine Gaussian kernel, with its shape
   determined by the spatial covariance matrix $\Sigma_{\varphi}$
   \begin{equation}
     \label{eq-2D-aff-gauss}
     g(x;\; \Sigma_{\varphi})
     = \frac{1}{2 \pi \sqrt{\det \Sigma_{\varphi}}}
         e^{-x^T \Sigma_{\varphi}^{-1} x/2}
    \end{equation}
    for $x = (x_1, x_2)^T \in \bbbr^2$.
\end{itemize}
Concerning time-dependent image data, spatio-temporal receptive fields
corresponding to simple cells
are, in turn, formulated according to
\begin{align}
  \begin{split}
    \label{eq-spat-temp-RF-model-der-norm-caus}
    T_{\text{simple}}(x_1, x_2, t;\; \sigma_{\varphi}, \sigma_t, \varphi, v, \Sigma_{\varphi}, m, n) 
  \end{split}\nonumber\\
  \begin{split}
   & = T_{{\varphi}^m, {\bar t}^n,\norm}(x_1, x_2, t;\; \sigma_{\varphi}, \sigma_t, v, \Sigma_{\varphi})
  \end{split}\nonumber\\
  \begin{split}
   &  = \sigma_{\varphi}^{m} \, 
          \sigma_t^{n} \, 
         \partial_{\varphi}^{m} \,\partial_{\bar t}^n 
          \left( g(x_1 - v_1 t, x_2 - v_2 t;\; \Sigma_{\varphi}) \,
           h(t;\; \sigma_t) \right),
  \end{split}
\end{align}
where, for the symbols not previously defined in connection
with Equation~(\ref{eq-spat-RF-model}), we have that:
\begin{itemize}
\item
  $\sigma_t$ represents the amount of temporal smoothing (in units of
  the temporal standard deviation),
\item
  $v = (v_1, v_2)^T$ represents a local motion vector in the
  direction $\varphi$ of the spatial orientation of the receptive field,
\item
  $\partial_{\bar t}^n = (\partial_t + v_1 \, \partial_{x_1} + v_2 \, \partial_{x_2})^n$
  represents an $n$th-order velocity-adapted temporal derivative
  operator, and
\item
  $h(t;\; \sigma_t)$ represents a temporal smoothing kernel with temporal
  standard deviation $\sigma_t$.
\end{itemize}
For the case of the temporal domain being non-causal
(meaning that the future relative to an temporal moment can be
accessed, as it can be on pre-recorded video data),
the temporal kernel can be chosen as the 1-D Gaussian kernel
\begin{equation}
  \label{eq-non-caus-temp-gauss}
  h(t;\; \sigma_t) = \frac{1}{\sqrt{2 \pi} \sigma_t} \, e^{-t^2/2\sigma_t^2},
\end{equation}
whereas in the case of the temporal domain being time-causal
(implying the more realistic real-time scenario, where the future cannot be accessed),
the temporal kernel can determined as the time-causal limit kernel
(Lindeberg \citeyear{Lin16-JMIV} Section~5;
Lindeberg \citeyear{Lin23-BICY} Section~3)
\begin{equation}
  \label{eq-time-caus-lim-kern}
  h(t;\; \sigma_t) = \psi(t;\; \sigma_t, c),
\end{equation}
characterized by having a Fourier transform of the form
\begin{equation}
  \label{eq-FT-comp-kern-log-distr-limit}
     \hat{\Psi}(\omega;\; \sigma_t, c) 
     = \prod_{k=1}^{\infty} \frac{1}{1 + i \, c^{-k} \sqrt{c^2-1} \, \sigma_t \, \omega}.
\end{equation}
This form of the temporal smoothing function corresponds to using
an infinite set of first-order integrators that are coupled in
cascade
\begin{equation}
  \label{eq-def-time-caus-lim-kern-inf-conv}
  \Psi(\cdot;\; \sigma_t, c) = *_{k=1}^{\infty} h_{\exp}(\cdot;\; \mu_k),
\end{equation}
with the time constants
\begin{equation}
  \mu_k = c^{-k} \sqrt{c^2-1} \, \sigma_t
\end{equation}
chosen so as to specifically enable temporal
scale covariance. The distribution parameter $c > 1$, often chosen
as $c = \sqrt{2}$ or $c = 2$, in this temporal
smoothing function reflects the ratio between adjacent discrete
temporal scale levels in the corresponding temporal
scale-space model. For practical implementation purposes, the infinite
number of first-order integrators can often be truncated after the 4
to 8 slowest time constants.

In Lindeberg (\citeyear{Lin21-Heliyon}), it was demonstrated
that idealized receptive field models of these types do
rather well model the qualitative shape of biological
simple cells, as obtained by neurophysiological measurements
by DeAngelis {\em et al.\/}\
(\citeyear{DeAngOhzFre95-TINS,deAngAnz04-VisNeuroSci}),
Conway and Livingstone (\citeyear{ConLiv06-JNeurSci}) and
Johnson {\em et al.\/}\ (\citeyear{JohHawSha08-JNeuroSci});
see Figures~12--18 in Lindeberg (\citeyear{Lin21-Heliyon})
for comparisons between biological receptive fields and idealized
models thereof, based on the generalized Gaussian derivative model for
visual receptive fields.

Specifically, this formulation of the idealized models of spatial and
spatio-temporal receptive fields implies that the shapes of the
receptive fields are expanded with respect to the degrees of freedom
of the corresponding geometric image transformations, as illustrated
in Figures~\ref{fig-1spatders-scale-var}--\ref{fig-timecaus-1tempders}.

\begin{figure*}[hbt]
    \begin{center}
    \begin{tabular}{cc}
      {\em\small (a) Non-covariant receptive fields\/}
      $\quad$ & $\quad$
     {\em\small (b) Covariant receptive fields\/} \\
      \includegraphics[width=0.35\textwidth]{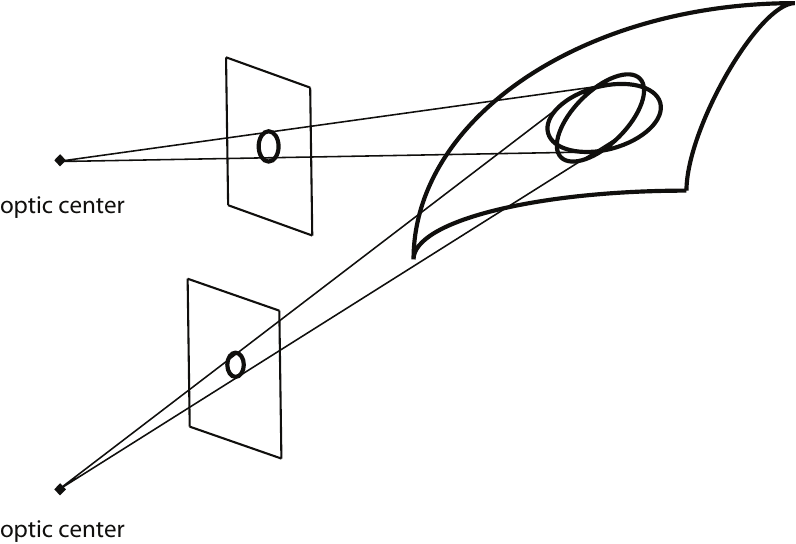}
      $\quad$ & $\quad$
      \includegraphics[width=0.35\textwidth]{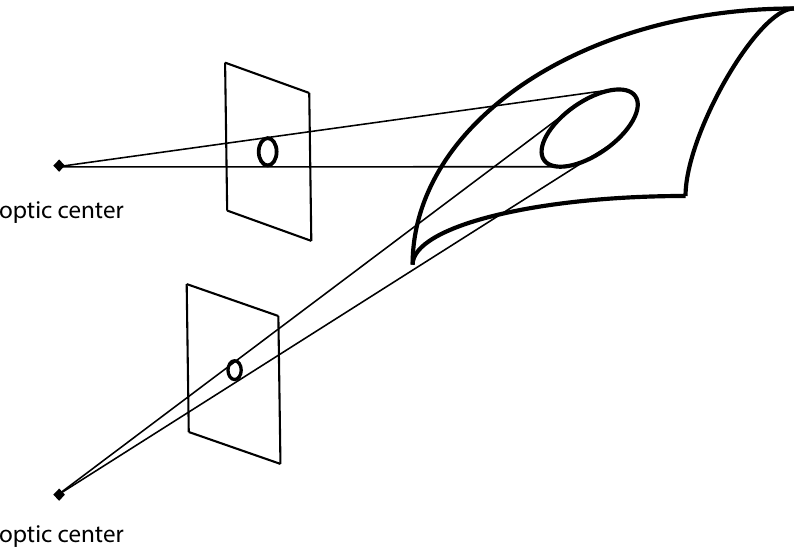} \\
    \end{tabular}
  \end{center}
  \caption{Illustration of the importance of covariance properties of
    the receptive fields when computing receptive field responses of a
    scene under different viewing conditions.
    (left) If the receptive field family is not covariant with respect
    to the appropriate class of geometric image transformations, then the
    backprojections of the receptive fields onto the tangent plane of
    an observed local surface patch will, in general, be different in the tangent
    plane of the surface. Thereby, if the receptive field responses
    are to be used for, to example, for computing local shape
    properties of the surface patch, then those shape estimates may be
    strongly biased because of effects of that mismatch between the
    backprojected receptive fields.
    (right) If the receptive field family is covariant with respect to
    the relevant class of geometric image transformations, then it is,
    on the other hand, possible to match the parameters of the
    receptive fields over the two image domains in such a way that the
    backprojected receptive fields do to first order of approximation
    coincide in the tangent plane of the surface. Thereby, the source
    to bias caused by a mismatch of the backprojected receptive fields
    can be substantially reduced, which enables to computation of more
    accurate estimates of the local surface shape.
    While this example concerns spatial receptive fields corresponding
    to two views with different viewing directions relative to a
    static scene, corresponding effects regarding the backprojections
    of the receptive fields will occur also when computing
    spatio-temporal receptive field responses for dynamic scenes.
    (Figure reproduced from Lindeberg
    \citeyear{Lin23-FrontCompNeuroSci} with permission (Open Access).)}
  \label{fig-ill-cov-rfs}
\end{figure*}

\section{Covariance properties of idealized models of simple cells
  under locally linearized geometric image transformations}
\label{sec-cov-props}

The notion of covariance, in some literature also referred to as
equivariance, means that the family of receptive fields is to be
well-behaved under a given class of geometric image transformations
${\cal G}$, see Figure~\ref{fig-ill-cov-rfs} for an illustration.

This property is specifically formulated
in the sense that the result of applying a receptive field, 
represented by the operator ${\cal R}$, to
geometrically transformed image data ${\cal G} \, f$
according to ${\cal R} \, {\cal G} \, f$ should be essentially equivalent to the
result of applying the same geometric transformation ${\cal G}$
to a closely related receptive field,
represented by the operator $\tilde{{\cal R}}$, applied to the
original image $f$, such that
\begin{equation}
  {\cal R} \, {\cal G} \, f = {\cal G} \, \tilde{{\cal R}} \, f.
\end{equation}
In this context, the ``closely related receptive field'' represented by
the operator $\tilde{{\cal R}}$ should either be a member of the same
family of receptive fields, as represented by the operator ${\cal R}$,
or constituting a sufficiently simple transformation thereof,
such as an amplitude scaling of the receptive field.

A very useful property of the receptive fields according the 
generalized Gaussian derivative model for visual receptive fields is
that (see Lindeberg (\citeyear{Lin23-FrontCompNeuroSci}) for details):
\begin{itemize}
\item
  for both the purely spatial model (\ref{eq-spat-RF-model}) for simple
  cells and the joint spatio-temporal model
  (\ref{eq-spat-temp-RF-model-der-norm-caus}) for
  simple cells, the receptive field responses are covariant under both uniform spatial scaling
  transformations of the form (\ref{eq-spat-sc-transf}) and spatial
  affine transformations of the form (\ref{eq-spat-aff-transf}), and
\item
  the joint spatio-temporal model (\ref{eq-spat-temp-RF-model-der-norm-caus})
  for simple cells is additionally covariant under Galilean transformations of
  the form (\ref{eq-gal-transf}) and temporal scaling transformations of
  the form (\ref{eq-temp-sc-transf}).
\end{itemize}
In the case of using the
non-causal Gaussian kernel (\ref{eq-non-caus-temp-gauss}) as the
temporal smoothing kernel in the idealized receptive field family,
the temporal scale covariance property holds for all non-negative temporal scaling
factors $S_t \in \bbbr_+$. In the case of using the time-causal limit
kernel (\ref{eq-time-caus-lim-kern}) as the temporal smoothing kernel,
for which the temporal scaling factors do not form a continuum but
are discrete, the temporal covariance property holds for all temporal
scaling factors $S_t$ that are integer powers of the distribution
parameter $c$ of the temporal smoothing kernel, according to
$S_t = c^i$ for $i \in \bbbz$.

These properties do thus imply that the receptive fields according to
generalized Gaussian derivative model for visual receptive field are
well-behaved under both (i)~uniform spatial scaling transformations,
(ii)~spatial affine transformations, (iii)~Galilean transformations,
and (iv)~temporal scaling transformations. In this way, the
generalized Gaussian derivative model can consistently process both
purely spatial and joint spatio-temporal image data, that are subject
to these individual geometric image transformations, as well as to joint
combinations thereof.

The subject of this section is to describe the main components in the
theory for covariant spatial and spatio-temporal receptive fields, in
such a way that it should be significantly easier to read for a reader
interested in the applications to neuroscience, compared to reading the
underlying publication (Lindeberg \citeyear{Lin25-JMIV}).
Specifically, this treatment will give (i)~the definitions of the
underlying concepts in relation to visual receptive fields,
(ii)~the properties that these notions obey in relation to covariance
properties under geometric image transformations, and
(iii)~interpretations of these results with regard to implications
regarding visual receptive fields in the primary visual cortex.
Thereby, we specifically do not include any explicit proofs, which the
interested reader will then be referred to by explicit references in
the text.

\subsection{Formal statement of the covariance properties for the pure
  spatial and spatio-temporal smoothing operations without
  spatial or temporal differentiation}

To express the joint covariance property in a compact manner, let us consider
the composed geometric transformation of the form in
Equations~(\ref{eq-x-transf})--(\ref{eq-t-transf}),
which models the joint effect of the four types of primitive
geometric image transformations
(\ref{eq-spat-sc-transf})--(\ref{eq-temp-sc-transf}) when observing a
possibly moving local surface patch with scaled orthographic
projection, from possibly different viewing distances and viewing
directions, in situations where there could be relative motions
between the object and the observer, and also a spatio-temporal event
may occur either faster or slower relative to a reference view.

Let us initially disregard the effects of the spatial and the
temporal derivative operators in the idealized spatio-temporal receptive field model
(\ref{eq-spat-RF-model}), by setting the differentiation orders to $m = 0$ and $n = 0$,
leading to the following form for the spatio-temporal smoothing
kernel%
\footnote{In this treatment, $\bbbs_+^2$ denotes the set
  of symmetric positive definite $2 \times 2$ matrices.}
$T \colon \bbbr^2 \times \bbbr \times \bbbr_+ \times \bbbs_+^2 \times \bbbr_+
\times \bbbr^2 \rightarrow \bbbr$
according to
\begin{equation}
  \label{eq-spat-temp-RF-model-again-cov-props-basic}
  T(x, t;\; s, \Sigma, \tau, v) 
  = g(x - v \, t;\; s, \Sigma) \, h(t;\; \tau)
\end{equation}
for the alternative parameterization of the spatial and temporal scale
parameters according to $s = \sigma_x^2$ and $\tau = \sigma_t^2$,
where we have here also for forthcoming use
redefined the spatial affine Gaussian kernel into the
following overparameterized form%
\footnote{In this context, it can be noted that the parameterization
  in terms of the spatial scale parameter $s$ and the spatial covariance
  matrix $\Sigma$ is not unique, since a multiplication of $s$ by a
  scalar factor $a$ and the multiplication of $\Sigma$ by $1/a$ leaves
  the resulting affine Gaussian kernel unchanged. When counting the
  degrees of freedom, we therefore count the total number of degrees
  of freedom (3) of the composed spatial covariance matrix $s \, \Sigma$.}
\begin{equation}
  \label{eq-gauss-fcn-2D}
  g(x;\; s, \Sigma)
  = \frac{1}{2 \pi \, s \sqrt{\det \Sigma}} \, e^{-x^T  \Sigma^{-1} x/2 s},
\end{equation}
in order to later more clearly be able to separate the degrees of
freedom between pure uniform spatial scaling transformations and more
general spatial affine transformations.

Let us also redefine the temporal smoothing kernel
$h \colon \bbbr \times \bbbr_+ \rightarrow \bbbr$
as either the non-causal 1-D Gaussian kernel according to
\begin{equation}
  \label{eq-non-caus-temp-gauss-var}
  h(t;\; \tau) = \frac{1}{\sqrt{2 \pi} \sqrt{\tau}} \, e^{-t^2/2\tau},
\end{equation}
or the time-causal limit kernel according to
(Lindeberg \citeyear{Lin16-JMIV} Section~5;
 Lindeberg \citeyear{Lin23-BICY} Section~3)
\begin{equation}
  \label{eq-time-caus-lim-kern-var}
  h(t;\; \tau) = \psi(t;\; \tau, c),
\end{equation}
characterized by having a Fourier transform of the form
\begin{equation}
  \label{eq-FT-comp-kern-log-distr-limit-var}
     \hat{\Psi}(\omega;\; \tau, c) 
     = \prod_{k=1}^{\infty} \frac{1}{1 + i \, c^{-k} \sqrt{c^2-1} \, \sqrt{\tau} \, \omega},
\end{equation}
where we will for all forthcoming use parameterize these kernels in terms of the
temporal variance $\tau = \sigma_t^2$, opposed to instead using the
temporal standard deviation $\sigma_t$ in
Equations~(\ref{eq-non-caus-temp-gauss})--(\ref{eq-FT-comp-kern-log-distr-limit}).

Next, let us for any 2+1-D spatio-temporal image data
$f \colon \bbbr^2 \times \bbbr \rightarrow \bbbr$
define spatio-temporally smoothed image data
$L \colon \bbbr^2 \times \bbbr \times \bbbr_+ \times \bbbs_+^2 \times \bbbr_+
\times \bbbr^2 \rightarrow \bbbr$
according to (Lindeberg \citeyear{Lin25-JMIV} Equation~(177))
\begin{equation}
  \label{eq-def-spat-temp-scsp}
  L(\cdot, \cdot;\; s, \Sigma, \tau, v) = T(\cdot, \cdot;\; s, \Sigma, \tau, v) * f(\cdot, \cdot).
\end{equation}
Let us also for geometrically transformed image data
\begin{equation}
  f'(x', t') = f(x, t)
\end{equation}
under the composed geometric image transformation according
to (\ref{eq-x-transf})--(\ref{eq-t-transf}) define correspondingly
spatio-temporally transformed image data according to
\begin{equation}
  \label{eq-def-spat-temp-scsp-prim}
  L(\cdot, \cdot;\; s', \Sigma', \tau', v') = T(\cdot, \cdot;\; s', \Sigma', \tau', v') * f'(\cdot, \cdot).
\end{equation}
Then, as shown in Section~5.2 in Lindeberg (\citeyear{Lin25-JMIV}), it
holds that these spatio-temporally smoothed image data are equal under
the composed geometric image transformation
of the form (\ref{eq-x-transf})--(\ref{eq-t-transf}) 
(Lindeberg \citeyear{Lin25-JMIV} Equation~(251))%
\footnote{This result holds for a continuous model of the joint
  spatio-temporal image domains and provided that the local
  linearization constitutes an accurate model of the spatio-temporal
  image deformations between the two domains. In the non-causal case,
  this result holds for all temporal scaling factors, whereas in the
  time-causal case, this result is only exact for temporal
  scaling factors $S_t$ that are integer powers of the distribution
  parameter $c$ of the time-causal limit kernel according to
  $S_t = c^j$ for some $j \in \bbbz$; otherwise it is an approximation.
  An additional degree of approximation
  then also arises when truncating the time-causal limit kernel to a finite
  number of truncated exponential kernels coupled of cascade.
  Due to the rapid decay of the time constants implied by the
  geometric distribution, it can for practical purposes of implementation often be
  sufficient to use as few as 4 to 8 truncated exponential kernels
  coupled in cascade. Additionally, this result
  depends on the spatio-temporal image intensities $f(x, t)$ and
  $f'(x', t')$ being bounded and
  locally integrable. See Section~\ref{sec-model-framework} for a more
  in-depth discussion about the validity of the underlying theoretical
  model.}
\begin{equation}
  \label{eq-joint-cov-prop-result-of-proof}
  L'(x', t';\; s', \Sigma', \tau', v') = L(x, t;\; s, \Sigma, \tau, v),
\end{equation}
provided that the parameters of the receptive fields over the two
spatio-temporal image domains are related
according to%
\footnote{Concerning the interpretations of these geometric
  transformation properties, it should be noted that for the purpose
  of the theoretical analysis in this paper, we throughout consider
  the forward effect of the geometric image transformation, and show
  that the receptive field responses can thereby be
  matched. Recovering the specific values of the parameters of the
  geometric image transformation in a given situation does on the
  other hand constitute an inverse problem, which may not be possible
  to solve from a single pair of image measurements, but may instead
  require multiple image measurements and/or complementary assumptions.}
(Lindeberg \citeyear{Lin25-JMIV} Equations~(252)--(255))
\begin{align}
  \begin{split}
    \label{eq-s-transf-result}
    s' & = S_x^2 \, s,
  \end{split}\\
  \begin{split}
    \label{eq-Sigma-transf-result}
    \Sigma' & = A \, \Sigma \, A^{T},
  \end{split}\\
  \begin{split}
    \label{eq-tau-transf-result}
    \tau' & = S_t^2 \, \tau,
  \end{split}\\
  \begin{split}
    \label{eq-v-transf-result}    
    v' & = \frac{S_x}{S_t} (A \, v + u).
  \end{split}
\end{align}
By restricting this result to the purely spatial case, when temporal
dependencies are disregarded, it holds that for purely spatial
smoothing kernels of the form
\begin{equation}
  \label{eq-spat-RF-model-cov-props-basic}
  T(x;\; s, \Sigma)  = g(x;\; s, \Sigma) 
\end{equation}
the corresponding purely spatially smoothed image representations are
related according to
\begin{equation}
  \label{eq-spat-cov-prop-result-of-proof}
  L'(x';\; s', \Sigma') = L(x;\; s, \Sigma),
\end{equation}
provided that the parameters of the purely spatial receptive fields
are related according to (\ref{eq-s-transf-result}) and
(\ref{eq-Sigma-transf-result}).

In this way, these results imply that the essential components of the
receptive fields in terms of either the purely spatial or the joint
spatio-temporal smoothing transformations can be perfectly matched
between the image data before and after the composed geometric image
transformation.
Thereby, these covariance properties provide a way of expressing an
identity operation between the spatial or spatio-temporal smoothing
effects of the idealized receptive field models, under compositions
of (i)~spatial scaling transformations, (ii)~affine image
transformations, (iii)~Galilean transformations and
(iv)~temporal scaling transformations.

A consequence of these results is, however, that in order to make it
possible to match the spatially or spatio-temporally smoothed image
data between two observations of the same object under different
viewing conditions, we have to expand the representation of the
receptive field responses over multiple values of the parameters of
the receptive fields; the set of parameters $(s, \Sigma)$ in the
purely spatial case or the set of parameters $(s, \Sigma, v, \tau)$
in the joint spatio-temporal case. In other words, the shapes of the visual
receptive fields should be expanded over the degrees of freedom of the
class of geometric image transformations. We will return to that topic
in Section~\ref{sec-span-vars} of this treatment.

\subsection{Transformation properties of the pure spatial and temporal
  derivative operators}

In addition to transforming the effect of the purely spatial or joint
spatio-temporal smoothing operation in the idealized receptive field
models (\ref{eq-spat-RF-model}) and
(\ref{eq-spat-temp-RF-model-der-norm-caus}), we do additionally have
to consider how to transform the effects of the spatial and the
temporal differentiation operators in the idealized receptive field
models according to (\ref{eq-spat-RF-model}) and
(\ref{eq-spat-temp-RF-model-der-norm-caus}).

Formally, under the classes of primitive geometric image
transformations in Equations~(\ref{eq-spat-sc-transf})--(\ref{eq-temp-sc-transf}),
we have 
the following transformation properties for the purely spatial and
temporal derivative operators:

\begin{description}
\item[\em Uniform spatial scaling transformations:]
  With $\nabla_x = (\partial_{x_1}, \partial_{x_2})^T$ denoting the
  spatial gradient operator, spatial derivatives between the two
  image domains transform according to
  \begin{equation}
    \nabla_{x'} = \frac{1}{S_x} \, \nabla_{x},
  \end{equation}
  implying that directional derivatives over the two image domains defined according
  to
  \begin{align}
 \begin{split}
     \partial_{\varphi} = e_{\varphi}^T \, \nabla_{x},
  \end{split}\\
  \begin{split}
     \partial_{\varphi'} = e_{\varphi'}^T \, \nabla_{x'},
  \end{split}
  \end{align}
  are related according to
  \begin{equation}
    \partial_{\varphi'} = \frac{1}{S_x} \, \partial_{\varphi}.
  \end{equation}

\item[\em Spatial affine transformations:]
  Spatial derivatives between the two image
  domains transform according to%
\footnote{Concerning the notation, we throughout this paper denote the
  transpose of an inverse matrix as $A^{-T} = (A^{-1})^T$.}
  \begin{equation}
    \nabla_{x'} = A^{-T} \, \nabla_{x}.
  \end{equation}
  
\item[\em Galilean transformations:]
  With $\nabla_{(x, t)} = (\partial_{x_1}, \partial_{x_2}, \partial_t)^T$ denoting the
  spatio-temporal gradient operator and with the $3 \times 3$ matrix $G$
  representing the effect of the Galilean
  transformation (\ref{eq-gal-transf}) on the form
  \begin{equation}
    \left(
      \begin{array}{c}
        x_1' \\
        x_2' \\
        t'
      \end{array}
    \right)
    =
    G
   \left(
      \begin{array}{c}
        x_1 \\
        x_2 \\
        t
      \end{array}
    \right)
    =
      \left(
      \begin{array}{c}
        x_1 - u_1 \, t \\
        x_2 - u_2 \, t \\
        t
      \end{array}
    \right),
  \end{equation}
  spatio-temporal derivatives
  between the two image domains transform according to
    \begin{equation}
    \nabla_{(x',  t')} = G^{-T} \, \nabla_{(x, t)},
  \end{equation}
  where
  \begin{equation}
    G =
    \left(
      \begin{array}{ccc}
        1 & 0 & -u_1 \\
        0 & 1 & -u_2 \\
        0 & 0 & 1
      \end{array}
    \right)
  \end{equation}
  and
    \begin{equation}
    G^{-T} =
    \left(
      \begin{array}{ccc}
        1 & 0 & 0 \\
        0 & 1 & 0 \\
        u_1 & u_2 & 1
      \end{array}
    \right).
  \end{equation}
  
\item[\em Temporal scaling transformations:]
  Temporal derivatives between the two
  image domains transform according to
  \begin{equation}
    \partial_{t'} = \frac{1}{S_t} \, \partial_{t}.
  \end{equation}
\end{description}

\noindent
A fundamental limitation of using such pure spatial and temporal
derivative operators in the idealized receptive field models, however,
is that the magnitudes of the corresponding receptive field responses
over the image domain after the geometric image transformation may be strongly
different from the magnitudes of the receptive field responses over the
image domain before the geometric image transformation.
Thereby, it would be very hard to establish a direct matching between the
receptive field responses before and after the geometric image
transformation, based on the magnitudes of the receptive field
responses, thus totally breaking the effect of the matching effects
established by covariance properties of the purely spatial smoothing
operation in (\ref{eq-spat-cov-prop-result-of-proof}) or the joint
spatio-temporal smoothing operation in
(\ref{eq-joint-cov-prop-result-of-proof}).

\subsection{Individual covariance properties for idealized
  models of receptive fields based on scale-normalized spatial and temporal
  derivatives}
\label{sec-indiv-cov-props-spat-temp-ders}

A powerful way of avoiding the problem described in the previous
section,
that the magnitudes of spatial and temporal
derivatives may be strongly influenced by the particular form of the
geometric image transformation, is by instead
introducing {\em scale-normalized derivative operators,\/} whose magnitudes
can be perfectly matched under the influence of geometric image
transformations.

\subsubsection{Scale-normalized spatial derivatives}
\label{sec-iso-norm-ders}

To handle the effect of uniform spatial scaling transformations on
spatial image data, we can introduce scale-normalized spatial
derivatives corresponding to the regular spatial gradient
operator $\nabla_x = (\partial_{x_1}, \partial_{x_2})^T$
according to (Lindeberg \citeyear{Lin97-IJCV}
Equation~(6)) (here simplified by setting the more general scale
normalization parameter $\gamma$ to $\gamma = 1$)
\begin{equation}
  \nabla_{x,\text{norm}} = s^{1/2} \, \nabla_x,
\end{equation}
where $s = \sigma_x^2$ denotes the spatial scale parameter of the here
assumed isotropic Gaussian kernel (with its covariance matrix being
equal to a unit matrix $\Sigma = I$) used for performing the spatial
smoothing. The corresponding scale-normalized directional derivative
operator in the direction $e_{\varphi} = (\cos \varphi, \sin
\varphi)^T$ then becomes
\begin{equation}
  \partial_{\varphi,\text{norm}}
  = s^{1/2} \, \partial_{\varphi}
  = s^{1/2} \, e^T \nabla_x.
\end{equation}
If we define the isotropic spatial scale-space representation
$L \colon \bbbr^2 \times \bbbr_+ \rightarrow \bbbr$ of any
purely spatial image $f \colon \bbbr^2 \rightarrow \bbbr$ according to
\begin{equation}
    \label{eq-def-iso-scsp}
  L(\cdot;\; s) = g(\cdot;\; s, I) * f(\cdot),
\end{equation}
then it can be shown
(Lindeberg \citeyear{Lin97-IJCV} Section~4.1;
Lindeberg \citeyear{Lin25-JMIV} Sections~3.1--3.2)
that under a uniform spatial scaling
transformation of the form (\ref{eq-spat-sc-transf})
\begin{equation}
  f'(x') = f(x) \quad\quad \mbox{for} \quad\quad x' = S_x \, x
\end{equation}
for matching values of
the spatial scale parameters according to (\ref{eq-s-transf-result})
\begin{equation}
  s' = S_x^2 \, s,
\end{equation}
the corresponding scale-normalized derivatives for
the transformed spatial scale-space representation
\begin{equation}
  L'(\cdot;\; s) = g(\cdot;\; s', I) * f'(\cdot),
\end{equation}
will be equal at
corresponding image points $x' = S_x \, x$
according to
\begin{align}
 \begin{split}
    \label{eq-eq-sc-norm-nabla-spat-scsp-spat-sc-transf}
     (\nabla_{x',\norm} L')(x';\; s') & = (\nabla_{x,\norm} L)(x;\; s),
   \end{split}\\
   \begin{split}
    \label{eq-eq-sc-norm-hess-spat-scsp-spat-sc-transf}
    (\nabla_{x',\norm} \nabla_{x',\norm}^T L')(x';\; s')
    & = (\nabla_{x,\norm} \nabla_{x,\norm}^T L)(x;\; s),
   \end{split}\\
  \begin{split}
    \label{eq-eq-sc-norm-dirders-spat-scsp-spat-sc-transf}    
     L'_{{\varphi'}^m,\norm}(x';\; s') & = L_{\varphi^m,\norm}(x;\; s).
  \end{split}
\end{align}
Here, the first expression
(\ref{eq-eq-sc-norm-nabla-spat-scsp-spat-sc-transf}) represents
(regular) scale-norm\-alized gradient operators over the domains after
and before the spatial scaling transformation.
The second expression
(\ref{eq-eq-sc-norm-hess-spat-scsp-spat-sc-transf}) represents
(regular) scale-normalized Hessian operators $({\cal H}_{x'} L')(x';\; s')$
and $({\cal H}_x L)(x;\; s)$ over the domains after and before the image
transformation.
The third expression
(\ref{eq-eq-sc-norm-dirders-spat-scsp-spat-sc-transf}) represents
(regular) scale-normalized directional derivative operators of order
$m$ over the domains after and before the geometric transformation.

By this use of (regular) scale-normalized spatial derivatives, the spatial and
the spatio-temporal receptive fields according to the generalized
Gaussian derivative model will be provably covariant under
uniform spatial scaling transformations of the form
(\ref{eq-spat-sc-transf}). In this way, a visual system built from
such computational primitives will be able to handle objects of
different size in the world as well as at different distances to the
visual observer in a similar manner, as previously extensively
explored to compute scale-covariant and scale-invariant image
representations in the area of classical computer vision
(Lindeberg \citeyear{Lin21-EncCompVis}).

\subsubsection{Affine-normalized spatial derivatives}
\label{sec-aff-norm-ders}

To handle the effect of more general spatial affine transformations
\begin{equation}
  f'(x') = f(x) \quad\quad \mbox{for} \quad\quad x' = A \, x
\end{equation}
on spatial image data, one needs to make use
of different spatial covariance matrices $\Sigma$ and $\Sigma'$ for the spatial receptive
fields before and after the spatial affine transformation.
For this purpose, three main notions%
\footnote{Notably, in the special case when the spatial covariance
  matrix $\Sigma$ in the affine Gaussian kernel is equal to a unit
  matrix $\Sigma = I$, the affine Gaussian kernel in
  (\ref{eq-def-aff-scsp}) reduces to the isotropic Gaussian matrix in
  (\ref{eq-def-iso-scsp}), in turn implying that the affine
  scale-normalized directional derivative operator in
  (\ref{eq-def-sc-norm-aff-grad-op}) reduces to the regular
  scale-normalized directional derivative operator used in
  (\ref{eq-eq-sc-norm-dirders-spat-scsp-spat-sc-transf}).
  Furthermore,  when $\Sigma = I$ the scale-normalized affine gradient operator
  in (\ref{eq-def-sc-norm-aff-grad-op}) similarly reduces to the regular
  scale-normalized gradient operator used in
  (\ref{eq-eq-sc-norm-nabla-spat-scsp-spat-sc-transf}). Similarly,
  when $\Sigma = I$ the scale-normalized affine Hessian operator
  in (\ref{eq-def-sc-norm-aff-hess-mat}) reduces to the regular
  scale-normalized Hessian operator used in
  (\ref{eq-eq-sc-norm-hess-spat-scsp-spat-sc-transf}).
  In these respects, the affine-normalized derivative operators in
  Section~\ref{sec-aff-norm-ders} constitute generalizations of the
  previously used regular (isotropic) scale-normalized derivative operators in
  Section~\ref{sec-iso-norm-ders}, from an isotropic spatial
  scale space representation generated by convolution with
  rotationally symmetric Gaussian kernels to an anisotropic affine
  Gaussian scale space generated by convolution with anisotropic
  affine Gaussian kernels.}
of affine-normalized spatial derivatives
have been proposed in Lindeberg (\citeyear{Lin25-JMIV})
Sections~3.3--3.8.

The theoretical background to these affine-normalized spatial
derivatives is that we consider
the spatial affine scale-space representation
$L \colon \bbbr^2 \times \bbbr_+ \times \bbbs_+^2 \rightarrow \bbbr$
of any 2-D purely spatial image $f \colon \bbbr^2 \rightarrow \bbbr$
defined according to
\begin{equation}
  \label{eq-def-aff-scsp}
  L(\cdot;\; s, \Sigma) = g(\cdot;\; s, \Sigma) * f(\cdot),
\end{equation}
generated by convolution with affine Gaussian kernels
$g \colon \bbbr^2 \times \bbbr_+ \times \bbbs_+^2 \rightarrow \bbbr$
according to (\ref{eq-gauss-fcn-2D}) and having spatial
covariance matrices $\Sigma$ not necessarily equal to a unit matrix,
with its corresponding transformed affine spatial scale-space
representation
\begin{equation}
  \label{eq-def-aff-scsp-prim}
  L'(\cdot;\; s', \Sigma') = g(\cdot;\; s', \Sigma') * f(\cdot),
\end{equation}
which obeys the affine covariance property
(Lindeberg and G{\aa}rding \citeyear{LG96-IVC} Equation~(29))
\begin{equation}
  L'(x';\; s', \Sigma') = L(x;\; s, \Sigma)
\end{equation}
for matching values of the scale parameters $s$ and $s'$ and the
spatial covariance matrices $\Sigma$ and $\Sigma'$ according to
 (Lindeberg \citeyear{Lin25-JMIV} Equation~(118))
 \begin{equation}
   s' \, \Sigma' = s \, (S_x \, A) \, \Sigma \, (S_x A)^T = s \, S_x^2 \, A \, \Sigma \, A^T.
 \end{equation}
Then, given these affine Gaussian scale-space representations over the
domains before and after the affine transformation, affine-normalized
spatial derivative operators can be defined as follows:

\begin{description}
\item[\em The affine scale-normalized directional
  derivative:]
  A straightforward way of defining an extension of scale-normalized
  derivatives when using a spatial covariance matrix not equal to the
  unit matrix is according to
  (Lindeberg \citeyear{Lin25-JMIV} Equation~(33)) 
  \begin{equation}
    \label{eq-aff-sc-norm-dir-der}
    \partial_{\varphi,\norm}^m
    = s^{m/2} \, (e_{\varphi}^T \, \Sigma \, e_{\varphi})^{m/2} \, \partial_{\varphi}^m,
  \end{equation}
  where the entity $e_{\varphi}^T \, \Sigma \, e_{\varphi}$ 
  reflects the amount of spatial smoothing in the direction
  $e_{\varphi}$, see Figure~\ref{fig-geom-ill-aff-sc-norm-dir-der} for
  an illustration.

  \begin{figure}[bt]
    \vspace{-10mm}
    \begin{center}
      \includegraphics[width=0.70\textwidth]{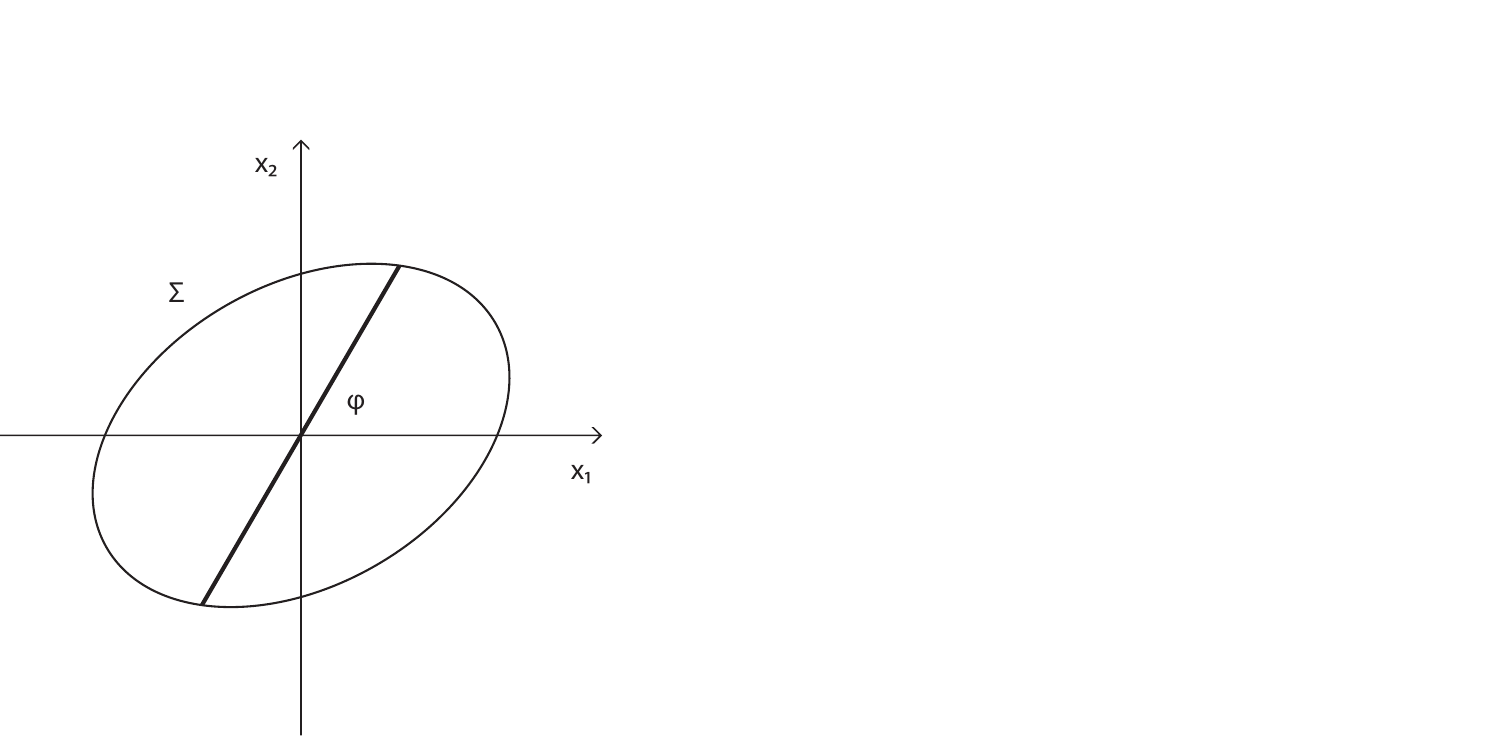}
    \end{center}
    \caption{Illustration of the definition of the scale-normalization
      factor $e_{\varphi}^T \, \Sigma \, e_{\varphi}$ in the
      definition of the affine scale-normalized derivative
      operator according to (\ref{eq-aff-sc-norm-dir-der}). For
      performing the scale normalization in the direction
      $e_{\varphi} = (\cos \varphi, \sin \varphi)$ of the spatial
      covariance matrix $\Sigma$, an ellipse representation of the
      spatial covariance matrix $\Sigma$ is intersected in that direction, thus
      projecting the spatial smoothing effect of corresponding affine
      Gaussian kernel in the direction of the directional derivative operator.
      (Figure reproduced from Lindeberg (\citeyear{Lin25-JMIV}) with
      permission (Open Access).)}
    \label{fig-geom-ill-aff-sc-norm-dir-der}
  \end{figure}

  In Lindeberg (\citeyear{Lin25-JMIV}) Section~3.4, it is shown that
  this notion of affine scale-normal\-ized directional derivatives is
  covariant under the similarity group, that is under combinations of
  uniform spatial scaling transformations and rotations.
  This notion of affine scale-normalized directional derivatives is
  also covariant in the special configuration when the affine
  transformation matrix $A$ and the spatial covariance matrix $\Sigma$
  have the same eigenvectors, with the geometric interpretation
  that such a configuration corresponds to varying the viewing
  direction along the tilt%
\footnote{The tilt direction is the projection of the local surface
  normal onto the image plane.}
  direction of an observed local surface
  patch. Thus, for these subgroups of the group
  of spatial affine transformations, the affine scale-normalized 
  directional derivatives will be equal over the domains before and
  after these special forms of spatial affine transformations:
  \begin{equation}
    \partial_{\varphi',\norm}^m L'(x';\; s', \Sigma')
    = \partial_{\varphi,\norm}^m L(x;\; s, \Sigma).
  \end{equation}
  As shown in Lindeberg (\citeyear{Lin25-JMIV}) Section~3.4.5,
  the affine scale-normalized
  directional derivatives are, however, not covariant
  under fully general affine transformations.
  \medskip

  \begin{figure}[bt]
    \begin{center}
      \includegraphics[width=0.40\textwidth]{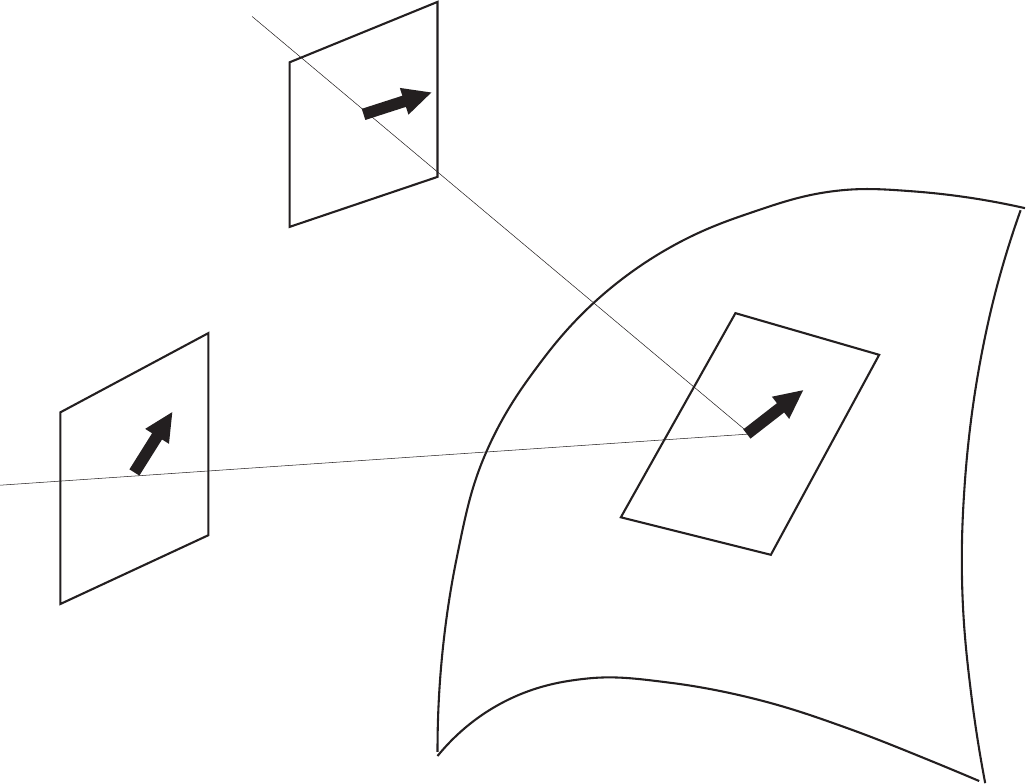}
    \end{center}
    \caption{Illustration of the covariance property
      (\ref{eq-equal-aff-scsp-repr-aff-cov-proof-again}) of the
      scale-normalized affine gradient operator according to
      (\ref{eq-def-sc-norm-aff-grad-op}) under
      general (non-singular) affine transformations.
      The interpretation of this covariance property is that, if we
      consider two cameras, that view the same local surface patch from
      general (non-degenerate) viewing conditions, then, to first order
      of approximation, the resulting affine gradient responses for the
      different views, here illustrated as arrows before the affine
      scale normalization, can, up
      to a rotation transformation $\tilde{\rho}$, be perfectly
      matched, provided that the scale parameters and the covariance
      matrices of the receptive fields are properly matched.
      (Figure reproduced from Lindeberg (\citeyear{Lin25-JMIV}) with
      permission (Open Access).)}
    \label{fig-geom-int-cov-prop-sc-norm-aff-grad}
  \end{figure}
  
\item[\em The scale-normalized affine gradient:]
  Given an eigenvalue decomposition of the $2 \times 2$
  symmetric and positive definite spatial
  covariance matrix $\Sigma$ of the form
  \begin{equation}
    \label{eq-eigen-decomp-Sigma-aff-grad}
    \Sigma
    = U \Lambda \, U^T,
  \end{equation}
  where $\Lambda = \diag(\lambda_1, \lambda_2)$ is a $2 \times 2$ diagonal matrix
  with positive elements, and $U$ is a $2 \times 2$  real unitary
  matrix, let us define the Cholesky-type principal square root of $\Sigma$ as
  \begin{equation}
    \label{eq-def-sqrt-of-Sigma}
    \Sigma^{1/2} = \Lambda^{1/2}\, U^T,
  \end{equation}
  where $\Lambda^{1/2} = \diag(\lambda_1^{1/2}, \lambda_2^{1/2})$.
  Let us, however, note that the definition of the
  Cholesky-type principal square root of a $2 \times 2$ matrix is
  not unique, since for any arbitrary $2 \times 2$ rotation
  matrix $\rho$, also the matrix
  \begin{equation}
    \label{eq-general-sq-root-mat}
    \Sigma^{1/2} = \rho \, \Lambda^{1/2}\, U^T,
  \end{equation}
  satisfies
  \begin{equation}
    (\Sigma^{1/2})^T (\Sigma^{1/2})
    = U \, \Lambda^{1/2} \, \rho^T \, \rho \, \Lambda^{1/2} \, U^T
    = U \Lambda \, U^T.
  \end{equation}
  Given this definition of the Cholesky-type principal square root of the spatial
  covariance matrix $\Sigma$, we can define the scale-normalized affine
  gradient operator as (Lindeberg \citeyear{Lin25-JMIV} Equation~(111))
  \begin{equation}
    \label{eq-def-sc-norm-aff-grad-op}
    \nabla_{x,\affnorm} = s^{1/2} \, \Sigma^{1/2} \, \nabla_x.
  \end{equation}
  Under a spatial affine transformation of the form
  (\ref{eq-spat-aff-transf}), it is shown in
  Lindeberg (\citeyear{Lin25-JMIV}) Section~3.6 that the scale-normalized
  affine gradient operator over the transformed domain
  $\nabla_{x',\affnorm}$ is up to a rotation matrix $\tilde{\rho}$
  related to the scale-normalized affine
  gradient operator $\nabla_{x,\affnorm}$ over the original domain
  according to (Lindeberg \citeyear{Lin25-JMIV} Equation~(132))
  \begin{equation}
    \nabla_{x',\affnorm} = \tilde{\rho} \, s^{1/2} \, \Sigma^{1/2} \, \nabla_x.
  \end{equation}
  Thereby, the scale-normalized affine gradient vectors
  $\nabla_{x,\affnorm} L$ and $\nabla_{x',\affnorm}  L'$
  computed from an affine scale-space representation of the form (\ref{eq-def-aff-scsp})
  over the domains before and after the affine transformation are
  related according to
  \begin{equation}
    \label{eq-equal-aff-scsp-repr-aff-cov-proof-again}
    (\nabla_{x',\affnorm}  L')(x';\; s', \Sigma') =
    \tilde{\rho} \,  (\nabla_{x,\affnorm} L)(x;\; s, \Sigma),
  \end{equation}
  see Figure~\ref{fig-geom-int-cov-prop-sc-norm-aff-grad} for an
  illustration.
  
  In the special case when the affine transformation $A$ is in the
  similarity group, it is shown in Lindeberg
  (\citeyear{Lin24-arXiv-UnifiedJointCovProps}) Section~3.6 that the
  rotation matrix $\tilde{\rho}$ reduces to a unit matrix.
  In this special case, the scale-normalized affine gradients over the
  two domains before and after the image transformation are
  therefore guaranteed to be equal.
  
  \medskip

  \begin{figure}[bt]
    \begin{center}
      \includegraphics[width=0.40\textwidth]{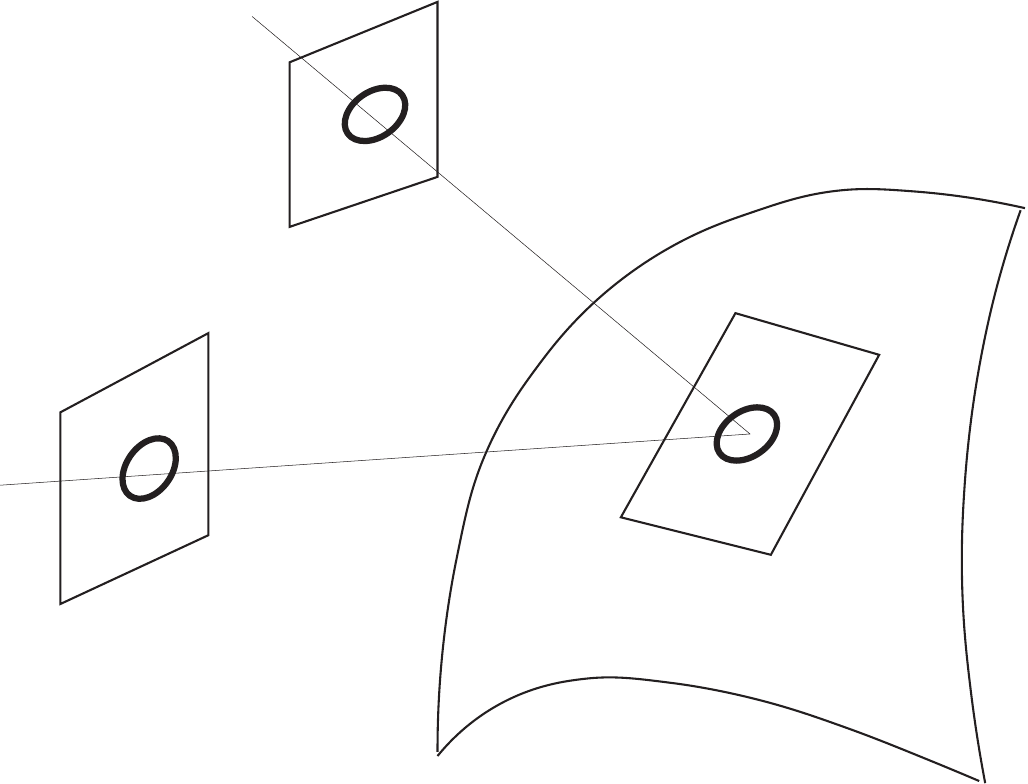}
    \end{center}  
    \caption{Illustration of the covariance property
      (\ref{eq-equal-aff-scsp-repr-aff-cov-proof-hess-again}) of the
      scale-normalized affine Hessian operator according to
      (\ref{eq-def-sc-norm-aff-hess-mat}) under
      general (non-singular) affine transformations.
      This covariance property means that, if we
      consider two cameras, that view the same local surface patch from
      general (non-degenerate) viewing conditions, then, to first order
      of approximation, the resulting affine Hessian responses for the
      different views, here illustrated as ellipses before the affine
      scale normalization, can, up
      to a combination of two (in this 2-D case related)
      rotation transformations $\tilde{\rho}$
      and $\tilde{\rho}^T$, be perfectly
      matched, provided that the scale parameters and the covariance
      matrices of the receptive fields are properly matched.
      (Figure reproduced from Lindeberg (\citeyear{Lin25-JMIV}) with
      permission (Open Access).)}
    \label{fig-geom-int-cov-prop-sc-norm-aff-hess}
  \end{figure}
  
\item[\em The scale-normalized affine Hessian:]
  To extend the above notion from first- to second-order spatial
  derivatives, we can define a corresponding scale-normalized
  affine Hessian operator ${\cal H}_{x,\affnorm}$ according to
  (Lindeberg \citeyear{Lin25-JMIV} Equation~(140))
  \begin{equation}
    {\cal H}_{x,\affnorm}
    = \nabla_{x,\affnorm} \, \nabla^T_{x,\affnorm},
  \end{equation}
  which, when expanded, then assumes the form
  \begin{align}
    \begin{split}
      \label{eq-def-sc-norm-aff-hess-mat}
      {\cal H}_{x,\affnorm}
      & = s \, (\Sigma^{1/2}) \, \nabla_x  \, \nabla_x^T (\Sigma^{1/2})^T.
    \end{split}
  \end{align}
  Under a spatial affine transformation of the form
  (\ref{eq-spat-aff-transf}), it can be shown that this operator
  transforms according to the following, between the domains before
  and after the image transformation
  (Lindeberg \citeyear{Lin25-JMIV} Equation~(152))
  \begin{equation}
    \label{eq-equal-aff-scsp-repr-aff-cov-proof-hess-again}
    {\cal H}_{x',\affnorm} = \tilde{\rho} \,  {\cal H}_{x,\affnorm} \,
    \tilde{\rho}^T,
  \end{equation}
  where again $\tilde{\rho}$ denotes a rotation matrix.
  Thereby, the scale-normalized affine Hessian matrices
  computed from the affine  Gaussian
  scale-space representations $L(x;\; s, \Sigma)$ and $L'(x';\; s', \Sigma')$
  over the domains before and after the image transformation are
  related according to
  \begin{multline}
    ({\cal H}_{x',\affnorm} L')(x';\; s', \Sigma') = \\
    = \tilde{\rho} \, ({\cal H}_{x,\affnorm} L)(x;\; s, \Sigma) \, \tilde{\rho}^T,
  \end{multline}
  thus showing that the set of second-order spatial derivatives before
  and after the image transformation can up
  to an indeterminacy with respect to a possibly unknown rotation
  matrix be perfectly matched, see
  Figure~\ref{fig-geom-int-cov-prop-sc-norm-aff-hess} for an illustration.

  Again, if the affine transformation matrix $A$ is
  in the similarity group, the rotation matrix $\rho$ reduces to a
  unit matrix.
\end{description}
In these ways, we can thus match receptive field responses
formulated in terms of spatial derivatives of covariant spatial
smoothing kernels
under the spatial affine transformations that arise when viewing the
same local surface patch from different viewing directions, with a few technical differences
depending on the types of spatial derivative expression and the
generality of the types of spatial affine transformations.

\subsubsection{Scale-normalized temporal derivatives}

To handle the effect of temporal scaling transformations, that model the
effect of temporal or spatio-temporal events occurring either faster or
slower relative to a reference view, we can introduce
scale-normalized temporal derivatives according to
Lindeberg (\citeyear{Lin17-JMIV}) Equation~(6)
\begin{equation}
  \label{eq-temp-der-def-sc-norm}
  \partial_{t,\norm}^n = \tau^{n/2} \, \partial_t^n,
\end{equation}
where $\tau = \sigma_t^2$ denotes the temporal scale parameter in the temporal
scale-space representation
$L \colon \bbbr \times \bbbr_+ \rightarrow \bbbr$
of a temporal signal $f \colon \bbbr \rightarrow \bbbr$ according to
\begin{equation}
  L(\cdot;\; \tau) = h(\cdot;\, \tau) * f(\cdot),
\end{equation}
with the temporal smoothing kernel
$h \colon \bbbr \times \bbbr_+ \rightarrow \bbbr$
being  either the non-causal 1-D Gaussian kernel according to
(\ref{eq-non-caus-temp-gauss-var}) or
the time-causal limit kernel according to
(\ref{eq-time-caus-lim-kern-var}).

With these definitions (and with the more general scale normalization power
$\gamma$ in Lindeberg (\citeyear{Lin17-JMIV}) set to $\gamma = 1$),
the resulting scale-normalized derivatives of the transformed
temporal scale-space representation
\begin{equation}
  L'_{{t'}^n,\text{norm}}(t';\; \tau')
  = {\tau'}^{n/2} \, \partial_{t^n} (h(\cdot;\, \tau') *  f'(\cdot))(t';\; \tau')
\end{equation}
do under a temporal scaling
transformation of the form (\ref{eq-temp-sc-transf})
\begin{equation}
  f'(t') = f(t) \quad\quad \mbox{for} \quad\quad t' = S_t \, t
\end{equation}
become equal%
\footnote{When using the time-causal limit kernel $\Psi(t;\; \tau, c)$
  for temporal smoothing, this relationship is, however, due to the
  discrete temporal scale levels of the time-causal receptive field representation,
  only exact for temporal scaling factors $S_t$ that are integer
  powers of the distribution parameter $c$ according to $S_t = c^j$
  for some $j \in \bbbz$. For other values of the temporal scaling
  factor $S_t$, the corresponding relation will instead only be approximate.}
at corresponding temporal moments $t' = S_t \, t$ according to
(Lindeberg \citeyear{Lin17-JMIV} Equation~(10))
\begin{equation}
   L'_{{t'}^n,\norm}(t';\; \tau') = L_{t^n,\norm}(t;\; \tau)
\end{equation}
for matching values of
the temporal scale parameters $\tau$ and $\tau'$ over the domains
before and after the temporal scaling transformation according to
(\ref{eq-tau-transf-result})
\begin{equation}
  \tau' = S_t^2 \, \tau.
\end{equation}
In this way, a vision system based on temporal filtering with
scale-normalized temporal derivatives of either the non-causal Gaussian kernel
or the time-causal limit kernel as the temporal
smoothing kernel will be able to handle spatio-temporal events that
occur either faster or slower between different views of an otherwise similar
type of spatio-temporal event.

\subsubsection{Scale-normalized velocity-adapted temporal derivatives}

To handle the effect of Galilean transformations on spatio-temporal
image data, we can extend the notion of scale-normalized temporal
derivatives according to (\ref{eq-temp-der-def-sc-norm}) into
scale-normalized velocity-adapted temporal derivatives according to
(Lindeberg \citeyear{Lin25-JMIV} Equation~(168))
\begin{equation}
  \label{eq-vel-adapt-der-def-sc-norm}
  \partial_{{\bar t},\norm}^n
  = \tau^{n/2} \,  (v^T \, \nabla_x + \partial_t)^n.
\end{equation}
Then, under the simultaneous application of a Galilean transformation
of the form (\ref{eq-gal-transf}) with potentially both spatial and temporal
scaling transformations according to (\ref{eq-spat-sc-transf})
and (\ref{eq-temp-sc-transf}) 
\begin{align}
  \begin{split}
     \label{eq-x-transf-noaff}
     x' = S_x \, (x + u \, t),
   \end{split}\\
  \begin{split}
     \label{eq-t-transf-noaff}
     t' = S_t \, t,
   \end{split}
\end{align}
it holds that if we define a joint spatio-temporal scale-space
representation
$L \colon \bbbr^2 \times \bbbr \times \bbbr_+ \times \bbbr_+
\times \bbbr^2 \rightarrow \bbbr$
by convolving any video sequence or video stream
$f \colon \bbbr^2 \times \bbbr$ with the spatio-temporal smoothing kernel
$T \colon \bbbr^2 \times \bbbr \times \bbbr_+ \times \bbbr_+
\times \bbbr^2 \rightarrow \bbbr$ according to
\begin{equation}
  \label{eq-spat-temp-RF-model-again-cov-props-basic-noaff}
  T(x, t;\; s, \tau, v) 
  = g(x - v \, t;\; s, I) \, h(t;\; \tau),
\end{equation}
then if we define the
spatio-temporal scale-space representations $L$ and $L'$
before and after the composed geometric image transformation
\begin{equation}
  f'(x', t') = f(x, t)
\end{equation}
according
to
\begin{align}
  \begin{split}
    L(\cdot, \cdot;\, s, \tau, v)
    = T(\cdot, \cdot;\, s, \tau, v) * f(\cdot; \cdot),
  \end{split}\\
  \begin{split}
    L'(\cdot, \cdot;\, s', \tau', v')
    = T(\cdot, \cdot;\, s', \tau', v') * f'(\cdot; \cdot),
  \end{split}
\end{align}
the corresponding scale-normalized velocity-adapted temporal derivatives
can be perfectly matched according to
\begin{equation}
  \label{eq-joint-cov-prop-result-of-proof-again}
  L'_{{\bar {t'}^n},\norm}(x', t';\; s', \tau' , v') = L_{{\bar t}^n,\norm}(x, t;\; s, \tau, v),
\end{equation}
provided that the filter parameters over the domains before and after
the composed geometric image transformation are matched according to
\begin{align}
  \begin{split}
    \label{eq-s-transf-result-noaff}
    s' & = S_x^2 \, s,
  \end{split}\\
  \begin{split}
    \label{eq-tau-transf-result-noaff}
    \tau' & = S_t^2 \, \tau,
  \end{split}\\
  \begin{split}
    \label{eq-v-transf-result-noaff}    
    v' & = \frac{S_x}{S_t} (v + u).
  \end{split}
\end{align}
In this way, a visual system based on spatio-temporal receptive fields
that comprise velocity-adapted receptive based on such a velocity
parameter $v$ in both the spatial smoothing kernel $T$ and the
velocity-adapted temporal derivative operators $\partial_{{\bar  t},\norm}^n$
will have the ability to handle different types of relative motions,
as parameterized by the velocity parameter $u$ between the viewing
direction and the observer.

Note, however, that it is, in general, not sufficient to include a
variability with respect to only the image velocity parameter $v$
in the model. Since the value of the velocity parameter may be changed during
the image transformation, also a variability is needed concerning the
transformation property of the ratio between the spatial and the temporal scale parameters
$s$ and $\tau$.
Thereby, it is necessary to also consider the potential
interaction effects between the different types of primitive geometric
image transformations
(\ref{eq-spat-sc-transf})--(\ref{eq-temp-sc-transf}) when modelling
the combined effect of composed spatio-temporal image transformations.

\subsection{Joint covariance properties for receptive fields in terms
  of spatial and spatio-temporal derivatives under the
  composed geometric transformation model}
\label{sec-joint-cov-props-spat-spattemo-rfs-comp-transf}

Let us next consider composed spatio-temporal image transformations
according to (\ref{eq-x-transf}) and (\ref{eq-t-transf}) for the
monocular {\em joint\/} projection model
\begin{align}
  \begin{split}
     \label{eq-x-transf-mon}
     x' = S_x \, (A \,  x + u \, t),
   \end{split}\\
  \begin{split}
     \label{eq-t-transf-mon}
     t' = S_t \, t.
   \end{split}
\end{align}
and according to (\ref{eq-sc-aff-vel-transf-alt-obs-model}) and
(\ref{eq-t-transf}) for the corresponding joint binocular projection model
\begin{align}
  \begin{split}
     \label{eq-x-transf-bin}
     x' = \tilde{B} \, x + \tilde{u} \, t,
   \end{split}\\
  \begin{split}
     \label{eq-t-transf-bin}
     t' = S_t \, t.
   \end{split}
\end{align}
Then, it follows that: 
\begin{itemize}
\item
  If we define spatio-temporal receptive fields defined based
  on to the affine scale-normalized directional
  derivative operator according to (\ref{eq-aff-sc-norm-dir-der})
    \begin{equation}
    \label{eq-aff-sc-norm-dir-der-again}
    \partial_{\varphi,\norm}^m
    = s^{m/2} \, (e_{\varphi}^T \, \Sigma \, e_{\varphi})^{m/2} \, \partial_{\varphi}^m,
  \end{equation}
  where $e_{\varphi}$ denotes the unit vector in the direction $\varphi$,
  and the scale-normalized velocity-adapted temporal derivative operator
  according to (\ref{eq-vel-adapt-der-def-sc-norm})
  \begin{equation}
    \label{eq-vel-adapt-der-def-sc-norm-again}
    \partial_{{\bar t},\norm}^n
    = \tau^{n/2} \,  (v^T \, \nabla_x + \partial_t)^n,
  \end{equation}
  then the composed
  spatio-temporal derivatives will for compositions of spatial
  transformations within the similarity group, for which the affine
  transformation matrix reduces to a rotation matrix $A = R_{\theta}$, Galilean
  transformations and temporal scaling transformations be equal at corresponding
  spatio-temporal image points
    \begin{multline}
    \label{eq-sc-norm-dir-der-joint-cov-prop-sc-norm-ders}
    \partial_{{\varphi'}^m,\norm} \, \partial_{\bar{t'},\norm}^n L'(x', t';\; s', \Sigma', \tau', v') = \\
    = \partial_{\varphi^m,\norm} \, \partial_{\bar{t},\norm}^n L(x, t;\; s, \Sigma, \tau, v),
  \end{multline}
    provided that the other parameters of the receptive fields are
  matched according to
  (\ref{eq-s-transf-result})--(\ref{eq-v-transf-result})
  such that
  (Lindeberg \citeyear{Lin25-JMIV} Equations~(277)--(281))
  \begin{align}
    \begin{split}
      s' & = S_x^2 \, s,
    \end{split}\\
    \begin{split}    
      \varphi' & = \varphi + \theta,
    \end{split}\\
    \begin{split}
      \Sigma' & = R_{\theta} \, \Sigma \, R_{\theta}^T,
    \end{split}\\
  \begin{split}
    \tau' & = S_t^2 \, \tau,
  \end{split}\\
  \begin{split}
    v' & = \frac{S_x}{S_t} (R_{\theta} \, v + u).
  \end{split}    
  \end{align}
\item
  If we consider the group of general
  affine transformation matrices $A$, and
  define the scale-normalized affine gradient vector
  according to (\ref{eq-def-sc-norm-aff-grad-op})
  \begin{equation}
    \nabla_{x,\affnorm} = s^{1/2} \, \Sigma^{1/2} \, \nabla_x,
  \end{equation}  
  with the scale-normalized affine Hessian operator ${\cal H}_{x,\affnorm}$
  defined from the regular Hessian operator $ {\cal H}_x = \nabla_x \nabla_x^T$
  according to (\ref{eq-def-sc-norm-aff-hess-mat})
  \begin{equation}
    {\cal H}_{x,\affnorm} = s \, (\Sigma^{1/2}) \, {\cal H}_x \, (\Sigma^{1/2})^T,
  \end{equation}
  and the scale-normalized velocity-adapted temporal derivative operator
  according to (\ref{eq-vel-adapt-der-def-sc-norm})
  \begin{equation}
    \label{eq-vel-adapt-der-def-sc-norm-again2}
    \partial_{{\bar t},\norm}^n
    = \tau^{n/2} \,  (v^T \, \nabla_x + \partial_t)^n,
  \end{equation}
  then under the composed geometric image transformation
  given by (\ref{eq-x-transf-mon}) and (\ref{eq-t-transf-mon}),
  the resulting composed spatio-temporal receptive field
  responses will be equal up to a rotation matrix $\tilde{\rho}$ according to
  \begin{multline}
    \label{eq-cov-prop-sc-norm-aff-grad-summ-overview}
    (\nabla_{x',\affnorm}  \, \partial_{\bar{t'},\norm}^n L')(x', t';\; s', \Sigma', \tau', v') = \\ =
    \tilde{\rho} \,  (\nabla_{x,\affnorm} \, \partial_{\bar{t},\norm}^n L)(x, t;\; s, \Sigma, \tau, v)
  \end{multline}
  and
  \begin{multline}
    \label{eq-cov-prop-sc-norm-aff-hess-summ-overview}
    ({\cal H}_{x',\affnorm}  \, \partial_{\bar{t'},\norm}^n L')(x', t';\; s', \Sigma', \tau', v') = \\
    = \tilde{\rho} \, ({\cal H}_{x,\affnorm} \, \partial_{\bar{t},\norm}^n L)(x, t;\; s, \Sigma, \tau, v) \, \tilde{\rho}^T,
  \end{multline}
  provided that the scale parameters $s$ and $s'$ as well as the
  spatial covariance matrices $\Sigma$ and $\Sigma'$ are matched
  according to (Lindeberg \citeyear{Lin25-JMIV} Equation~(118))
  \begin{equation}
     s' \, \Sigma' = s \, (S_x \, A) \, \Sigma \, (S_x A)^T = s \, S_x^2 \, A \, \Sigma \, A^T,
   \end{equation}
  and provided that the other parameters of the receptive fields are
  matched according to
  (\ref{eq-tau-transf-result})--(\ref{eq-v-transf-result})
  \begin{align}
    \begin{split}
      \label{eq-tau-transf-result-again-expl-stat}
      \tau' & = S_t^2 \, \tau,
    \end{split}\\
    \begin{split}
      \label{eq-v-transf-result-again-expl-stat}
      v' & = \frac{S_x}{S_t} (A \, v + u).
    \end{split}
  \end{align}
\item
  Irrespective of any restrictions on the family of affine
  transformation matrices $A$, the velocity-adapted temporal
  derivative operators according to (\ref{eq-vel-adapt-der-def-sc-norm})
  will be equal
  (Lindeberg \citeyear{Lin25-JMIV} Equation~(291))
  \begin{multline}
    \label{eq-equal-veladapt-ders-composed-transf-main-result-sc-norm}
    \partial_{{\bar t}',\norm}^n L' (x', t';\; s', \Sigma', \tau', v')  = \\
    =  \partial_{{\bar t},\norm}^n L(x, t;\; s, \Sigma, \tau, v),
  \end{multline}
  provided that the parameters $s$, $s'$, $\Sigma$, $\Sigma'$,
  $\tau$, $\tau'$, $v$ and $v'$ of the receptive fields are matched
  according to Equations~(\ref{eq-s-transf-result})--(\ref{eq-v-transf-result}).
\end{itemize}
While the above results have been formulated based on the monocular
projection model (\ref{eq-x-transf-mon})--(\ref{eq-t-transf-mon}),
corresponding results for the binocular projection model
(\ref{eq-x-transf-bin})--(\ref{eq-t-transf-bin}) can be obtained by
setting%
\footnote{Please note, however, that this algebraic similarity does
  not imply that there is no scaling in the binocular case. Instead,
  the scaling factor is absorbed in the matrix $B$ for the locally
  linearized binocular projection model.}
the uniform spatial scaling factor to $S_x = 1$ and then
replacing the affine transformation matrix $A$ by the affine
transformation matrix $\tilde{B}$ in
Equations~(\ref{eq-cov-prop-sc-norm-aff-grad-summ-overview})--(\ref{eq-equal-veladapt-ders-composed-transf-main-result-sc-norm}).

\begin{figure*}[hbtp]
  \[
    \begin{CD}
       \hspace{0mm}\nabla_{x,\affnormtiny} \partial_{{\bar t},\normtiny} L(x, t;\; s, \Sigma, \tau, v)
       @>{\footnotesize\begin{array}{c} x' = S_x (A \, x + u \, t)  \\ t' =
                         S_t \, t  \\ s' = S_x^2 \, s \\ \Sigma' = A \,
                         \Sigma \, A^{T} \\ \tau' = S_t^2 \, \tau \\
                         v' = \frac{S_x}{S_t} (A \, v + u) \\
                         \nabla_{x,\affnormtiny} = s^{1/2} \,
                         \Sigma^{1/2} \, \nabla_x \\ \nabla_{x',\affnormtiny} = {s'}^{1/2} \,
                         {\Sigma'}^{1/2} \, \nabla_{x'} \\ \nabla_{x',\affnormtiny} =
    \tilde{\rho} \, \nabla_{x,\affnormtiny} \\ \partial_{{\bar t}',\normtiny} = 
                         \partial_{{\bar t},\normtiny} \end{array}}>>
                     \nabla_{x',\affnormtiny} \partial_{{\bar t}',\normtiny} L'(x', t';\; s', \Sigma', \tau', v') \\
       \Big\uparrow\vcenter{\rlap{$\scriptstyle{{*
               (\nabla_{x,\affnormtiny} \partial_{{\bar t},\normtiny} T)(x, t;\; s,
               \Sigma, \tau, v)}}$}} & &
       \Big\uparrow\vcenter{\rlap{$\scriptstyle{{*(
               \nabla_{x',\affnormtiny} \partial_{{\bar t}',\normtiny} T)(x', t';\; s',
               \Sigma', \tau', v')}}$}} \\
       f(x, t) @>{\footnotesize \begin{array}{c} x' = S_x (A \, x + u \, t)
                                  \\ t' = S_t \, t \end{array}}>> f'(x', t')
    \end{CD}
  \]
\caption{Commutative diagram for scale-normalized spatio-temporal derivative operators
  defined from the joint spatio-temporal receptive field model
  (\ref{eq-spat-temp-RF-model-scnormaff-veladapt}) under the
  composition of
  (i)~a spatial scaling transformation, (ii)~a spatial affine transformation,
  (iii)~a Galilean transformation and (iv)~a temporal scaling transformation
  according to (\ref{eq-x-transf}) and (\ref{eq-t-transf}).
  This   commutative diagram, which should be read from the lower left
  corner to the upper right corner, means that irrespective of whether the input video 
  sequence or video stream $f(x, t)$ is first subject to the composed transformation
  $x' = S_x (A \,  x + u \, t)$ and $t' = S_t \, t$
  and then filtered with a scale-normalized spatio-temporal derivative kernel
  $(\nabla_{x',\affnorm} \partial_{t',\norm} T)(x', t';\; s', \Sigma', \tau', v')$,
  or instead directly convolved
  with the scale-normalized spatio-temporal smoothing kernel
  $(\nabla_{x,\affnorm} \partial_{t,\norm} T)(x, t;\; s, \Sigma, \tau, v)$ and then
  subject to the same joint spatio-temporal transformation, we do
  then, up to a possibly unknown rotation transformation $\tilde{\rho}$, get the same
  result, provided that the parameters of the spatio-temporal
  smoothing kernels are related
  according to $s' = S_x^2 \, s$, $\Sigma' = A \, \Sigma \, A^{T}$,
  $\tau' = S_t^2 \, \tau$ and $v' = \frac{S_x}{S_t} (A \, v + u)$.
  (Adapted from Lindeberg (\citeyear{Lin25-JMIV}) (Open Access).)}
\label{fig-comm-diag-spat-temp-ders-sc-norm}

\bigskip

  \[
    \begin{CD}
       \hspace{0mm}\nabla_{x,\affnormtiny} \partial_{{\bar t},\normtiny} L(x, t;\; \tilde{\Sigma}, \tilde{\tau}, \tilde{v})
       @>{\footnotesize\begin{array}{c} x' = \tilde{B} \, x + \tilde{u} \, t  \\ t' =
                         S_t \, t  \\ \tilde{\Sigma}' = \tilde{B} \,
                         \Sigma \, \tilde{B}^{T} \\ \tilde{\tau}' = S_t^2 \, \tilde{\tau} \\
                         v' = \frac{1}{S_t} (\tilde{B} \, \tilde{v} +
                         \tilde{u}) \\
                         \nabla_{x,\affnormtiny} = \tilde{\Sigma}^{1/2} \, \nabla_x \\
                         \nabla_{x',\affnormtiny} = \tilde{\Sigma'}^{1/2} \, \nabla_{x'} \\    \nabla_{x',\affnormtiny} =
    \tilde{\rho} \, \nabla_{x,\affnormtiny} \\ \partial_{{\bar t}',\normtiny} = 
                         \partial_{{\bar t},\normtiny} \end{array}}>>
                     \nabla_{x',\affnormtiny} \partial_{{\bar t}',\normtiny} L'(x', t';\; \tilde{\Sigma}', \tilde{\tau}', \tilde{v'}) \\
       \Big\uparrow\vcenter{\rlap{$\scriptstyle{{*
               (\nabla_{x,\affnormtiny} \partial_{{\bar t},\normtiny} T)(x, t;\; 
               \tilde{\Sigma}, \tilde{\tau}, \tilde{v})}}$}} & &
       \Big\uparrow\vcenter{\rlap{$\scriptstyle{{*(
               \nabla_{x',\affnormtiny} \partial_{{\bar t}',\normtiny} T)(x', t';\; 
               \tilde{\Sigma}', \tilde{\tau}', \tilde{v}')}}$}} \\
       f(x, t) @>{\footnotesize \begin{array}{c} x' = \tilde{B} \, x + \tilde{u} \, t
                                  \\ t' = S_t \, t \end{array}}>> f'(x', t')
    \end{CD}
  \]
  \caption{Commutative diagram for scale-normalized spatio-temporal derivative operators
    defined from the joint spatio-temporal receptive field model
    (\ref{eq-spat-temp-RF-model-scnormaff-veladapt})
    under the composition of
    (i)~a spatial affine transformation, (ii)~a Galilean
    transformation and a (iii)~temporal scaling transformation
    according to (\ref{eq-x-transf-bin}) and (\ref{eq-t-transf-bin}) 
    between different pairwise views of the same local surface patch.
    This commutative diagram, which should be read from the lower left corner to the
    upper right corner, means that irrespective of whether the input video 
    sequence or video stream $f(x, t)$ is first subject to the composed transformation
    $x' = \tilde{B} \,  x + \tilde{u} \, t$ and $t' = S_t \, t$
    and then filtered with a scale-normalized spatio-temporal derivative kernel
    $(\nabla_{x',\affnormtiny} \partial_{t',\normtiny} T)(x', t';\; \tilde{\Sigma}', \tilde{\tau}', \tilde{v}')$,
    or instead directly convolved
    with the scale-normalized spatio-temporal smoothing kernel
    $(\nabla_{x,\affnormtiny} \partial_{t,\normtiny} T)(x, t;\; \tilde{\Sigma}, \tilde{\tau}, \tilde{v})$
    and then subject to the same joint spatio-temporal transformation, we do
    then, up to a possibly unknown rotation transformation $\tilde{\rho}$, get the same
    result, provided that the parameters of the spatio-temporal
    smoothing kernels are related
    according to $\tilde{\Sigma}' = \tilde{B} \, \tilde{\Sigma} \, \tilde{B}^{T}$,
    $\tilde{\tau'} = S_t^2 \, \tilde{\tau}$ and
    $\tilde{v}' = \frac{1}{S_t} (\tilde{B} \, \tilde{v} + \tilde{u})$.
  (Adapted from Lindeberg (\citeyear{Lin25-JMIV}) (Open Access).)}
  \label{fig-comm-diag-spat-temp-ders-sc-norm-pairwise-views}
\end{figure*}

Figures~\ref{fig-comm-diag-spat-temp-ders-sc-norm}--\ref{fig-comm-diag-spat-temp-ders-sc-norm-pairwise-views}
illustrate these results in terms of commutative diagrams for
spatio-temporal receptive field responses under geometric image
transformations for the specific spatio-temporal receptive field model
\begin{multline}
  \label{eq-spat-temp-RF-model-scnormaff-veladapt}
  T_{x,\affnorm,{\bar t},\norm}(x, t;\; s, \Sigma, \tau, v) = \\
  = \nabla_{x,\affnorm} \, \partial_{{\bar t},\norm} \, T(x, t;\; s, \Sigma, \tau, v).
\end{multline}
Corresponding commutative diagrams can also be formulated for the other
combinations of spatio-temporal receptive field operators with the
general types of composed geometric image transformations.

In this way, we thus have a general framework for how spatio-temporal
receptive field responses can be matched under compositions of
(i)~uniform spatial scaling transformations,
(ii)~spatial affine transformations,
(iii)~Galilean transformations and
(iv)~temporal scaling transformations.

Corresponding results for purely spatial receptive fields can in turn be obtained
by fully removing all the explicit temporal dependencies from the above
relationships, that is by removing all the occurrences of scale-normal\-ized
velocity-adapted temporal derivative operators $\partial_{{\bar t},\norm}$
as well as removing all the explicit
dependencies on time $t$, the temporal scale $\tau$, the temporal
scaling factor $S_t$, as well as the velocity parameters $u$ and $v$.

\subsection{Explicit examples of covariant receptive field families}
\label{sec-expl-ex-cov-props-rfs}

Given the above theoretical results in the previous section, and stated more
explicitly, these results thus mean that:
\begin{itemize}
\item
  If purely spatial image data $f \colon \bbbr^2 \rightarrow \bbbr$
  are filtered with the family of spatial receptive fields
  \begin{equation}
    T_{\varphi^m,\norm}(x;\; s, \Sigma)
    = \partial_{\varphi,\norm}^m \, g(x;\; s, \Sigma),
  \end{equation}
  with the affine scale-normalized directional derivative operator
  $\partial_{\varphi,\norm}^m$ according to (\ref{eq-aff-sc-norm-dir-der}),
  then the resulting spatial receptive field responses are covariant
  under the spatial similarity group, that is under combinations of
  spatial scaling transformations and spatial rotations.
\item
  If joint spatio-temporal image data $f \colon \bbbr^2 \times \bbbr \rightarrow \bbbr$
  are filtered with the family of spatio-temporal receptive fields
  \begin{multline}
    \label{eq-varphi-tbar-cov-sep-list}
    T_{\varphi^m,\norm,{\bar t}^n,\norm}(x, t;\; s, \Sigma, \tau, v) = \\
    = \partial_{\varphi,\norm}^m \, \partial_{{\bar t},\norm}^n \, (g(x - v t;\; s, \Sigma) \, h(t;\; \tau)),
  \end{multline}
  with the affine scale-normalized directional derivative operator
  $\partial_{\varphi,\norm}^m$ according to
  (\ref{eq-aff-sc-norm-dir-der})
  and the scale-normalized velocity-adapted temporal derivative
  operator $\partial_{{\bar t},\norm}^n$ according to (\ref{eq-vel-adapt-der-def-sc-norm}),
  then the resulting spatial receptive field responses are covariant
  under the spatial similarity group, combined with joint covariance
  properties under Galilean
  transformations and temporal scaling transformations.%
\footnote{In the time-causal case, the temporal scale covariance 
  for the spatio-temporal receptive fields of the form
  (\ref{eq-varphi-tbar-cov-sep-list}) does, however, only hold exactly
  provided that the temporal scale factor $S_t$ is an integer power of the 
  distribution parameter $c$ of the time-causal limit kernel according
  to $S_t = c^j$ for some $j \in \bbbz$. Such a constraint
  also holds for the spatio-temporal receptive fields of the forms
  (\ref{eq-nabla-affnorm-tbar-cov-sep-list}),
  (\ref{eq-hess-affnorm-tbar-cov-sep-list}) and
  (\ref{eq-tbar-cov-sep-list}). For other values of temporal
  scaling factor $S_t$, the corresponding result will
  only constitute an approximation.}
\item
  If purely spatial image data $f \colon \bbbr^2 \rightarrow \bbbr$
  are filtered with the family of spatial receptive fields
  \begin{equation}
    \label{eq-nabla-affnorm-cov-sep-list}
    T_{\nabla,\affnorm}(x;\; s, \Sigma)
    = \nabla_{x,\affnorm} \, g(x;\; s, \Sigma),
  \end{equation}
  with the scale-normalized affine gradient operator
  $\nabla_{x,\affnorm}$ according to (\ref{eq-def-sc-norm-aff-grad-op}),
  then the resulting spatial receptive field responses are covariant%
\footnote{In this context, it should be noted that the notion of affine
  covariance of receptive fields of the form  (\ref{eq-nabla-affnorm-cov-sep-list})
  incorporates equality of the receptive field responses up
  to a possibly unknown rotation matrix $\tilde{\rho}$ according to
  (\ref{eq-equal-aff-scsp-repr-aff-cov-proof-again}).
  Such an indeterminacy up to a possibly unknown rotation matrix
  $\tilde{\rho}$ also holds for the affine-covariant receptive fields
  of the forms (\ref{eq-nabla-affnorm-tbar-cov-sep-list}),
  (\ref{eq-hess-affnorm-cov-sep-list}) and
  (\ref{eq-hess-affnorm-tbar-cov-sep-list}),
  extended to combined pre- and post-multiplication of the rotation matrix
  $\tilde{\rho}$ for the Hessian-based receptive fields according to
  (\ref{eq-hess-affnorm-cov-sep-list}) and
  (\ref{eq-hess-affnorm-tbar-cov-sep-list}).}
  under both spatial scaling transformations and spatial affine
  transformations.
\item
  If joint spatio-temporal image data $f \colon \bbbr^2 \times \bbbr \rightarrow \bbbr$
  are filtered with the family of spatio-temporal receptive fields
  \begin{multline}
    \label{eq-nabla-affnorm-tbar-cov-sep-list}
    T_{\nabla,\affnorm,{\bar t}^n,\norm}(x, t;\; s, \Sigma, \tau, v) = \\
    = \nabla_{x,\affnorm} \, \partial_{{\bar t},\norm}^n  (g(x - v t;\; s, \Sigma) \, h(t;\; \tau)),
  \end{multline}
  with the scale-normalized affine gradient operator
  $\nabla_{x,\affnorm}$ according to
  (\ref{eq-def-sc-norm-aff-grad-op})
  and the scale-normalized velocity-adapted temporal derivative
  operator $\partial_{{\bar t},\norm}^n$ according to (\ref{eq-vel-adapt-der-def-sc-norm}),
  then the resulting spatial receptive field responses are covariant
  under combinations of spatial scaling transformations,
  spatial affine transformations, Galilean transformations
  and temporal scaling transformations.
\item
  If purely spatial image data $f \colon \bbbr^2 \rightarrow \bbbr$
  are filtered with the family of spatial receptive fields
  \begin{equation}
    \label{eq-hess-affnorm-cov-sep-list}
    T_{{\cal H},\affnorm}(x;\; s, \Sigma)
    = {\cal H}_{x,\affnorm} \, g(x;\; s, \Sigma),
  \end{equation}
  with the scale-normalized affine Hessian operator
  $ {\cal H}_{x,\affnorm}$ according to (\ref{eq-def-sc-norm-aff-hess-mat}),
  then the resulting spatial receptive field responses are covariant
  under both spatial scaling transformations and spatial affine
  transformations.
\item
  If joint spatio-temporal image data $f \colon \bbbr^2 \times \bbbr \rightarrow \bbbr$
  are filtered with the family of spatio-temporal receptive fields
  \begin{multline}
    \label{eq-hess-affnorm-tbar-cov-sep-list}
    T_{{\cal H},\affnorm,{\bar t}^n,\norm}(x, t;\; s, \Sigma, \tau, v) = \\
    = {\cal H}_{x,\affnorm} \, \partial_{{\bar t},\norm}^n  (g(x - v t;\; s, \Sigma) \, h(t;\; \tau)),
  \end{multline}
  with the scale-normalized affine Hessian operator
  $ {\cal H}_{x,\affnorm}$ according to
  (\ref{eq-def-sc-norm-aff-hess-mat})
  and the scale-normalized velocity-adapted temporal derivative
  operator $\partial_{{\bar t},\norm}^n$ according to (\ref{eq-vel-adapt-der-def-sc-norm}),
  then the resulting spatial receptive field responses are covariant
  under combinations of spatial scaling transformations,
  spatial affine transformations, Galilean transformations
  and temporal scaling transformations.
\item
  If joint spatio-temporal image data $f \colon \bbbr^2 \times \bbbr \rightarrow \bbbr$
  are filtered with the family of spatio-temporal receptive fields
  \begin{multline}
    \label{eq-tbar-cov-sep-list}
    T_{{\bar t}^n,\norm}(x, t;\; s, \Sigma, \tau, v) = \\
    = \partial_{{\bar t},\norm}^n  (g(x - v t;\; s, \Sigma) \, h(t;\; \tau)),
  \end{multline}
  with the scale-normalized velocity-adapted temporal derivative
  operator $\partial_{{\bar t},\norm}^n$ according to (\ref{eq-vel-adapt-der-def-sc-norm}),
  then the resulting spatial receptive field responses are covariant
  under combinations of spatial scaling transformations,
  spatial affine transformations, Galilean transformations
  and temporal scaling transformations.
\end{itemize}
Notably, these theoretical results comprise combinations of spatial derivatives up to
order 2 with temporal derivatives for any order of temporal
differentiation (including the special case when the temporal order of
differentiation $n = 0$).
In these ways, we can thus formulate a rich set of both purely spatial
and joint spatio-temporal receptive field models, that are provably
covariant under combinations of 4 main types of primitive
geometric image transformations according to
Equations~(\ref{eq-spat-sc-transf})--(\ref{eq-temp-sc-transf}),
as summarized in the composed geometric image transformations 
according to (\ref{eq-x-transf}), (\ref{eq-t-transf}) and
(\ref{eq-sc-aff-vel-transf-alt-obs-model}).

\subsection{Relationships to the influence of illumination variations}
\label{sec-illum-var}

Regarding variabilities in image data caused by natural image
transformations, we do in this treatment focus on the influence due to
geometric image transformations. Regarding the influence of
illumination variations, which also constitute a large source to
variability in image data, it is, however, interesting to note
that according to the theory in
Lindeberg (\citeyear{Lin13-BICY}) Section~2.3 condensedly summarized in
Lindeberg (\citeyear{Lin21-Heliyon}) Section~3.4, it holds that
if the image data $f$ used as input to the receptive fields are
parameterized in terms of the logarithm of the intensities in
the dimension of the incoming energy $\log I(x, y)$ or
$\log I(x, y, t)$ according to
\begin{equation}
  f(x, y) \sim \log I(x, y)
  \quad\quad\mbox{or}\quad\quad
  f(x, y, t) \sim \log I(x, y, t),
\end{equation}
then the computed spatial or spatio-temporal receptive field responses
in terms of either spatial derivatives, temporal derivatives or both
will be automatically invariant under local multiplicative intensity
transformations%
\footnote{Note that this invariance property holds exactly if and only
  if the multiplication factor $C$ is constant over the entire support
  region of the receptive field. If the multiplication factor $C$
  on the other hand varies within the support region, then the corresponding
  relationship will only be approximate.}
of the form
\begin{equation}
   I(x, y) \mapsto C \,  I(x, y)
\end{equation}
or
\begin{equation}
   I(x, y, t) \mapsto C \,  I(x, y, t),
\end{equation}
for any strictly positive local multiplication factor $C \in \bbbr_+$.
A similar invariance result holds concerning the influence of global
exposure compensation mechanisms of a similar multiplicative form.

In this way, a substantial component of the influence due to illumination
variations and exposure compensation mechanism can be directly handled in a 
straightforward manner.

In this context, it is interesting to note that the retinex theory of
early vision (Land \citeyear{Lan74-RoyInst,Lan86-VR})
also makes use of a logarithmic brightness scale.

An exposure mechanism on the retina that adapts the diameter of the
pupil and the sensitivity of the photopigments, in such a way that the
relative range in the variability of the signal divided by the mean
illumination is held constant, can also be seen as implementing an
approximation of the derivative of a logarithmic transformation
\begin{equation}
  d(\log z) = \frac{dz}{z},
\end{equation}
see  {\em e.g.\/}\ Peli (\citeyear{Pel90-JOSA}).
Furthermore, in the area of psychophysics, the {\em Weber-Fechner law\/} 
states that the ratio
\begin{equation}
  \label{eq-Weber-ratio}
  \frac{\Delta I}{I} = k,
\end{equation}
between the threshold $\Delta I$ corresponding to a just
noticeable difference in image intensity and the background intensity $I$
is constant over large ranges of magnitude variations,
see {\em e.g.\/} Palmer (\citeyear{Pal99-Book}) pages~671--672,
thus providing further support for the relevance of a logarithmic
brightness scale.

\begin{figure*}[hbt]
  \begin{center}
    \includegraphics[width=0.99\textwidth]{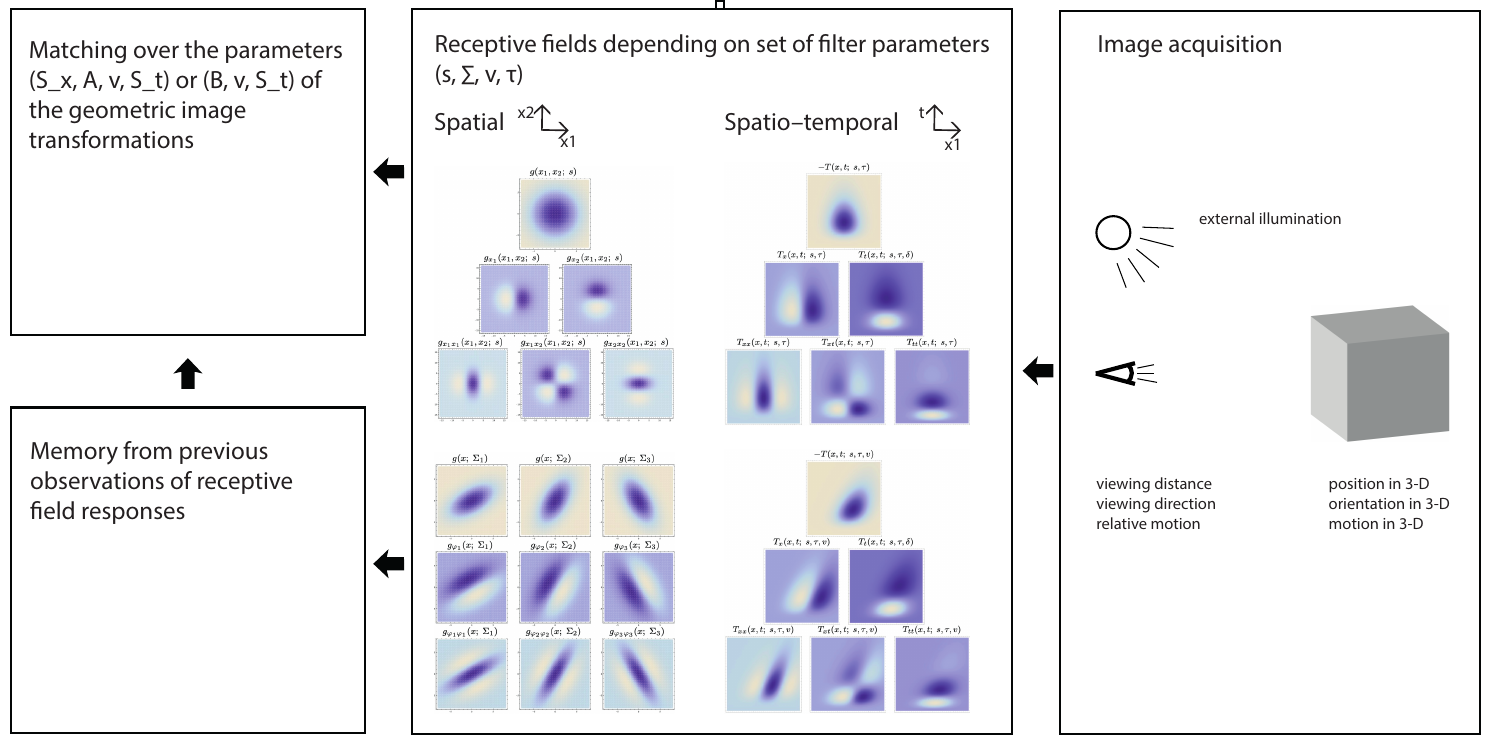}
  \end{center}
  \caption{Conceptual illustration of how sets of spatial and/or
    spatio-temporal receptive field responses computed over different
    values of the filter
    parameters $(s, \Sigma, v, \tau)$ of the receptive fields can be matched between
    different views of the same scene or spatio-temporal event under
    different viewing conditions, by making use of the matching
    relations in
    Equations~(\ref{eq-s-transf-result})--(\ref{eq-v-transf-result})
    between the filter parameters for the two image domains
    before and after the geometric image transformation, 
    under variabilities of the parameters
    $(S_x, A, v, S_t)$ or $(B, v, S_t)$ of the
    primitive geometric image transformations in
    Equations~(\ref{eq-spat-sc-transf})--(\ref{eq-temp-sc-transf}), as
    combined into composed geometric image transformations according
    to either Equation~(\ref{eq-x-transf}) or
    Equation~(\ref{eq-sc-aff-vel-transf-alt-obs-model}) in combination with
    Equation~(\ref{eq-t-transf}).
    Such matching of the receptive field responses under geometric
    image transformations is possible for receptive fields that obey provable
    covariance properties as exemplified in Section~\ref{sec-expl-ex-cov-props-rfs}.
    (The arrows between the boxes indicate the information
    flow from the image acquisition stage to the matching stage.
    Regarding the visualized receptive fields in the middle box, for the spatial
    receptive fields in the left column, the coordinates are the
    purely spatial coordinates $(x_1, x_2) \in \bbbr^2$, whereas for
    the spatio-temporal receptive fields, only one of the spatial
    coordinates is shown, thus with the spatio-temporal image
    coordinates $(x_1, t) \in \bbbr \times \bbbr$.
    The receptive fields in the top parts of the middle box are
    separable over image space or joint space-time and are based on
    isotropic Gaussian smoothing over the spatial domain with the
    spatial covariance matrix $\Sigma$ equal to a unit matrix $I$.
    The purely spatial receptive fields in the left bottom part of the
    middle box are based on spatial smoothing with non-isotropic
    affine Gaussian kernels for which the spatial covariance matrix is
    not equal to a unit matrix. The joint spatio-temporal receptive
    fields in the right bottom part of the middle box are based on
    velocity-adapted spatio-temporal smoothing kernels in combination
    with velocity-adapted temporal derivatives. All the
    spatio-temporal receptive fields in this figure are based on
    temporal smoothing with the time-causal limit kernel
    (\ref{eq-time-caus-lim-kern}).)}
  \label{fig-match-rf-resp-pars}
\end{figure*}

\section{Do the shapes of the simple cells in the primary visual
  cortex of higher mammals span the variabilities of geometric image
  transformations to support explicitly covariant families of visual receptive fields?}
\label{sec-span-vars}

A main result of the above presented theory is that, according to the model, the output from
both purely spatial and joint spatio-temporal receptive fields according to
generalized versions of the idealized receptive field models
(\ref{eq-spat-RF-model}) and
(\ref{eq-spat-temp-RF-model-der-norm-caus}) can be matched under the
composed geometric image transformations according to both the
monocular projection model in
Equations~(\ref{eq-x-transf-mon})--(\ref{eq-t-transf-mon})
and the binocular projection model in
Equations~(\ref{eq-x-transf-bin})--(\ref{eq-t-transf-bin}).
This result holds provided
that we allow for the sets of the parameters $(s, \Sigma, \tau, v)$ and
$(s', \Sigma', \tau', v')$ of the receptive fields
$T(s, \Sigma, \tau, v)$ and $T'(s', \Sigma', \tau', v')$
to be varied between the image domains
before and after the geometric image transformation, and specifically
having the values of the receptive field parameters being matched
according to Equations~(\ref{eq-s-transf-result})--(\ref{eq-v-transf-result}) 
as functions of the parameters $(S_x, A, u, S_t)$ of the composed
geometric image transformation.

Since the parameters $(S_x, A, u, S_t)$ of the geometric image transformation cannot be
expected to be {\em a priori\/} known to a vision system, that is to
analyze an {\em a priori\/} unknown scene, a general purpose strategy for a vision
system could therefore be to expand the receptive field measurements over a rich
family of receptive fields, with the shapes of the receptive fields
expanded over the degrees of freedom of the corresponding
geometric image transformations.
Thereby, it would be
possible to match the outputs from populations of receptive fields, to
establish a corresponding matching of the receptive field responses
obtained from a particular viewing condition in relation to a learned memory of
receptive field responses computed from previous views of similar objects and
spatio-temporal events under different sets of viewing conditions,
see Figure~\ref{fig-match-rf-resp-pars} for an illustration.

According to the most idealized version of the theory, a direct
prediction is hence that the receptive fields of the simple cells in
the primary visual cortex ought to have their receptive fields
expanded over the following degrees of freedom:
(i)~the spatial size (spatial scale),
(ii)~the preferred orientation in image space,
(iii)~the degree of elongation,
(iv)~a fourth spatial degree of freedom,
(v)~the preferred direction in joint space-time,
(vi)~the preferred speed in joint space-time and
(vii)~the temporal duration (temporal scale).
Alternatively, and as we will discuss further below,
due to the cascade smoothing properties over spatial and temporal
scales of the spatial smoothing kernel over the spatial scale
parameter and the temporal smoothing kernel over the temporal scale
parameter, it may, however, not be formally necessary for an idealized
vision system to represent more than the finest spatial and temporal
scale levels in the sensorium.

One may hence consider an idealized vision system as representing a
set of receptive fields over a range of shape parameters over the
degrees of freedom of the receptive fields.
How wide the spans are regarding the variabilities of the receptive
fields over each parameter of the receptive fields will hence
determine how large geometric image transformations can be fully captured in the
respect of matching the receptive field responses before and after the
geometric image transformation.
How wide and how densely sampled parameter ranges are explicitly represented in an actual
implementation of the theory could hence determine the operational
range of the vision system, which could then, in particular, be
considered to be different for different species.

Additionally, it should be noted that
regarding an actual implementation of a matching scheme of
receptive field responses of this form, one could also conceive that
such a matching could be performed, not directly on the
receptive fields responses in the primary visual cortex itself,
but instead based on derived receptive field responses at higher levels in the visual
hierarchy. This would indeed be possible based on the presented theory,
if the receptive field responses at the higher levels are computed
from the receptive field responses using the responses of the simple
cells in a causal feed-forward manner that respects the covariance properties.
A further outline about tentative ways of extending the covariance
properties to higher layers in the visual hierarchy, and using such
higher layers for explicit matching, is given in
Appendix~\ref{app-prop-cov-props-high-layers}.

Given this idealized theory,
one may therefore ask if biological vision has
evolved similar mechanisms
to be able to handle the influence of geometric image
transformations on the receptive field responses in a way that is
closely related to the results from the presented theory.
Specifically, one may ask if the shapes of the receptive
fields of the simple cells in the primary visual cortex would be expanded over
the degrees of freedom of (i)~uniform spatial scaling transformations,
(ii)~non-isotropic spatial affine transformations, (iii)~Galilean
transformations and (iv)~temporal scaling transformations,
and if so for what species?

Given evolutionary arguments, it could be hypothesized that for at least
higher species with more developed general purpose vision systems
for addressing a wide range of visual tasks, like primates and cats,
it would constitute an evolutionary advantage to adapt their visual
system to the structure of the external environment, and thus to the
influence of geometric image transformations on the image data that
reach their visual sensors.

An underlying motivation for this argument is that for an animal, that
depends strongly on its vision system for its survival, it ought to
constitute an evolutionary advantage for the vision system to evolve
in such a way that the vision system is well adapted to the
geometric properties that arise in the projections of a 3+1-D dynamic
environment to 2+1-D measurements on the retinas.
In a competition between
different species or between different individuals of a given
species, one could then argue that an individual that has better
ability to perceive the 3+1-D structure of the environment better
than a competitor, predator or prey ought to
have a higher likelihood of surviving under or coping with critical circumstances,
such as in flight-or-fight scenarios.

One may therefore ask if there for such higher
species under a sufficient evolutionary pressure
would be variabilities in the shapes of their visual receptive
fields in the visual area corresponding to the primary visual cortex, as implied by what would
be the result of expanding the receptive field shapes with respect to
the degrees of freedom of the 4 main classes of primitive geometric
image transformations, and thereby leading to
variabilities in the shapes of the receptive fields over
the following degrees of freedom: (i)~the spatial size (spatial scale),
(ii)~the preferred orientation in image space,
(iii)~the degree of elongation,
(iv)~a fourth spatial degree of freedom,
(v)~the preferred direction in joint space-time,
(vi)~the preferred speed in joint space-time and
(vii)~the temporal duration (temporal scale).
In the following, we will consider either explicit variabilities
and/or indirect covariance properties over each such degree of freedom.

\subsection{Purely spatial variabilities in the shapes of spatial
  and spatio-temporal receptive fields}
\label{sec-variabil-spat}

In Lindeberg (\citeyear{Lin25-BICY}), the theoretical problem
regarding variabilities in receptive field shapes as implied by
covariance properties under geometric image transformations is addressed in
detail in relation to the
first degrees of freedom concerning the combined effect of (i)~uniform
spatial scaling transformations and (ii)~non-isotropic spatial affine
transformations.
In brief, and extended from a purely spatial domain to also encompass the joint
spatio-temporal domain, the results from that treatment are that:

\subsubsection{Variability under uniform spatial scaling transformations}
\label{sec-var-scale-trans}

Regarding the degree of freedom corresponding to uniform spatial
scaling transformations, the corresponding degree of freedom in
terms of the spatial scale parameter $\sigma_x = \sqrt{s}$ is
special in the sense that the spatial affine Gaussian kernel obeys a
semi-group property over spatial scales
(see Lindeberg (\citeyear{Lin93-Dis}) Equation~(15.35))
\begin{equation}
  g(\cdot;\; \Sigma_1) * g(\cdot;\; \Sigma_2) = g(\cdot;\; \Sigma_1 + \Sigma_2),
\end{equation} from which it follows that
any receptive field response at a coarser spatial scale,
here as represented by the spatial gradient vector $\nabla_x L$
or the Hessian matrix ${\cal H}_x L$
for the spatial covariance matrix $\Sigma_2$, can be computed by affine
Gaussian smoothing of the receptive field responses for any finer spatial
scale, here as represented by the spatial gradient vector $\nabla_x
L$ or the Hessian matrix ${\cal H}_x L$ defined for the
spatial covariance matrix $\Sigma_1$
\begin{align}
  \begin{split}
    \label{eq-cascade-smooth-gradient}
    (\nabla_x \, L)(\cdot;\; \Sigma_2)
    = g(\cdot;\; \Sigma_2 - \Sigma_1)  * (\nabla_x \, L)(\cdot;\;
    \Sigma_1),
  \end{split}\\
  \begin{split}
    \label{eq-cascade-smooth-hessian}
    ({\cal H}_x \, L)(\cdot;\; \Sigma_2)
    = g(\cdot;\; \Sigma_2 - \Sigma_1)  * ({\cal H}_x \, L)(\cdot;\;
    \Sigma_1),
  \end{split}
\end{align}
provided that the difference between the spatial covariance matrices
\begin{equation}
  \Delta \Sigma = \Sigma_2 - \Sigma_1
\end{equation}
is a symmetric positive definite matrix.%
\footnote{Please, note that this cascade smoothing property makes it
  possible to cascade affine Gaussian receptive fields with different
  orientations and different degrees of elongation, as long as the
  difference covariance matrix $\Delta \Sigma = \Sigma_2 - \Sigma_1$
  is positive semi-definite. In this context it should, however, be
  noted that not all possible covariance matrices may be possible to
  generate by cascade smoothing from a single seed. Instead, multiple
  seeds may be necessary with different shape parameters for each
  seed. Thus, the expansion of receptive field shapes with respect to
  the spatial scale parameter could be intimately coupled with the
  expansion of the receptive field shapes with respect to the other
  spatial degrees of freedom in terms of (i)~the orientation, (ii)~the
  degree of elongation and (iv)~the fourth spatial degree of freedom. An additional
  extension of this principle to also cascading spatio-temporal
  receptive fields with different image velocities $v$ is given in
  Lindeberg (\citeyear{Lin26-JMIV}).}

Thereby, a vision system could in principle choose to
compute the earliest layers of spatial receptive fields at only the finest
spatial scale and nevertheless be able to compute the coarser
spatial receptive fields at higher layers in the visual hierarchy.%
\footnote{For example, regarding variabilities in receptive field
  size in higher layers of the visual hierarchy,
  Orban (\citeyear{Orb97-ExtrStriCortPrim}) (page~371) reports
  that in the visual area MT, ``... at all eccentricities there is a tenfold range in RF size''.}
Thus, irrespective of whether the spatial receptive fields
corresponding to the simple cells are
expanded over the spatial scales, the theory suggests that the
vision system could have the ability to compute visual operations
corresponding to spatial scale covariance.

Figure~\ref{fig-1spatders-scale-var} shows an example of such a
variability under spatial scaling variations for first-order spatial
directional derivatives computed based on isotropic Gaussian
smoothing. Figure~5 in Lindeberg (\citeyear{Lin25-BICY}) shows an
example of such a variability extended to second-order spatial
directional derivatives.

\subsubsection{Variability under spatial rotations in the image plane}

\begin{figure}[hbtp]
  \begin{center}
    \begin{tabular}{c}
      \includegraphics[width=0.35\textwidth]{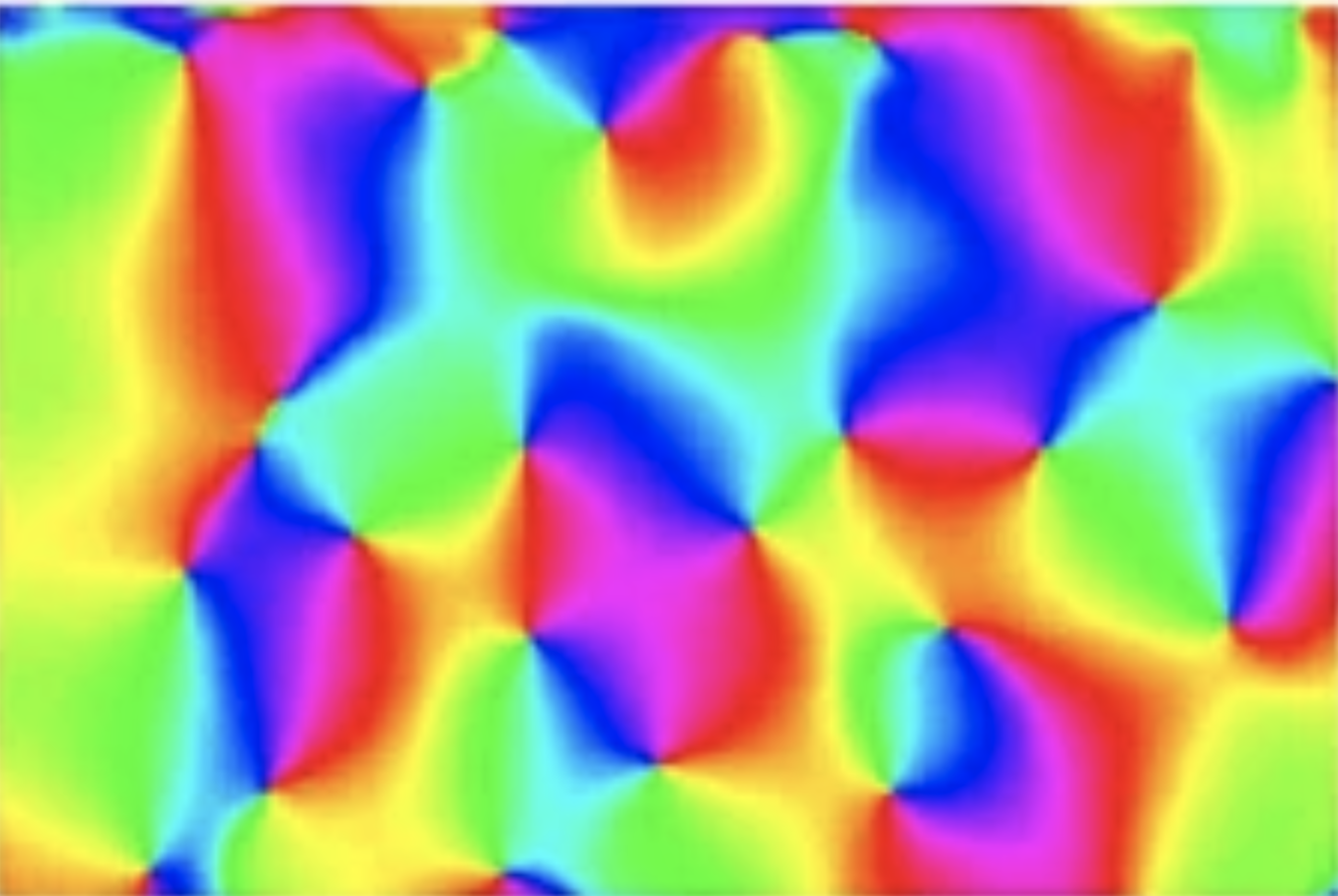}
    \end{tabular}
  \end{center}
  \caption{Orientation map in the primary visual cortex of cat, 
    as recorded by Koch {\em et al.\/} (\citeyear{KocJinAloZai16-NatComm})
    (OpenAccess), and demonstrating that the
    visual cortex of higher mammals can be regarded as performing an explicit
    expansion of the receptive field shapes over spatial image
    orientations, as would be the result of combining the notion of
    covariance over the subgroup of pure image rotations.}
  \label{fig-koch-ori-map-rot}
\end{figure}

From the structure of orientation maps in the primary visual cortex
of higher mammals, as studied by
Bonhoeffer and Grinvald (\citeyear{BonGri91-Nature}),
Blasdel (\citeyear{Bla92-JNeuroSci}),
Koch {\em et al.\/} (\citeyear{KocJinAloZai16-NatComm}) and others
(see Figure~\ref{fig-koch-ori-map-rot} for an illustration),
it is clear that we can interpret these orientation maps as an
explicit expansion of the receptive field shapes over the image orientations,
corresponding to the parameter $\varphi$ in the idealized receptive
field models (\ref {eq-spat-RF-model}) and
(\ref{eq-spat-temp-RF-model-der-norm-caus}).
Thereby, we could regard the vision system of higher mammals to have
the ability to compute visual operations corresponding to rotation
covariance.

Combined with spatial scale covariance, this means that the
vision system of higher mammals could have the ability to be covariant
under spatial similarity transformations, that is to 
combinations of spatial scaling transformations and spatial
rotations.

Figure~6 in Lindeberg (\citeyear{Lin25-BICY}) shows examples of such
a variability under spatial rotations for first- and second-order spatial directional
derivatives computed based on non-isotropic affine Gaussian smoothing.
\medskip

\subsubsection{Variability under the degree of elongation of the receptive
  fields}

\begin{figure}[hbtp]
  \begin{center}
    \includegraphics[width=0.20\textwidth]{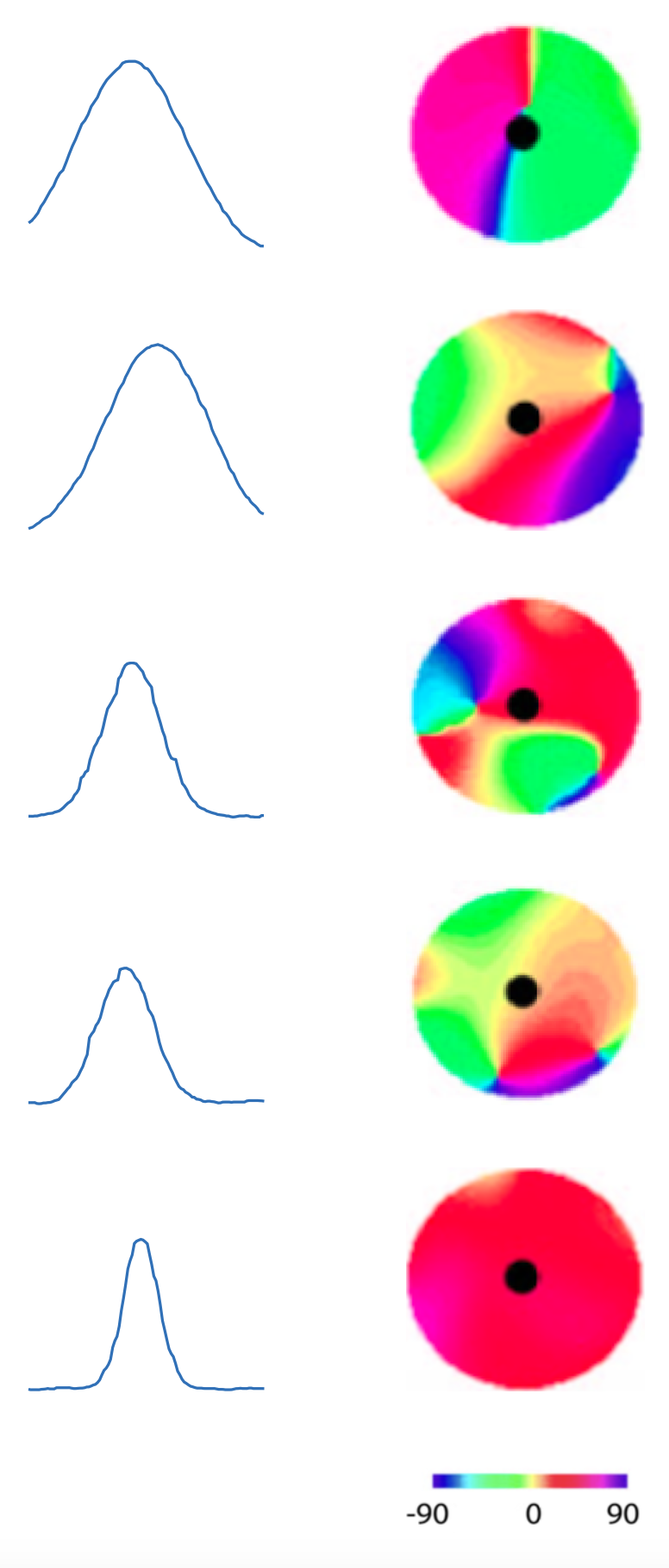}
  \end{center}
  \caption{Schematic depiction of results from measurements of
    the orientation tuning of neurons, at
    different positions in the visual cortex, adapted from
    Nauhaus {\em et al.\/}\ (\citeyear{NauBenCarRin09-Neuron}),
    showing how the orientation tuning
    changes from wide to narrow, and thus higher degree of orientation
    selectivity, with increasing distance from the pinwheels, consistent
    with the qualitative behaviour that would be obtained if the ratio
    $\kappa$, between the scale parameters in underlying affine Gaussian
    smoothing step in the idealized models of spatial and spatio-temporal
    receptive fields, would increase when moving away from the centers of
    the pinwheels on the cortical surface.
    (From Lindeberg (\citeyear{Lin25-JCompNeurSci-spanelong}) with
    permission (OpenAccess).)}
  \label{fig-nauhaus-ori-tun-meas}
\end{figure}

From studies of the orientation selectivity properties of simple
cells established from neurophysiological cell recordings by
Nauhaus {\em et al.\/}\ (\citeyear{NauBenCarRin09-Neuron})
and Goris {\em et al.\/}\ (\citeyear{GorSimMov15-Neuron}),
it is clear that the simple and complex cells of primates and cat
have a substantial variability in orientation selectivity
properties, ranging from narrow to wide orientation selectivity
properties, as illustrated in Figure~\ref{fig-nauhaus-ori-tun-meas}.

\begin{figure*}[hbtp]
  \begin{center}
    \begin{tabular}{ccc}
      {\em\footnotesize First-order simple cells\/}
      &       {\em\footnotesize Second-order simple cells\/}
      &       \hspace{-14mm} {\em\footnotesize Complex cells\/} \\
      \includegraphics[height=0.18\textwidth]{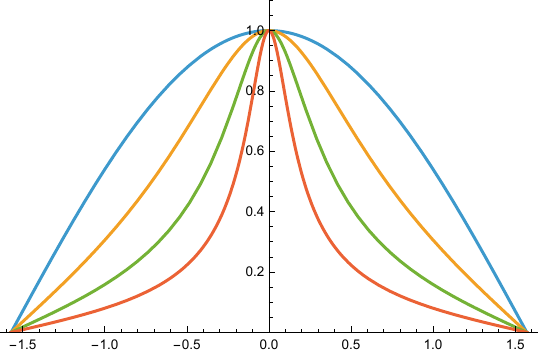}
      & \includegraphics[height=0.18\textwidth]{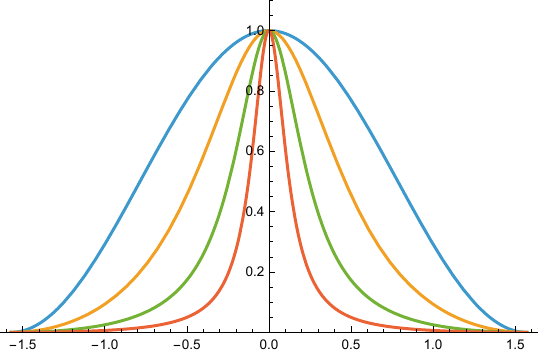}
      & \includegraphics[height=0.18\textwidth]{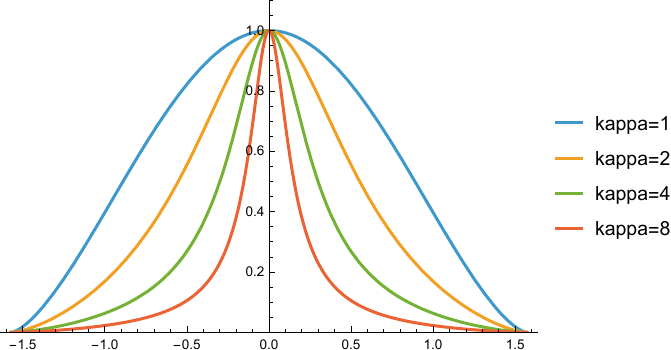}
    \end{tabular}
  \end{center}
  \caption{Graphs of the orientation selectivity for the idealized models of
    (top left) simple cells in terms of first-order
    directional derivatives of affine Gaussian kernels, (top middle) simple
    cells in terms of second-order directional derivatives of affine Gaussian
    kernels, (top right) complex cells in terms of directional
    quasi-quadrature measures that combine the first- and second-order
    simple cell responses in a Euclidean way.
    The parameter $\kappa$ represents the degree of elongation of the
    receptive fields, and corresponds to the ratio between the
    square roots of the eigenvalues of the covariance matrix $\Sigma$ of the affine
    Gaussian derivative-based receptive fields.
    (Horizontal axes: orientation $\theta \in [-\pi/2, \pi/2]$.
    Vertical axes: Amplitude of the receptive field response relative
    to the maximum response obtained for $\theta = 0$.)
    (Adapted from Lindeberg (\citeyear{Lin25-JCompNeurSci-spanelong}) with
    permission (OpenAccess).)}
  \label{fig-ori-sel}
\end{figure*}

From a theoretical analysis of the orientation selectivity
properties of the affine Gaussian derivative and affine Gabor models
of visual receptive fields in Lindeberg (\citeyear{Lin25-JCompNeurSci-orisel}),
we have established a connection that the
degree of orientation selectivity increases with the degree of
elongation of the receptive fields, as illustrated in
Figure~\ref{fig-ori-sel}, and meaning the a highly elongated receptive
field will have more narrow orientation selectivity properties than
a less elongated receptive field. For the affine Gaussian
derivative and affine Gabor models, that degree of elongation
corresponds to the ratio between the eigenvalues of the affine
Gaussian kernel used in these idealized models.

In Lindeberg (\citeyear{Lin25-JCompNeurSci-spanelong}), we have
combined these two sources of biological and theoretical knowledge
to propose that these results are consistent with the receptive
fields of primates and cats spanning a variability in the degree of
elongation of the receptive fields. In Lindeberg
(\citeyear{Lin25-BICY}), we further show that this degree of freedom
of the receptive fields corresponding to the degree of freedom
spanned by the ratio between the singular values obtained from a
singular value decomposition of the affine transformation matrix
$A$.

In Meshcherina {\em et al.\/} (\citeyear{MesAufLin20-bioRxiv}), by
reconstructing the receptive fields of simple cells in the primary
visual cortex of mice based on calcium imaging experiments
performed by the Allen Institute, we have shown that these results
reveal a variability in the degree of elongation of the spatial
components of the receptive fields of a factor of~5.

In modelling of biological direction selectivity curves
recorded Livingstone and Conway (\citeyear{LivCon03-JNeuroPhys}) in
terms of idealized direction selectivity curves derived from
Galilean-adapted affine Gaussian derivatives of the form
(\ref{eq-spat-temp-RF-model-der-norm-caus}), we have in
Lindeberg (\citeyear{Lin25-arXiv-dirsel}) demonstrated that the
biologically recorded direction selectivity curves are better described
by assuming a variability in the degree of elongation of the spatial
component of the joint spatio-temporal receptive fields. In this
respect, also that joint spatio-temporal model supports a variability
in the degree of elongation of the receptive fields.

In summary, the analysis of these three types of comparisons to
results of neurophysiological measurements is consistent with an
explicit variability in the degree of elongation of the
receptive fields of simple cells in the primary visual cortex.

Figure~7 in Lindeberg (\citeyear{Lin25-BICY}) shows examples of such
a variability over the degree of elongation for first- and
second-order spatial directional derivative operators computed based
on non-isotropic affine Gaussian smoothing.
Figure~\ref{fig-1dir-gaussder} in this paper shows an example of a
combined variability over the degree of elongation of the receptive
fields with spatial rotations in the image plane for first-order
spatial directional derivatives based on non-isotropic affine Gaussian smoothing.
  
\subsubsection{Variability under a 4th purely spatial degree of
  freedom}

To span all the 4 degrees of freedom corresponding to the
combination of uniform spatial scaling transformations with non-isotropic spatial affine
transformations, there is one additional degree of freedom
that corresponds to generalizing the idealized receptive field models
(\ref {eq-spat-RF-model}) and
(\ref{eq-spat-temp-RF-model-der-norm-caus})
to not necessarily having the orientation $\varphi$ for computing the
directional derivatives $\partial_{\varphi}^m$ being parallel to any
of the eigendirections of the spatial covariance matrix $\Sigma_{\varphi}$.

As argued in Lindeberg (\citeyear{Lin25-BICY}) Section~7.4,
there have been examples of biological receptive fields
recorded by Yazdanbakhsh and Livingstone (\citeyear{YazLiv06-NatNeuroSci})
(see Figure~6 in that paper) that
appear to be more similar to first- or second-order directional
derivatives of Gaussian kernels in directions different from
the principal directions of an affine Gaussian kernel compared to
directional derivatives of such kernels in directions that coincide
with the principal directions of affine Gaussian kernels.

Figure~8 in Lindeberg (\citeyear{Lin25-BICY}) shows an example of
such a variability over the angle between the orientation for
computing spatial directional derivatives relative to the principal
eigendirections of the affine Gaussian kernel used for spatial smoothing.

\subsubsection{Summary of spatial variabilities of the receptive fields}

In these ways, the presented theory and the comparisons to
available biological data are consistent in different respects with the
hypothesis that the receptive fields of simple cells in the primary
visual cortex of higher mammals ought to have the ability to be covariant
under the combination of uniform spatial scaling transformations,
rotations in the image plane and non-isotropic spatial affine transformations.

\begin{figure*}[hbtp]
  \begin{center}
    Spatial differentiation order $m = 1$

    \bigskip
    
    \begin{tabular}{ccc}
      $\theta = 0$ & $\theta = \pi/3$ & \hspace{-12mm} $\theta = 2\pi/3$ \\ 
      \includegraphics[height=0.19\textheight]{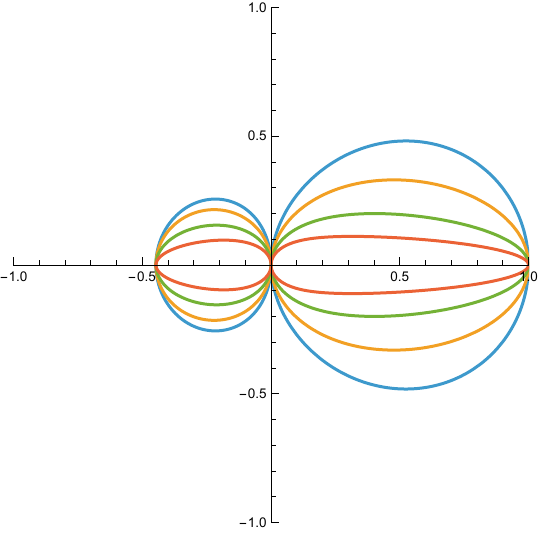}
      & \includegraphics[height=0.19\textheight]{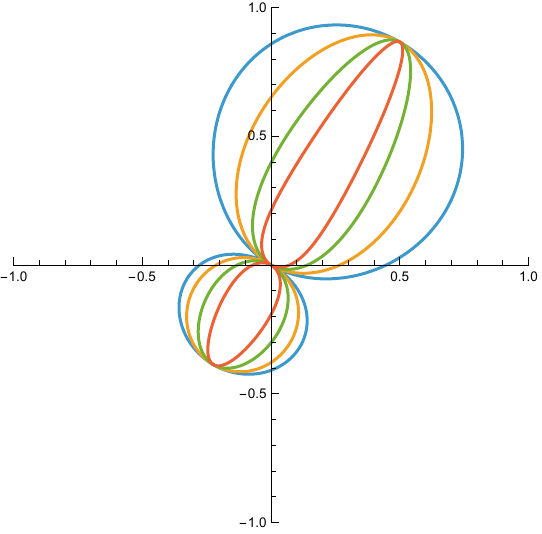}
      & \includegraphics[height=0.19\textheight]{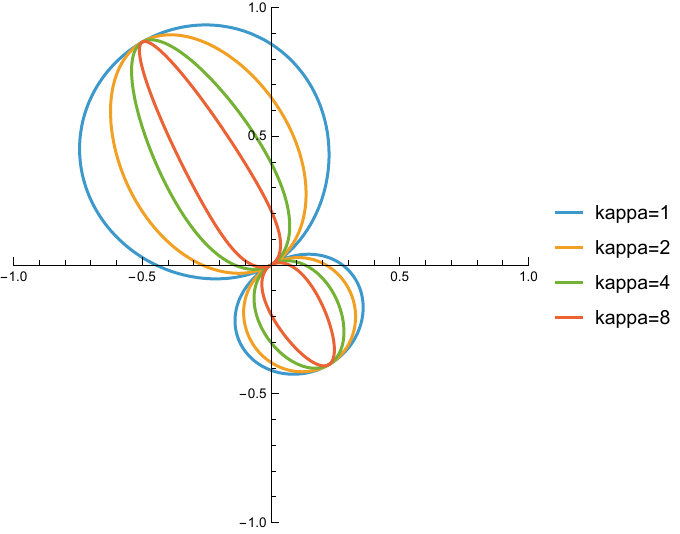}   \\
      $\theta = \pi$ & $\theta = 4\pi/3$ & \hspace{-12mm} $\theta = 5\pi/3$ \\ 
       \includegraphics[height=0.19\textheight]{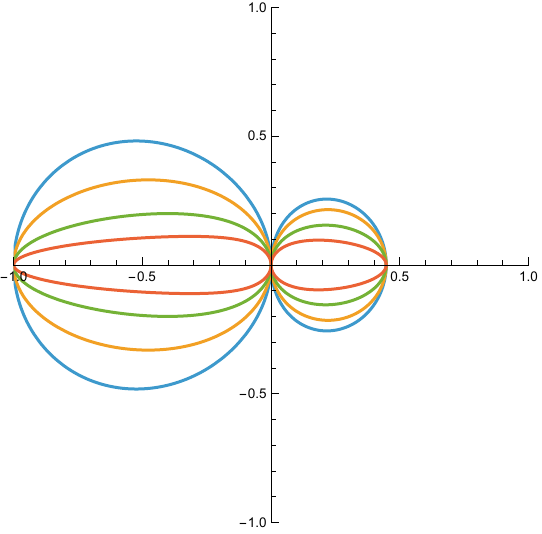}
      & \includegraphics[height=0.19\textheight]{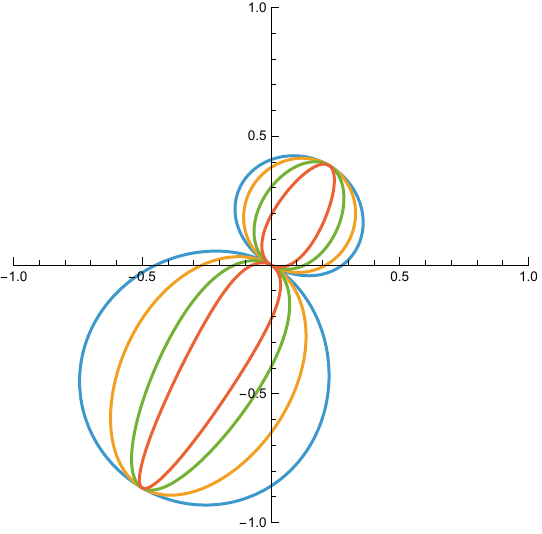}
      & \includegraphics[height=0.19\textheight]{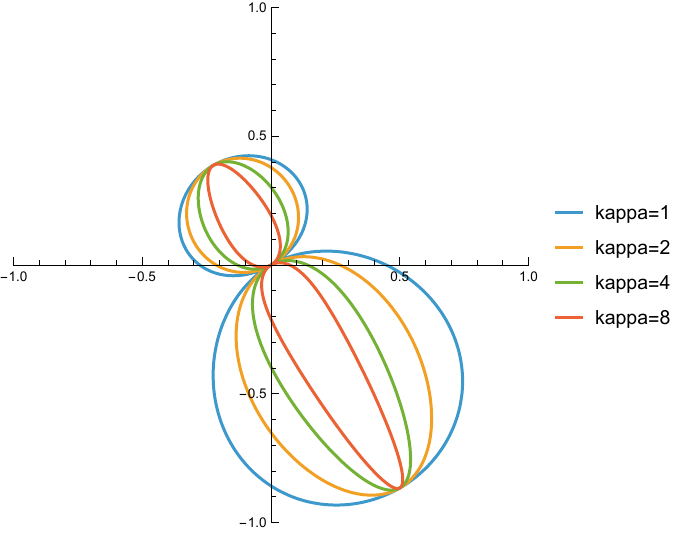}   \\   
    \end{tabular}

    \bigskip
    
    Spatial differentiation order $m = 2$

    \bigskip
    
    \begin{tabular}{ccc}
      $\theta = 0$ & $\theta = \pi/3$ & \hspace{-12mm} $\theta = 2\pi/3$ \\ 
      \includegraphics[height=0.19\textheight]{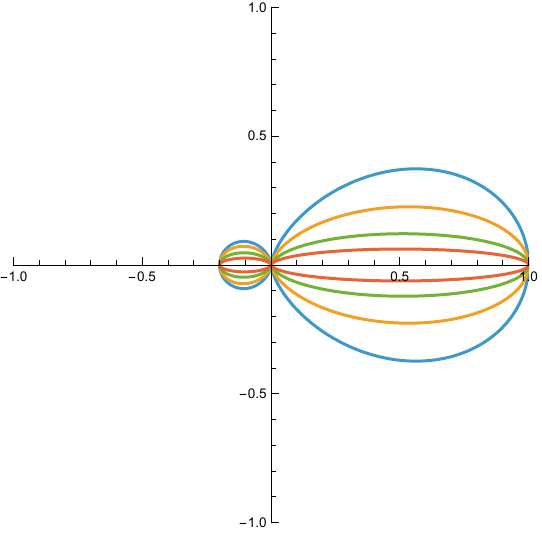}
      & \includegraphics[height=0.19\textheight]{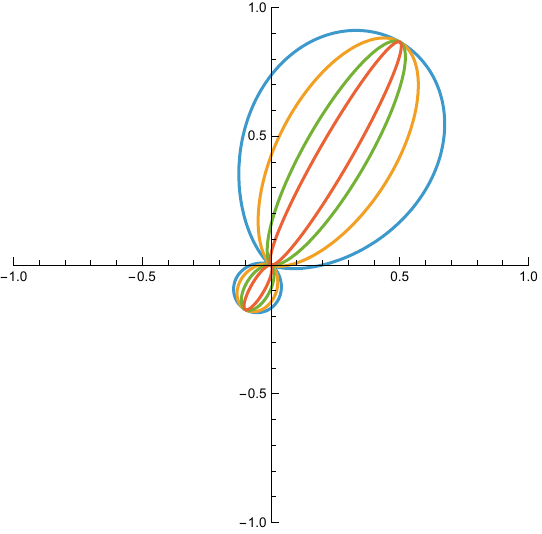}
      & \includegraphics[height=0.19\textheight]{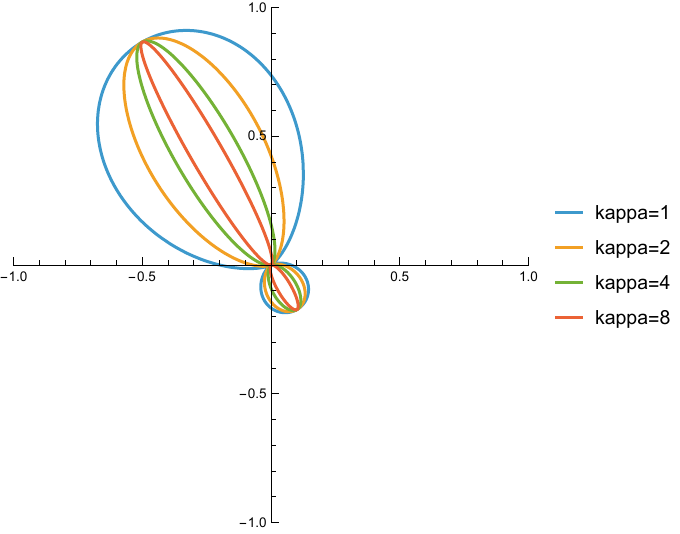}   \\
      $\theta = \pi$ & $\theta = 4\pi/3$ & \hspace{-12mm} $\theta = 5\pi/3$ \\ 
       \includegraphics[height=0.19\textheight]{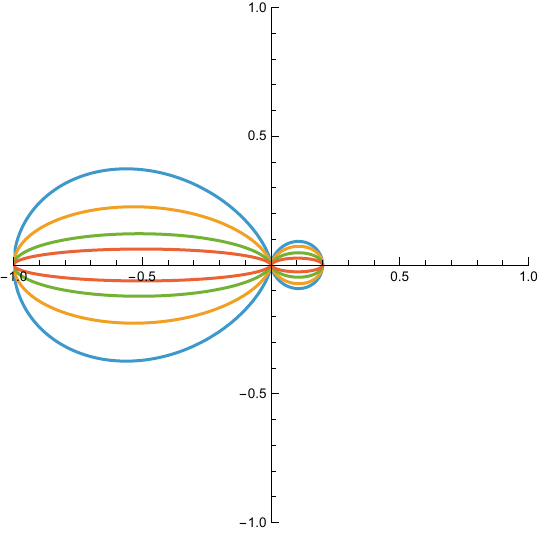}
      & \includegraphics[height=0.19\textheight]{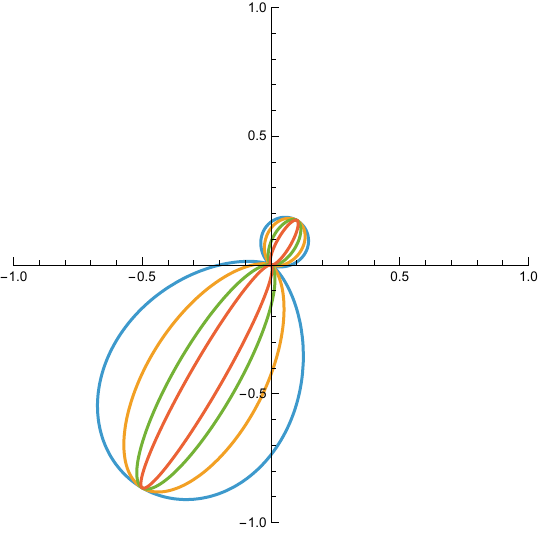}
      & \includegraphics[height=0.19\textheight]{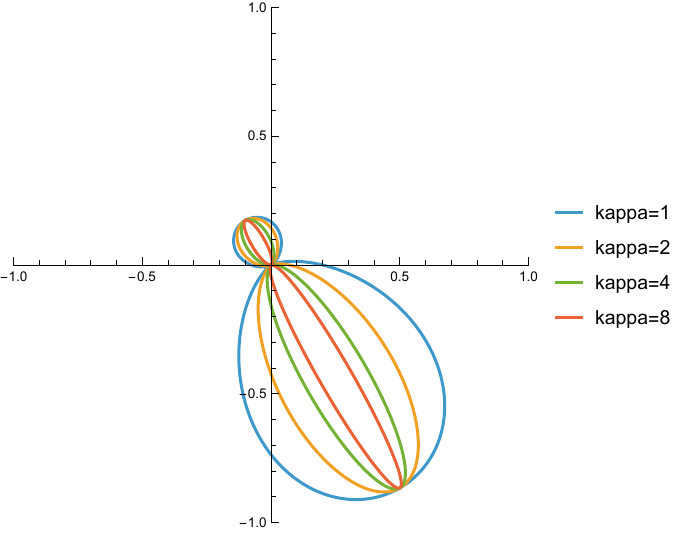}   \\   
    \end{tabular}
  \end{center}
  \caption{Variability in the orientation selectivity properties of
    idealized models of simple cells of the form
    (\ref{eq-spat-temp-RF-model-der-norm-caus})
    in terms of velocity-adapted affine
    Gaussian kernels for spatial derivative orders $m = 1$ and $m = 2$
    and temporal derivative order $n = 0$ over the direction
    $\theta = \arg v$ of the
    image velocity $v$ for a set of velocity-adapted spatio-temporal
    receptive fields in the directions
    $\theta \in \{0, \pi/3, 2\pi/3, \pi, 4\pi/3, 5\pi/3 \}$.
    The different colours of the orientation selectivity curves show
    the results for different degrees of elongation
    $\kappa \in \{ 1, 2, 4, 8 \}$, with the directional selectivity
    curves becomes more narrow for larger values of $\kappa$.
    All the orientation selectivity curves have been computed using the
    same magnitude $|v|$ of the image velocity, only by varying the
    direction $\theta = \arg v$. The magnitudes of each directional
    selectivity curve have, in turn, been normalized such that they
    assume the maximum value $1$ when velocity of the stimulus is equal
    to the velocity parameter of the receptive field. For other velocity
    magnitude values of the stimulus, the peak value of the directional
    selectivity curve will be lower.}
  \label{fig-dir-sel-simpl}
\end{figure*}

\subsection{Additional variabilities involving the temporal domain}
\label{sec-variabil-temp}

Additionally, by extending the arguments in Lindeberg
(\citeyear{Lin23-FrontCompNeuroSci}), we can predict the following
variabilities with regard to time-dependent image data:

\subsubsection{Variability under Galilean transformations}

Regarding processing of time-dependent image data in the primary
visual cortex and the middle temporal visual area MT, we have the
following facts (Lindeberg (\citeyear{Lin25-arXiv-dirsel}):
\begin{itemize}
\item
  the ability of the simple cells in the visual system of higher
  mammals to compute spatio-temporal receptive
  field responses similar to velocity-adapted temporal derivatives
  (DeAngelis {\em et al.\/}\
  \citeyear{DeAngOhzFre95-TINS,deAngAnz04-VisNeuroSci};
  de~Valois  {\em et al.\/}\ \citeyear{ValCotMahElfWil00-VisRes};
  Lindeberg \citeyear{Lin21-Heliyon} Figure~18 bottom part),
\item
  visual neurons in the primate visual cortex being known
  to be direction selective
  (Hubel \citeyear{Hub59-JPhys}, Orban {\em et al.\/} \citeyear{OrbKenNul86-JNeurPhys},
  Churchland {\em et al.\/}\ \citeyear{ChuPriLis05-JNeuroPhys}),
  and ``... organized into subcolumns within an iso-orientation
  column, with each subcolumn preferring motion in
  a different direction'' (Elstrott and Feller \citeyear{ElsFel09-CurrOpNeurBiol}),
\item
  the receptive
  fields in the middle temporal visual area MT being able to compute
  to a rich variety of direction-selective responses
  (Orban \citeyear{Orb97-ExtrStriCortPrim},
  Born and Bradley \citeyear{BorBra05-AnnRevNeurSci}),
\item
  with the output from the primary visual cortex providing 
  input to the middle temporal visual area MT (Movshon and Newsome \citeyear{MovNew98-JNeuroSci})
  and
\item
  with the direction-selective neurons in MT organized into direction
  columns (Albright {\em et al.\/}\
  \citeyear{AlbDesGro84-JNeuroPhys}).
\end{itemize}
From these results, it appears as if we can regard the primary visual
cortex and middle temporal visual area MT as performing expansions of the visual
representations over the directions and speeds of local Galilean
motions.

Such an overall view can specifically also be supported by
the following explicit variabilities over the Galilean parameters in
idealized models of simple cells based on the generalized Gaussian
derivative model:
\begin{itemize}
\item
  {\em Variability over the preferred speeds of motion-sensitive
    neurons:} 
  By modelling biological speed selectivity curves for monkeys
  recorded by Orban {\em et al.\/} (\citeyear{OrbKenNul86-JNeurPhys})
  in terms of idealized models of speed selectivity curves derived from
  our generalized Gaussian derivative model for visual receptive fields
  for the inclination angle $\theta = 0$ of the motion stimulus in
  relation to the preferred direction of the idealized model of a simple
  cell, we have found (Lindeberg \citeyear{Lin25-arXiv-dirsel})
  that the preferred speed of the biological
  neurons vary over a factor of 70 (from 0.68~deg/s to 48~deg/s)
  among the neurons reported by
  Orban {\em et al.\/}\ (\citeyear{OrbKenNul86-JNeurPhys}).
  In this way, there is biological evidence for a variability over
  the preferred speed of simple cells in monkeys, corresponding to the
  magnitude $|v| = \sqrt{v_1^2 + v_2^2}$ of the velocity parameter $v$ in
  the idealized model (\ref{eq-spat-temp-RF-model-der-norm-caus}).
\item
  {\em Variability over the preferred direction of motion-sensitive neurons:}
  By modelling biological direction selectivity curves for monkeys
  recorded by Livingstone and Conway (\citeyear{LivCon03-JNeuroPhys})
  in terms of idealized direction selectivity curves derived from
  idealized models of simple cells based on the Galilean-adapted
  affine Gaussian derivative within our generalized Gaussian
  derivative theory for visual receptive fields,
  we have found (Lindeberg \citeyear{Lin25-arXiv-dirsel}) that
  these biologically recorded direction selectivity curves can be well
  explained by corresponding idealized models based on the generalized
  Gaussian derivative theory. The underlying biological data
  additionally show a substantial variability over the preferred
  directions of the motion-sensitive neurons, as do also the
  corresponding biological results of direction selectivity properties
  for complex cells.
  In this way, (i)~the raw underlying biological
  results support the prediction about an explicit variability in the
  preferred direction of the simple cells, corresponding to the
  direction $\operatorname{arctan2}(v_1, v_2)$ of the velocity parameter $v$ in the idealized model
  (\ref{eq-spat-temp-RF-model-der-norm-caus}). Additionally, (ii)~the
  very good qualitative similarity between the explicit modelling results
  and the results of the neurophysiological measurements
  strongly support the use of the generalized
  Gaussian derivative model for modelling the functional properties of
  the motion-sensitive simple cells recorded in this experiment.
\end{itemize}
Thereby, the comparisons to these biological results are consistent
with the predictions about explicit variabilities in the preferred direction
and speed of motion-sensitive simple cells in the primary visual
cortex, based on which it seems plausible that:
\begin{itemize}
\item
  the visual system could be able to compute Galilean-covariant
  receptive field responses, so as to be able to process visual
  information over a wide range of motion speeds%
  \footnote{The range of velocities may, however, be strongly dependent
  on the distance from the center of the fovea, as characterized in
  detail by
  Orban {\em et al.\/} \citeyear{OrbKenNul86-JNeurPhys}, and probably
  depending on both the spatial size of the receptive field, which roughly increases
  linearly from the distance from the center of the fovea, and the
  minimum temporal scale of the neuron.}
and motion directions,
\item
  this motion processing hierarchy could be based on
  Galilean-covariant receptive fields in the primary visual cortex.
\end{itemize}
Figure~18 in
Lindeberg (\citeyear{Lin21-Heliyon}) together with a corresponding idealized model
of the visual receptive field with explicit Galilean motions, shows
an example of a receptive field of a simple cell that is clearly
sensitive to a particular direction and speed of the stimulus.
Figure~\ref{fig-1spat1tempdir-timecaus-spattempscsp} in this paper shows an
example of such a variability under Galilean transformations for
spatio-temporal receptive fields over a 1+1-D spatio-temporal
domain, based on a second-order spatial derivative of a Gaussian
kernel and a zero-order temporal derivative of the time-causal
limit kernel.
  
Figure~\ref{fig-dir-sel-simpl} shows the variability in the
directional selectivity properties of simple cells, that such a
variability in the image velocity $v$ of the receptive fields gives
rise to, when also including a variability over the degree of
elongation $\kappa$ of the receptive fields, as based on the
theoretical analysis in Lindeberg (\citeyear{Lin25-arXiv-dirsel}).

\subsubsection{Variability under temporal scaling transformations}
\label{sec-var-temp-scaling}

Concerning the degree of freedom corresponding to temporal scaling
transformations, the corresponding degree of freedom in
terms of the temporal scale parameter $\sigma_t= \sqrt{\tau}$ is
also special in the sense that both the non-causal temporal Gaussian
kernel and time-causal limit kernel used for temporal smoothing in
the idealized model (\ref{eq-spat-temp-RF-model-der-norm-caus}) for
spatio-temporal receptive fields obey cascade properties over
temporal scales of the form (Lindeberg \citeyear{Lin23-BICY} Equation~(2))
\begin{equation}
  \label{eq-casc-rel-temp-scsp}
  h(\cdot;\; \tau_2) = (\Delta h)(\cdot;\; \tau_1 \mapsto \tau_2) * h(\cdot;\; \tau_1)
\end{equation}
for any pair of temporal scales $(\tau_1, \tau_2)$ with $\tau_2 > \tau_1$
and for some family of transformation kernels
$(\Delta h)(t;\; \tau_1 \mapsto \tau_2)$.
This implies that any receptive field response, here represented as
the temporal derivative $L_{t^n}(\cdot;\; \tau)$ at 
any coarser temporal scale $\tau_2$, can be computed by an additional temporal
filtering operation being applied to the receptive field responses,
here represented as the temporal derivatives $L_{t^n}(\cdot;\; \tau)$ 
at any finer temporal scale $\tau_1$, according to
\begin{equation}
  L_{t^n}(\cdot;\; \tau_2) =
  (\Delta h)(\cdot;\; \tau_1 \mapsto \tau_2) * L_{t^n}(\cdot;\; \tau_1).
\end{equation}
In the case of a non-causal temporal domain, the difference kernel
$\Delta h$ is a Gaussian kernel
\begin{equation}
  (\Delta h)(t;\; \tau_1 \mapsto \tau_2) = g(t;\; \tau_2 - \tau_1)
\end{equation}
because of the semi-group property
of the Gaussian kernel. In the case of a time-causal domain, the
difference kernel is either a single truncated exponential kernel
\begin{equation}
  (\Delta h)(t;\; \tau_1 \mapsto \tau_2) = h_{\exp}(t;\; \mu_k) 
  = \left\{
    \begin{array}{ll}
      \frac{1}{\mu_k} e^{-t/\mu_k} & t \geq 0, \\
      0         & t < 0,
    \end{array}
  \right.
\end{equation}
or the convolution of a set of such truncated exponential kernels,
with appropriate time constants $\mu_k$ to fit the relationship between
the temporal scale levels $\tau_1$ and $\tau_2$, according to the
recurrence relation in Lindeberg (\citeyear{Lin23-BICY}) Equation~(28).
  
This, in turn, means that a vision system could, in
principle, choose to only implement the earliest layers of temporal
receptive fields at the finest temporal scale and nevertheless
have the ability to compute the representations at coarser temporal
scales, based on additional temporal smoothing applied to the
temporal or spatio-temporal receptive field representations at the finest temporal
scales.
Thus, irrespective of whether the temporal receptive fields are
expanded over the temporal scales, the theory suggests that the
vision system could have the ability to compute visual operations
corresponding to temporal scale covariance.

Figure~\ref{fig-timecaus-1tempders} shows an example of such a
variability under temporal scaling transformations for the
purely temporal time-causal limit kernel.

An extended theory for how an idealized vision system,
based on receptive fields in terms of the generalized Gaussian
derivative theory for visual receptive fields, could compute receptive
field responses at coarser spatial and temporal scales, based on
cascade smoothing properties from finer spatial and temporal scales, is
given in Lindeberg (\citeyear{Lin26-JMIV}).

\subsection{Parameter domains for the shape parameters of the
  idealized models of simple cells}

In this section, we describe the ranges of the shape parameters in the
idealized models of simple cells according to the generalized Gaussian
derivative theory for visual receptive fields, as defined in
(\ref{eq-spat-temp-RF-model-der-norm-caus}).

\subsubsection{Variability of the spatial scale parameter}

According to the most idealized version of the theory, the spatial
scale parameter $\sigma_{\text{x,geom}}$, which describes the spatial extent of the
receptive fields, and which can be defined from the determinant of the
spatial covariance matrix
\begin{equation}
  \Sigma_{\varphi}
  =R_{\varphi} \, \diag(\lambda_{\varphi}, \lambda_{\orth}) \, R_{\varphi}^T
\end{equation}
according to
\begin{equation}
  \sigma_{\text{x,geom}}
  = \sqrt[4]{\det \Sigma_{\varphi}}
  = \sqrt[4]{\lambda_{\varphi} \, \lambda_{\orth}}
  = \sqrt{\sigma_{\varphi} \, \sigma_{\orth}}
\end{equation}
for $\lambda_{\varphi} = \sigma_{\varphi}^2$ and
$\lambda_{\orth} = \sigma_{\orth}^2$,
thus representing the geometric average of the two eigenvalues
$\lambda_{\varphi}$ and $\lambda_{\orth}$ of $\Sigma_{\varphi}$
in units of standard deviations as opposed to variances,
should assume values over the entire positive axis
according to
\begin{equation}
  \sigma_{\text{x,geom}} \in \bbbr_+.
\end{equation}
When to implement this theory in an actual physical or biological
vision system, this scale range must, however, be delimited to a
finite scale range. The resulting minimum and maximum scale values
will hence, determine the working capacity of the system to handle
different types of image structures over a corresponding scale range
\begin{equation}
  \sigma_{\text{x,geom}} \in [\sigma_{x,\min}, \sigma_{x,\max}].
  \end{equation}
Specifically, the minimum scale value $\sigma_{x,\min}$ will impose
a fundamental limit on the effective spatial resolution, concerning how small
image structures can at all be resolved, as measured in units of
degrees of visual angle.
We may hence consider that different species may have different values
of the spatial scale parameter $\sigma_{\text{x,geom}}$, as depending on
the ecological niche within which they operate.

Higher mammals with foveated sensors do specifically handle this
choice of scale range by letting the minimum scale value vary over the
fovea and the visual cortex in the sense that minimum scale value is
smallest in a narrow central part of the visual field denoted the
fovea, and then with the minumum receptive field size increasing
linearly with the distance from the centre of the fovea towards more
peripheral points in the visual field.

As we have discussed earlier in Section~\ref{sec-var-scale-trans}, an
idealized vision system could specifically choose to only represent a
small range of finest scale values in the sensorium, and by the use of
cascade smoothing properties of the forms
(\ref{eq-cascade-smooth-gradient}) and (\ref{eq-cascade-smooth-hessian})
compute image representations at coarser scales in a more implicit
manner at higher layers in the visual hierarchy.

In a recent study
(Meshcherina {\em et al.\/} \citeyear{MesAufLin20-bioRxiv})
of variabilities of simple cells in mice, based on reconstructed receptive
fields from the results of a calcium imaging study performed by
the Allen Institute, it was found that the spatial size of
the receptive fields varied over a rather
narrow scale range corresponding to about a factor of~2.
An open problem still, however, concerns if this limited spatial scale range
constitutes a final answer or if the range of observed set of scales was
fundamentally restricted by the combination of imaging technique and the choice of
stimuli in the underlying neurophysiological experiment,
see Meshcherina {\em et al.\/} (\citeyear{MesAufLin20-bioRxiv})
for a further discussion about this.

\subsubsection{Variability of the spatial orientation angle}

To enable unbiased processing of image structures with different
orientations in the image plane, the orientation angle $\varphi$
should assume its values over the entire circle, that is
\begin{equation}
  \varphi \in [0, 2\pi[.
\end{equation}
For receptive fields corresponding to derivatives of even
differentiation order along the preferred direction of the receptive
fields, it does, however, hold that the receptive fields will be
symmetric under the mapping $\varphi \mapsto \varphi + \pi$.

The basic question concerning the handling of the range of this
parameter then rather concerns how densely this range of
orientations is to be sampled, which may also be considered to be possibly
different for different species. In this context, a denser sampling of
the space of spatial orientations would make it possible for the vision
system to compute more accurate estimates of visual information than
possible with a more sparse sampling.

\subsubsection{Variability over the degree of elongation}

With $\sigma_{\varphi}$ and $\varphi_{\orth}$ denoting the spatial scale
parameters in the directions parallel with {\em vs.\/} orthogonal to
the preferred direction $\varphi$ of the receptive field,
the degree of elongation is defined as
\begin{equation}
  \kappa = \frac{\sigma_{\orth}}{\sigma_{\varphi}}
\end{equation}
and should according to the most idealized version of the theory
assume its values over the entire positive axis
\begin{equation}
  \kappa \in \bbbr_+.
\end{equation}
For a physical or biological vision system, this range does, however,
need to be delimited to a finite range
\begin{equation}
  \kappa \in [\kappa_{\min}, \kappa_{\max}],
\end{equation}
where the minimum elongation $\kappa_{\min}$ and the maximum
elongation $\kappa_{\max}$ then determine the operational range of
the vision system with respect to this dimension of variability.
Specifically, these parameters determine how steep
slant angles towards the fundamental limit $\pi/2$ that the vision
system can accurately handle.

Again, one may consider that different species could have different ranges
regarding the degrees of elongation that they are able to represent.

In the study by Meshcherina {\em et al.\/}
(\citeyear{MesAufLin20-bioRxiv}) over 31 verified simple cells from
mice, elongation values were observed within the range [0.41, 2.15],
corresponding to a variability over a factor of 5.
Due to possible limitations in the underlying calcium image
experiments, it, however, remains as an open problem whether this
limited range is fully representative, or if receptive fields with possibly
higher degree of elongation could not be properly captured by the
combination of the calcium imaging methodology in combination with the
stimuli used for the underlying neurophysiological experiments.

Contrary, in the modelling of the direction selectivity curves
of simple cells in the primary visual cortex of monkeys recorded
by Livingstone and Conway (\citeyear{LivCon03-JNeuroPhys}),
the results were better explained by much larger values of the degree
of elongation, up to 4 or 8 (Lindeberg \citeyear{Lin25-arXiv-dirsel}).
Given that the vision system of monkeys is regarded as much more
developed than the vision system of mice, one may hence speculate
whether the range $[\kappa_{\min}, \kappa_{\max}]$ of elongation
values is substantially wider for monkeys than for mice, to enable
more accurate estimates of visual information under substantial
variabilities in the geometry, specifically in relation to the slant
angles of the observed objects.

Notably, the evolutionary pressure on the vision system could be
expected to be comparably higher towards accurate geometric inference
for higher mammals, such as primates and cats, compared to the
evolutionary pressure on mice, for which more basic detection
properties of the vision system, to avoid becoming prey,
may be more cost-effective in terms of the metabolic costs associated
with a more developed vision system.

\subsubsection{The angle between the angle of spatial differentiation
  and the angles of the principal directions of the spatial covariance
  matrix}

Given the idealized model of simple cells of the form
(\ref{eq-spat-temp-RF-model-der-norm-caus}),
with the direction of spatial differentiation parallel with
one of the eigendirections of the spatial covariance matrix
$\Sigma_{\varphi}$, it holds that the resulting receptive field
representations are closed under a 3-D submanifold of
the 4-D set of general spatial affine transformations.
Since the receptive field responses are thereby not closed
under general spatial affine transformations, 
this means that if a binocular visual observer views a local surface
patch in general position, it will not be possible to match the
receptive field responses computed from two general views of the
same local surface patch. If we on the other hand allow the differentiation
direction $\varphi_{\text{diff}}$ to be different than the direction
of nearest eigendirection $\varphi_{\text{cov}}$ of the spatial
covariance matrix $\Sigma_{\varphi}$, then we can, on the other hand,
obtain a manifold of receptive field responses of genuine
4-D variability, that will allow general views of the
same surface to be matched, to first order of approximation, 
in terms of their receptive field responses.

From this viewpoint, the difference angle
\begin{equation}
  \Delta \varphi = \varphi_{\text{diff}} - \varphi_{\text{cov}}
\end{equation}
could then be regarded as a straightforward measure of
how large deviations can be resolved in relation to the
restricted 3-D submanifold.
Currently, we have no biological data concerning the possible
variability of this descriptor, except for the illustration in
Figure~6 in
Yazdanbakhsh and Livingstone (\citeyear{YazLiv06-NatNeuroSci}),
which shows an example of a receptive field that does not look like any
receptive fields from the 3-D submanifold of receptive fields
of the form (\ref{eq-spat-temp-RF-model-der-norm-caus}),
with the direction $\varphi_{\text{diff}}$ for spatial differentiation parallel with one of
eigendirections of the spatial covariance matrix $\Sigma_{\varphi}$,
but rather like a receptive field for which these directions are
different, and as illustrated further in Figure~8 in Lindeberg (\citeyear{Lin25-BICY}).

Given that the difference angle $\Delta \varphi$ could in the most
idealized version of the theory be allowed to vary over the range
\begin{equation}
  \Delta \varphi \in [-\tfrac{\pi}{2}, \tfrac{\pi}{2}],
\end{equation}
we leave it as an open problem to characterize the possible
variability of this difference angle in biological vision.

\subsubsection{The spatio-temporal speed parameter}

By parameterizing the Galilean velocity vector $v = (v_1, v_2)^T$
in the idealized model (\ref{eq-spat-temp-RF-model-der-norm-caus})
for spatio-temporal receptive fields in terms of polar coordinates,
we obtain the speed parameter from the magnitude
$|v| = \sqrt{v_1^2 + v_2^2}$.
According to the most idealized version of the generalized Gaussian
derivative theory for visual receptive fields, this speed
parameter should assume its values over the entire non-negative axis
\begin{equation}
  |v| \in [0, \infty[.
\end{equation}
In an actual physical or biological implementation of the theory,
this range must, however, be delimited to some finite range
\begin{equation}
  |v| \in [|v|_{\min}, |v|_{\max}].
\end{equation}
for some minimum speed value $|v|_{\min}$ and some
maximum speed value $|v|_{\max}$.
How large the maximum speed value $|v|_{\max}$ is
in practice will then impose a limit on how fast moving visual stimuli can be
observed in a sufficiently accurate manner.
How small the minimum speed value $|v|_{\min}$ is could,
on the other hand, impose a limit on how slowly moving visual patterns
can be reliably distinguished from static visual patterns.

Hence, one may specifically assume that different values of
$|v|_{\min}$ and $|v|_{\max}$ may occur for different species.
For example, it seems likely that $|v|_{\max}$ could be significantly larger
for a cheetah than for a sloth.

In the modelling of biological speed selectivity curves for monkeys
recorded by Orban {\em et al.\/} \citeyear{OrbKenNul86-JNeurPhys},
it was found in Lindeberg (\citeyear{Lin25-arXiv-dirsel}) that the
fitted speed values varied over a factor of 70
(from 0.68~deg/s to 48~deg/s) among the 5~neurons
considered in that study, that is over two orders of magnitude.
To characterize more firmly the range of this variability, a
substantially larger set of motion-sensitive neurons would, however,
need to be investigated.

\subsubsection{The parameter for the preferred spatio-temporal
  direction of the receptive field}

When parameterizing the Galilean velocity vector $v = (v_1, v_2)^T$,
in terms of polar coordinates, the preferred direction for any
idealized model (\ref{eq-spat-temp-RF-model-der-norm-caus})
of a spatio-temporal receptive field is given by the angle
$\operatorname{arctan2}(v_1, v_2)$.
To enable similar unbiased processing of all motion directions, this
parameter should assume its values over the entire circle
\begin{equation}
  \operatorname{arctan2}(v_1, v_2) \in [0, 2\pi[
\end{equation}
Additionally, the distribution of the receptive fields over this
parameter should be uniform over the entire interval.

With respect to practical implementation of the theory, one could
conceive that a denser sampling of these preferred directions of the
spatio-temporal receptive fields could enable more accurate estimates
from the motion of visual patterns, and that different species could
have different degrees of denseness regarding the sampling of this parameter.

\subsubsection{The temporal scale parameter}

According to the most idealized version to the theory, the temporal
scale parameter $\sigma_t$ should assume its values over the entire
positive axis
\begin{equation}
  \sigma_t \in \bbbr_+.
\end{equation}
In a physical or biological implementation of the theory, this
temporal scale interval does, however, need to be delimited to
a finite scale range
\begin{equation}
  \sigma_t \in [\sigma_{t,\min}, \sigma_{t,\max}],
\end{equation}
where the minimum temporal scale level $\sigma_{t,\min}$
will impose a limit on the temporal resolution of the vision system,
concerning how fast changing visual patterns can at all be observed.

Again, it seems plausible that the
minimum temporal scale level $\sigma_{t,\min}$
may be different for different species, expecting that
$\sigma_{t,\min}$ could be significantly lower for a cheetah than for
a sloth.

As we have also described earlier in
Section~\ref{sec-var-temp-scaling},
it may not be necessary for a vision system to explicitly represent
the coarser temporal scale levels in the earliest stages of the sensorium.
Instead, due to the cascade smoothing property over temporal scales,
the visual representations at coarser temporal scale levels could be
computed in an implicit manner from the corresponding representations
at finer temporal scales at later stages in the visual hierarchy.

\subsection{Outlines to further research to characterize
  variabilities of visual receptive fields with regard to
  variabilities in relation to geometric image transformations}
\label{sec-outlines-neurophys}

Given the above theoretically motivated predictions, main questions for
further research concern:
\begin{itemize}
\item
  For which of the above variabilities can neurophysiological support
  be found regarding expansions of the shapes of the receptive fields
  over the corresponding degrees of freedom for different types of species?
\item
  Are the degrees of freedom corresponding to spatial scaling
  transformations and temporal scaling transformations special, because
  of the properties that the smoothing operations over the spatial and
  the temporal scale parameters can be performed in terms of cascade
  smoothing operations, implying that only representations of explicit
  receptive field responses at the finest spatial and/or temporal
  scales would be needed to compute  the receptive field responses at
  coarser and/or temporal scales?

  Does efficiency call for the visual system to only implement the
  finest spatial and temporal scales in the sensorium, or does the
  vision system implement redundancies in this respect?
\item
  If there would be expansions over both the degrees of freedom of the
  affine group beyond uniform scaling transformations and rotations as
  well as an expansion over the degrees of freedom of the Galilean
  group, are those expansions then performed jointly or separately for
  different species?
  In other words, are such expansions performed over the tensor
  product of the corresponding affine and Galilean parameters for some
  species, or
  separately for each subset of these degrees of freedom for other species?

  For efficiency reasons, separate expansions over the affine and the
  Galilean degrees of freedom would need much less wetware to span the
  degrees of freedom of the corresponding geometric image
  transformations. A joint expansion over these degrees of freedom
  could on the other hand allow for the computation of more accurate
  estimates of the 3-D structure of dynamic scenes.
\item
  How well do the idealized models for
  spatial and spatio-temporal receptive fields explicitly listed in
  Section~\ref{sec-expl-ex-cov-props-rfs} model the biological
  receptive fields for simple cells in the primary visual cortex for
  different species? 
\item
  For which combinations of spatial and temporal derivative operations
  according to the theory for generalized Gaussian derivative based
  spatial and spatio-temporal receptive fields can
  biological receptive fields be found with qualitatively similar
  shapes for different species?
\item
  Specifically, what fractions of the receptive
  fields in the primary visual cortex for higher mammals can be
  reasonably well modelled by idealized receptive fields according to
  the here described generalized Gaussian derivative theory for visual
  receptive fields?
\item
  How wide ranges in the parameter spaces of the receptive fields are
  spanned for the respective degrees of freedom for different species?
  Does the vision system span dense sets over the multi-dimensional
  parameter space, or does the vision system only span sparser
  submanifolds, with possibly coupled dependencies between different
  subsets of the parameters?
\end{itemize}
Answering these questions could provide valuable cues about how well different types
of species have adapted and organized their visual systems to handle the influence
of different types of geometric image transformations on the image data.

Unfortunately, previous neurophysiological studies of visual receptive
fields may not have been designed to answer the questions raised from
this theoretical background. Answering the question about
variabilities in the shapes of the visual receptive fields
corresponding to the simple cells in the primary cortex would call for
the accumulation of shape parameters of reconstructed receptive fields
for a relatively large number of visual neurons, as well as using
visual stimuli that would probe against the listed types of
variabilities with regard to the 4 main types of geometric image
transformations as leading to a total number of 7 degrees of freedom.

Currently, it seems very hard to get access to raw
reconstructed receptive field profiles from previous recordings of the
visual receptive fields in V1 of higher mammals. Publicly available
receptive fields, as published in scientific publications, typically
only show a rather small number of visual receptive fields.
Concerning the hypothesized variabilities with regard to the degree of
elongation of visual receptive fields, the neurophysiological evidence,
that we have based our previous hypothesis on, are based on
orientation selectivity properties of higher mammals
(primates or cats) with a clear pinwheel structure. Our support for
Galilean covariance is correspondingly largely based on measurements
of direction and speed selectivity properties of motion-sensitive
neurons.

Recordings from mice, for which there is the
largest number of public data, may on the other hand not be able to
fully answer this
question, because the vision system of mice is far less developed
compared the visual systems of higher mammals
(see {\em e.g.\/}\ Huberman and  Niell (\citeyear{HubNie11-TINS})),
and does specifically lack pinwheel structures.
Indeed the results about variabilities in the degree of elongation
that we have obtained by reconstructing receptive fields from
calcium imaging studies of mice
(Meshcherina {\em et al.\/} \citeyear{MesAufLin20-bioRxiv})
correspond to a lower variability than could be inferred from
our results of modelling direction selectivity curves recorded
for monkeys (Lindeberg \citeyear{Lin25-arXiv-dirsel}).
In addition, recent results by
Fu {\em et al.\/}
(\citeyear{FuPieWilBasMuhDiaFroResPonDenSinTolFra24-CellRep})
of reconstructing most exciting inputs (MEIs) for early visual receptive
fields in mice suggest that complex spatial features emerge earlier in the
visual pathway of mice compared to primates, and that the receptive
fields of mice are therefore more complex than a Gabor or Gaussian
derivative model, as used in the theoretical model for stating the
predictions in this paper.

For these overall reasons,
in Lindeberg (\citeyear{Lin23-FrontCompNeuroSci}) Sections~3.2.1--3.2.2,
Lindeberg (\citeyear{Lin25-JCompNeurSci-spanelong}) Sections~4.2--4.3 and
Lindeberg (\citeyear{Lin25-arXiv-dirsel}) Section~8,
sets of additional suggestions for more specific neurophysiological experiments
are proposed
to investigate the underlying hypotheses in more detail, and to characterize the
structure of biological receptive fields in the primary visual cortex
with respect to the influence of parameters of the receptive fields
corresponding to the different degrees of freedom of the 4 main types
of geometric image transformations.
Complementary suggestions for further research in relation to the
influence of geometric image transformations on early vision are also outlined in
Lindeberg (\citeyear{Lin25-BICY}) Section~8.1.

Additionally, concerning the extension to non-linear visual neurons,
in Lindeberg (\citeyear{Lin25-PONE}) it is shown
that that the results concerning an expansion over the degree of
elongation for simple cells do also appear to extend to complex cells,
with the modelled degree of elongation varying over the range
[0.65, 5] corresponding to a variability over a factor of the order of 7 to 8.
The results of modelling speed selectivity and direction selectivity
for complex cells in the primary visual cortex in
(Lindeberg \citeyear{Lin25-arXiv-dirsel}) also support that the
variabilities in the speed parameter and the preferred direction of
spatio-temporal receptive fields do appear to extend to complex cells.

\section{On the framework for theoretical modelling used in the
  presented theory about visual receptive fields}
\label{sec-model-framework}

When a visual observer views general types of objects and
spatio-temporal events in the environment, the perspective image
transformations from the environment to the retinal surface or the
image plane are generally described by non-linear projection equations.
So are the image transformations between different views of the same
object or event, as observed by the two eyes or cameras in a binocular
viewing system or from views at different time instances of the same
scene, also described by non-linear geometric image transformations.

Modelling such non-linear geometric transformations fully accurately, as originating
from views of general curved objects in the environment, can indeed
constitute quite a hard problem, or even be regarded as intractable if one would aim
at a theoretical analysis in closed form, with desirable explicit
expressions for the involved entities.

In this section, we give a summary of the different types of major
types of conceptual simplifications and assumptions that are used
in the theoretical modelling underlying this work, and why these
simplifications and assumptions can be regarded as appropriate, and
not overly restrictive, with regard to the overall conclusions that
are drawn from the presented theoretical framework.

The general principle that we follow in this work, is that we base the
treatment on an {\em idealized model\/} of visual receptive fields with their
relationship to geometric image transformations. This approach is
similar to approaches in theoretical physics and theoretical biology,
where one focuses the analysis on modelling those aspects of a more
complex system that are to be analyzed, while abstracting away other
phenomena that are judged to be of lower relevance with respect to the
overall goals of a specific analysis.

Hence, the presented theory does not in
any way aim at constituting a fully detailed and quantitatively highly
accurate model of the full vision system with all the external factors
that may influence the neural computations. Clearly, formulating and
analysing such a detailed model for an as complex object as the
visual system would be an extremely of even intractably hard task.

Instead, the treatment
focuses on how variabilities in the shapes of projected image
structures under different types of imaging conditions will affect the
receptive field responses, and how those geometrically induced
variabilities in the image structures call for the vision system to
have mechanisms for handling such variabilities, which will then lead
to corresponding variabilities in the shapes of the receptive fields in the primary
visual cortex.

Thereby, we argue that it is not necessary to in detail model the
multitude of fine-grained mechanisms in the primary visual cortex, in
order to draw the main conclusions in the article. In the following,
we explain why, regarding some of the main additional factors that may affect the
receptive field responses. In addition, we outline how the
theoretical model could be extended in different ways,
both regarding predicting corresponding variabilities for complex cells
and computing more accurate receptive field responses near
discontinuities in depth or surface orientation.

\subsection{Local linear approximations}
\label{sec-loc-lin-approx}

Fortunately, the geometric modelling problem can be substantially simplified,
if approximating the non-linear perspective or projective projection
equations by {\em local linearizations\/}. This amounts to locally
approximating each local surface patch with its local tangent plane,
and then performing mappings between the tangent planes, as
constituting a standard methodology in differential geometry,
see {\em e.g.\/}\ O'Neill (\citeyear{Nei66}) for an introductory 
overview.

While such local approximations may not be as numerically accurate as
the original non-linear projection mappings, a very useful property
of such local linearizations is that they make the theoretical
analysis much more manageable to carry out in closed form.
The use of such local linearizations is indeed a standard methodology
in a wide variety of scientific fields that make use of mathematical
analysis as a main tool.

In this context, it should also be emphasized that while the use of
{\em global\/} linear approximation may in many cases not be a very good
approximation, by the use of {\em local\/} linear approximations a new model
of the local linear approximation is used for each image position $(x, y)$
corresponding to the appropriate backprojection to the world, and for
each time moment $t$. In this respect, the use of local linear
approximations should not be regarded as fundamentally restrictive.
Instead, by the use of local approximations, we argue that the
proposed theoretical framework should be able to be able to reflect
the overall {\em gross\/} properties concerning how variabilities in the
geometric viewing conditions combined with geometric covariance
properties by necessity map to variabilities in the shapes of the
visual receptive fields over the joint spatio-temporal image domain.

For the local linearizations to constitute accurate approximations of
the local image geometry,
the form of local linearization should vary {\em slowly\/} within the spatial or
spatio-temporal support region of each receptive field. This means that
the form of the model used for the local linearization should not be
too different between different points and time moments within the
support region of a visual receptive field based on which an actual
decision is to be made. If on the other
hand the form of the local linearization would vary significantly
within the support region of the receptive field, then that should be
taken as an indication that the variations in the image geometry are
too fast relative to the current levels of spatial and temporal scales.
Therefore, to be able to reasonably well reproduce the properties
of the local imaging geometry, finer scale spatial and/or temporal
scale levels would then need to be chosen.

If one would instead aim at quantitatively potentially more accurate
modelling of how the underlying fully non-linear projection models
interact with curved surfaces in the environment, one would most
likely by necessity have to resort to numerical simulations, which
would then also imply a substantial data analysis stage to form
concrete abstractions from the necessary large number of simulation
results, concerning how different shape parameter of the
viewed objects interact with different parameters of the geometric
viewing conditions in a compact form. While it would indeed constitute
an interesting problem to quantitatively characterize the accuracy of the local
linear approximations in relation to such genuinely non-linear ground
truths, it would constitute a substantial effort by itself to carry
out such project in a rigorous manner, why we leave that topic
to future work.

In relation to the theoretical model in terms of local linear 
approximations, it should, however, be emphasized that due to the incorporation
of local Galilean motions in the model, it is not in any way necessary
that the relationship between the object and the model remains
constant during the temporal integration duration of the
spatio-temporal receptive fields. Rather, the model is, because of the
temporal scale covariance property, able to handle both shorter and
longer temporal integration durations. Similarly, by the incorporation
of scale covariance, affine covariance and Galilean covariance properties, the model is
also able to handle different viewing conditions as arising from
observing possibly moving objects or spatio-temporal events at
different time moments.

This property thereby means that both (i)~the object, (ii)~the observer and (iii)~the
viewing direction may be moving during the temporal integration
duration of the spatio-temporal receptive fields. Specifically, the
receptive field model has been designed so as to be able to handle
such relative motions in a consistent manner, to be able to handle
the variabilities in visual perception that may arise for a moving
visual observer in a dynamic environment.

\subsection{Smooth local surface patches}

Another modelling assumption underlying the geometric projection
models in Equations~(\ref{eq-x-transf})
and~(\ref{eq-sc-aff-vel-transf-alt-obs-model}) is that the surfaces of
the viewed objects are considered as locally smooth, so as to be able
to define local tangent planes of the viewed objects. 

Obviously, such a model breaks down at occlusions, where there may be
discontinuities in depth, as well as when there are discontinuities in
the local surface orientation.
The proposed model is therefore generally concerned with {\em interior
points\/} of the viewed surfaces, that are assumed to be a sufficient
distance away from any discontinuities or occlusions.
Seen from the perspective of an entire
scene, we could, however, expect that when viewing a nearby object,
the interior points, for which the occlusions are sufficiently far
away, these interior points ought to represent a sufficient number of
image points. Applying the normative arguments to these interior
points does thereby by itself impose a necessary structure regarding
variabilities in the shapes of the receptive fields, so as to be able to handle
different viewing conditions in a provably covariant manner, and as
postulated as a main underlying assumption for the presented theory.

When observing occlusions under the influence of geometric image
transformations, there will, in general, be qualitative different
types of variabilities in the image structures, as caused by either new physical points 
becoming visible or previous physical points becoming invisible from the field of
view, as caused the temporal variabilities of the occlusion.
In contrast to the views of smooth local surface patches,
there may also be discontinuities in the corresponding
deformation fields across the different sides of the
discontinuities. These types of spatio-temporal image structures
thereby go beyond what can be covered by the overall underlying geometric
model in Equations~(\ref{eq-x-transf}) 
and~(\ref{eq-sc-aff-vel-transf-alt-obs-model}).
  
Any additional explicit model of occlusions, or other discontinuities, could only imply further
possibilities regarding complementary variabilities and types of receptive
field, for example, concerning end-stopping neurons. In no way,
however, is the presented theory for the receptive fields of simple
cells intended to be regarded as exclusive, since there are also other
types of visual neurons in the primary visual cortex.
Indeed, it seems likely that a vision system ought to also
comprise more specialized types of receptive fields, dedicated to
handling the qualitatively different types of spatio-temporal image
structures that arise near occlusions and discontinuities.
  
From this theoretical background, the exclusion of occlusions from the
model for the visual receptive fields does not in any way constitute
any restriction for drawing the main conclusions presented in this
study, regarding the proposed 7 types of variabilities in the shapes
of the receptive fields of simple cells in the primary visual cortex.

Due to the non-infinitesimal size of
the receptive fields, however, when applying the receptive fields
according to our theoretical model
at or sufficiently near such occlusions and discontinuities,
there will be a resulting mixing of the information
from the different sides of the discontinuities. Near occlusions, the
receptive field responses may additionally be affected by certain
image points becoming either visible or invisible, as caused by the
temporal variability of the occlusion. Thereby, the
receptive field responses will be smeared out between the different
sides of the occlusions or discontinuities. Additionally, the
assumptions underlying the geometric model in
Equations~(\ref{eq-x-transf}) 
and~(\ref{eq-sc-aff-vel-transf-alt-obs-model})
will no longer hold. This then specifically means that the
receptive field responses would no longer be covariant under the
geometric image transformations.
  
To handle the influence of occlusions or discontinuities in surface
orientation on the receptive field responses in an actual
implementation of the receptive field model, there is, however,
a straightforward way to extend the model,
by instead of using the explicit shapes of the receptive fields for the
different parameters of the receptive field, considering an alternative
implementation approach, by computing the
receptive field responses based on the local differential and difference
equations that emulate the computational function of the receptive fields under
increasing values of the spatial and temporal scale parameters.

Specifically, for the purely spatial component of the receptive field
model, which is the component of the receptive field model mostly
affected by discontinuities in depth and/or surface orientation,
the evolution over the scale parameter $s$ is determined by the affine
diffusion equation (Lindeberg \citeyear{Lin21-Heliyon} Equation~(19))
\begin{equation}
  \label{eq-vel-adapt-scsp-diff-eq}
  \partial_s L 
  = \frac{1}{2} \, \nabla^T \left( \Sigma \, \nabla L \right) - \delta^T \nabla L,
\end{equation}
where the spatial covariance matrix $\Sigma$ determines the amount of
diffusion over the spatial domain and the parameter $\delta$ can be
adapted to the Galilean motion velocity $v$, by multiplication with a
local time variable $\Delta t$, in the correspondingly
extended spatio-temporal receptive field model.

Specifically, be reformulating that underlying diffusion equation
into a non-linear diffusion equation, in a way so as to prevent smoothing across
discontinuities in depth or surface orientation, while allowing for
near linear smoothing within the interior regions of the visual field
that are sufficiently far away from the occlusions or discontinuities,
should enable the computation of more accurate receptive field
responses at either sides of the discontinuities. These receptive
field responses would then obey the geometric model in
Equations~(\ref{eq-x-transf}) 
and~(\ref{eq-sc-aff-vel-transf-alt-obs-model}),
although with different parameters at either sides of the occlusions
or discontinuities, and in this respect then also being covariant
under the geometric image transformations.

In this way, the domain of applicability of the
model in the image space could be substantially enlarged to points
much closer to discontinuities or occlusions. Yet, sufficiently far away
from the discontinuities or occlusions, the behaviour of the non-linear diffusion
model should be essentially the same as for the linear model for
simple cells used in the submitted paper. In this way,
the extension to a non-linear diffusion model for the
spatio-temporal receptive fields could significantly extend the
domain of the regions in image space where the receptive field
responses obey the geometric covariance properties, thus implying that
it should be possible to compute geometrically consistent while
numerically different receptive field responses at the different sides
of occlusions and/or discontinuities. 

In the area of computer vision, the underlying linear scale-space
theory based on smoothing with Gaussian kernels has indeed been
extended to a variety of different non-linear diffusion methods,
see {\em e.g.\/}\ the monograph edited
ter~Haar Romeny (\citeyear{Haa94-GDDbook}) for an introductory
overview.
Additional both theoretical and experimental work is, however,
necessary to extend these ideas about extended formulations in terms
of non-linear diffusion for the considered generalized Gaussian
derivative model, why we leave that topic to future work.

\subsection{Exact continuous models of the receptive fields}

In the theoretical treatment, we have throughout modelled both the
measurement so of the image intensities and the receptive fields
as continuous functions.%
\footnote{Additionally, we assume that the image intensities are
  bounded and locally integrable, which does, however, not constitute
  any serious limitation for real-world image data.}
In a physical or biological implementation of
the theory, however, both the image intensities and the spatial domain
need to be discretized. This implies that the covariance properties, that
are assumed to be exact in the theory, cannot hold exactly after
discretization. Still, however, the presented theory can be
approximated numerically in a physical or biological implementation,
with different possible degrees of accuracy, as depending on the
available hardware/wetware and the computational ability of the
corresponding computational mechanisms. Specifically, one may conceive
that different species could approximate the ideal covariance
properties to different degrees of approximation, and lay out
distributions of receptive field shapes with different ranges for the
shape parameters of the receptive fields, as well as different
densities concerning how sparsely the parameter space
is sampled.

\subsection{Disregarding the effect of illumination variations}

Throughout this treatment, we have been concerned with how
variabilities in the geometry of different viewing conditions interact
with the receptive field responses. Thereby, we have not included any
explicit models for illumination, or for how illumination effects interact
with the surface properties of the viewed objects. Obviously, the
latter can be very important for fine-grained modelling of how images
are formed under different viewing conditions.

For this work, we have primarily made this theoretical simplification
to obtain a manageable theoretical model that amends itself to
closed-form theoretical analysis. A fundamental theoretical result
that we have, however, presented in this context is to in
Section~\ref{sec-illum-var} describe how the theoretical model in
terms of spatial or spatio-temporal derivatives of spatial or
spatio-temporal smoothing kernel is {\em invariant\/} to local multiplicative
illumination variations or multiplicative variations of exposure
parameters, provided that the image intensities are measures on a
logarithmic brightness scale. In this respect, the theoretical model
has a very strong invariance property under a very important class of
gross illumination variations and variabilities in the exposure parameters.

Furthermore, if the local surface patches can be well modelled by a
Lambertian reflection model, and if the illumination field is uniform
and constant over time, then the intensity values of the corresponding
image points on the retina or the image
plane will remain constant for the different viewing conditions.
In this special case, the presented theory is thereby directly
applicable and provides a corresponding strong
constraint for how variabilities in the viewing conditions by
necessity have to lead to variabilities in the shapes of the receptive
fields by the postulated assumption of covariance properties.

Arguing that if an idealized vision system is to be able to handle the
influence of geometric image transformation on the receptive field
responses under a more general illumination and reflection model,
it follows that 
the vision system must also be able to handle the influence of
geometric image transformation on the receptive field
responses under this specific illumination and reflection model.
Thus, we should thereby be able to infer that the derived types of
variabilities derived for the purely geometric model in this treatment
do therefore also ought to hold if the geometric model is combined
with a more general and fine-grained combined illumination and
reflection model. Any more complex illumination and reflection models
can only add additional variabilities to the study.
Thereby, the omission of more developed illumination
or reflection models from the overall analysis in this paper
does not constitute any limitation with regard to
drawing the conclusions about the implied 7~variabilities in the
shapes of the receptive fields.

Studying the influence on the receptive field responses for more
general illumination variations, as obtained for more general
reflectance models for the viewed surfaces, would in this context
notably lead to much more complex analysis, where again one would then have to
resort to numerical simulations to carry out the investigations in
practice. While such studies could possibly
also be interesting to consider, the task of summarizing the results
obtained from such (again very extensive) simulations would imply a substantial amount of
work, why we leave that topic to future work.

\subsection{Handling noise in the image data}

The presented theory is based on relationships between image geometry
and receptive field responses. For this purpose, we have modelled the
image intensities before and after the geometric image
transformations, treating the image intensities as they are and not
considering explicit models for sources of noise in the data. 
Instead, any noise in the image data is treated as an integral part of
the input data and not as the result of complementary perturbations.

The use of Gaussian smoothing operations over the spatial image domain
and the use of either Gaussian smoothing over a non-causal temporal
domain or smoothing with truncated exponential kernels over a
time-causal temporal domain, as used as the underlying computational
primitives in the proposed idealized models for visual receptive
fields, does, however, lead to receptive field
responses that are very robust to the influence of noise in the image
data.

First of all, if we isolate the smoothing operations with either
Gaussian kernels or truncated exponential kernels over one-dimensional
signal domains, then it can be shown that these smoothing operations
are so-called variation-diminishing smoothing operations, that are
guaranteed to never increase the number of local extrema or the number
of zero-crossings in any signal. This follows from
theoretical results by Schoenberg (\citeyear{Sch50}) also
restated in Lindeberg (\citeyear{Lin23-BICY}) and imply a very strong
smoothing property, that will effectively suppress the influence of
spurious noise in the image data.

Secondly, receptive fields based on Gaussian
smoothing and Gaussian derivatives have also been very successfully
used for computing image features and image descriptors from
real-world image data based on scale-space theory in classical
computer vision, see {\em e.g.\/} Lindeberg (\citeyear{Lin93-Dis})
and ter~Haar Romeny (\citeyear{Haa04-book}) for overviews
and the biannual series of conferences on Scale-Space Theory 1997--2005 and Scale Space
and Variational Methods in Computer Vision 2007--2025 for more
specific treatments.

Hence, the combined use of spatial and temporal smoothing with spatial
and temporal derivative operators can provide the ability to compute
robust receptive field responses, provided that the spatial and the
temporal scale levels are appropriated determined in relation to the
actual noise level in the image data. In practice, this means that
coarser spatial and temporal scale values will be needed with
increasing noise levels in the data.

\subsection{Focusing on the computational function of simple cells}

Throughout this treatment, we have solely focused on the computational function
of simple cells in the primary visual cortex of higher mammals,
as modelled in terms of linear receptive fields based
on idealized models of visual receptive fields according to the
generalized Gaussian derivative model for visual receptive fields.
According to the taxonomy by Hubel and Wiesel
(\citeyear{HubWie59-Phys,HubWie62-Phys,HubWie68-JPhys,HubWie05-book}),
there is, however, also a complementary category of complex cells.

Our main reason for focusing on simple cells in this treatment is that
the simple cells constitute an important class of neurons that
appear to be both (i)~easier to understand in terms
of their functional properties and (ii)~substantially easier to model
theoretically, because of their linearity properties.

This restriction of scope made in the theoretical analysis should,
however, not be taken as restricting the notion of covariance
properties to simple cells only.
Contrary, we propose that it ought to
be possible to extend corresponding notions about geometric covariance
properties to complex cells. Specifically, if we would consider
modelling complex cells as operating on the output from neural computations
similar to simple cells, it seems plausible that the covariance
properties for simple cells ought to extend to complex cells.
Thereby, the shapes of the receptive
fields underlying the computations in complex cells do also ought to be
expanded over corresponding degrees of freedom as for the simple
cells,
so as to be able to consistently
handle variabilities in the image data induced by the same classes of
geometric image transformations. Such a view, is specifically
consistent with the generalized view regarding propagating the
covariance properties of the simple cells to handling the influence of
geometric image transformations in higher visual areas,
as proposed in
the introduction of Section~\ref{sec-span-vars}
and in Appendix~\ref{app-prop-cov-props-high-layers}.

Indeed, for one way of modelling complex cells,
in terms of quasi quadrature combinations of simple cells,
in the case of complex cells over a purely spatial image domain
modelled of the form
\begin{equation}
  \label{eq-quasi-quad-dir}
  {\cal Q}_{\varphi,12,\text{pt}} L
  = \sqrt{L_{\varphi,\norm}^2+ C_{\varphi} \, L_{\varphi\varphi,\norm}^2},
\end{equation}
that is in terms of an energy model of combined first- and
second-order oriented spatial derivative responses,
we have in Lindeberg (\citeyear{Lin25-PONE}) demonstrated that
variabilities in the orientation selectivity properties of complex
cells recorded by Goris {\em et al.\/}\
(\citeyear{GorSimMov15-Neuron})
are consistent with the receptive fields having a variability over the
degree of elongation of the receptive fields.

For a spatio-temporal extension of that idealized quasi quadrature
model for complex cells, we have additionally in
Lindeberg (\citeyear{Lin25-arXiv-dirsel}) shown that idealized models
of speed selectivity curves and direction selectivity curves do very
well model biologically recorded speed selectivity curves recorded
for hypothesized motion-sensitive complex cells
by Orban {\em et al.\/} (\citeyear{OrbKenNul86-JNeurPhys}) as well as
biologically recorded direction selectivity curves for verified
complex cells recorded by Livingstone and Conway
(\citeyear{LivCon03-JNeuroPhys}).
Specifically, these modelling results are very well consistent with
explicit variabilities over the speed parameter $|v|$ and the preferred
direction $\operatorname{arctan2}(v_1, v_2)$ of complex cells.
Additionally, the modelling results for
the direction selectivity curves are also consistent with an explicit
variability in the degree of elongation $\kappa$ of the underlying receptive
fields in complex cells.

Generalizing from this observation, we therefore propose the
hypothesis that the variabilities over spatial scale, orientation,
elongation, image velocity, and temporal scales
described in Sections~\ref{sec-variabil-spat}
and~\ref{sec-variabil-temp} regarding simple cells could in a
corresponding manner be extended to also comprise complex cells.
To answer this question firmly, access to new neurophysiological
recordings, as well as more quantitative modelling work,  would,
however, be needed, since previous work on recording properties of as
well as modelling the functional properties of complex cells may not
have been performed in a way designed so as the answer the questions raised
from this new theoretical background.

\subsection{Disregarding the influence of a foveated visual sensor,
  saccades, adaptation, attention, learning and task context}

Many higher mammals, like primates and cats, have a foveated retina, with a
substantially higher density of photoreceptors in the center, and a
gradually decrease in the density of the effective receptive fields
towards the periphery. Quantitative measurements indicate that the
minimim receptive field size increases essentially linearly from the
central region of the fovea towards the periphery, see
Koenderink and van Doorn (\citeyear{KoeDoo78-BC})
and Lindeberg (\citeyear{Lin13-BICY}) Section~7.

In this treatment, we have, however, not explicitly modelled that type
of influence on the spatial scale levels due to a foveated image
sensor. Instead, we consider the receptive field responses as being
defined for any scale level. Extension to a foveated image sensor can,
however, be performed by extending the spatial scale-space model
underlying the treatment in this paper with the foveal scale-space model
in Lindeberg and Florack (\citeyear{CVAP166}), also summarized with respect
to some of the major components in Lindeberg (\citeyear{Lin13-BICY}) Section~7.

To handle the position-dependent resolution on the retina of a
foveated sensor, animals with such a foveated retina make use of saccades, to
direct the attention to different foci of interest. While we humans in
our consciousness perceive the world as coherent, the underlying
neural representations in the earliest layers in the visual hierarchy
would exhibit substantial discontinuities over time at the time
instants of the saccades, which may occur after a few couples of
100~ms. For animals with foveated retinas that perform saccades, the
presented theory would therefore be applicable at the lowest neural
levels for the time periods in between the saccade movements.
Thus, there should be a reset of the temporal integration mechanisms in the
spatio-temporal receptive fields at the time instances the saccades.

Let us, however, state that due to the incorporation of local Galilean
motions into the receptive field model, the receptive field model is
able to handle both objects that have a constant relationship between
the object and the viewing conditions, as will be the case if the
viewing direction follows a moving object by smooth pursuit,
as well as objects that have explicit non-zero relative motion in relation to
the viewing direction in between the saccades.

More generally,
given the computational functionalities of the vision system, that
enable a stable world view across multiple cascades, one could,
however, also consider applying corresponding normative requirements
at coarser temporal scales regarding covariance properties to the
visual representations in an abstracted stable world
view, at a higher level where the influence of the
saccades has then been abstracted away.
Furthermore, for animals that do not perform saccades,
corresponding normative arguments regarding covariance properties
could, without such restrictions, be applied straight on at coarser temporal
scales.

Let us also remark that it is well-known that mechanisms
such as adaptation, attention, learning and task context can
also dynamically modify the characteristics of V1 neurons.
Since the main focus of this treatment is to study the influence of
geometric image transformations on the receptive field responses, we
have in this treatment decided to not model such more fine-grained
mechanisms.
This choice is similar to the types of choices made in theoretical physics and
theoretical biology, where the scientific methodology focuses on
modelling the phenomena of interest to be study, while isolating the
studies from other external mechanisms that are not of primary
interest with respect to a given study.

By adding such mechanisms one obtains inductive biases in the system,
and thereby a more specialized vision system. The theory in this paper
is formulated from the viewpoint of a general baseline, where no specific
biases are introduced into the system.

Following the general above arguments regarding other sources of
variabilities above, such complementary mechanisms could again, however,
only add further sources to variabilities to the neurons.
Still, the influence of geometric image transformations on the image
data should constitute an environmental drive for variabilities in the
shapes of the receptive field that ought to persist also when including
adaptation, attention, learning and task context into the analysis,
specifically since the vision system ought to be able
to handle the influence due to image transformations at time scales
much faster than the time scales of adaptation, learning or task context.
The established relationships between predictions from the theory with
neurophysiological measurements are also consistent with such a view.


\subsection{Relations to other types of theories and models for visual receptive
  fields}

Beyond the normative theory for visual receptive fields, that this
treatment builds upon, other types of theories have also been proposed
for explaining biological receptive fields, such as sparse coding
(Olshausen and Field \citeyear{OlsFie96-Nature,OlsFie97-VR},
Rehn and Martin \citeyear{RehSom07-JCompNeuroSci},
Zylberberg {\em et al.\/} \citeyear{ZylMurDeW11-PlosCompBio},
King {\em et al.\/} \citeyear{KinZylDeW13-JNeuroSci})
and predictive coding
(Singer {\em et al.\/} \citeyear{SinTerWilSchKinHar18-Elife};
Kwan and Park \citeyear{KwoPar19-ICCV};
Lotter {\em et al.\/} \citeyear{LotKreCox20-NatMachIntell}).
Models of visual receptive fields have also been proposed based based on
convolutional neural networks (CNNs)
(Keshishian {\em et al.\/} \citeyear{KesAkbKhaHerMehMes20-Elife})
although there have also been raised issues concerning such approaches
(Bae {\em et al.\/} \citeyear{BaeKimKim21-FrontSystNeuroSci},
Bowers {\em et al.\/}\
 \citeyear{BowMalDujMonTsvBisPueAdoHumHeaEvaMitBly22-BehBrainSci},
Heinke {\em et al.\/}\ \citeyear{HeiLeoLee22-VisRes},
Wichmann and Geirhos \citeyear{WichGei23-AnnRevVisSci}).

While both the sparse coding and predictive coding approaches have been
demonstrated to be able to extract receptive fields from training data
with good qualitative similarities to biological receptive fields, and
furthermore models based on CNNs have been demonstrated to be able to
be well learned from biological measurements, the normative theory for
visual receptive fields used as fundament for this work has been
formulated from a different perspective.

The overall underlying thesis is that the structure of the geometric vision
problem in relation to geometric image transformations and receptive
field responses calls for that there are specific shapes of
idealized receptive fields that are very natural to handle the
image structures that arise from viewing a sufficiently rich set of
objects and scenes from different viewing conditions, such as
variations in depth, surface orientation, relative motion and speed
(faster/slower).

An underlying explanation to this result is that if we
consider a vision system that is exposed to a sufficiently rich
variety of natural image structures, as obtained from observing a
sufficiently rich variability of objects and scenes under a
sufficiently rich variability of viewing conditions, then because of
the nature of the thereby arising geometric image transformations,
the resulting image structures will exhibit variabilities in the
space of possible image data.

A follow-up argument based on this reasoning is that if
a sufficiently general and powerful
learning-based paradigm, such as sparse coding or predictive coding,
is then exposed to such image data,
we could then expect that the learned receptive fields obtained from that
data-driven approach ought to be able to pick up those variabilities
in terms of resulting variabilities in the receptive fields.
Hence, provided that the training data contains a sufficiently rich
variety of image structures, collected from image data of a
sufficiently large number of objects and spatio-temporal events under
a sufficiently large variety in variabilities of the viewing
conditions, then the receptive fields obtained from such a
learning-based mechanism ought to show similar types of variabilities
as predicted from this purely theoretical study.

Hence, the theoretical framework for receptive fields used as the
conceptual foundation for the studies in this paper
predicts a set of receptive field shapes that a sufficiently good
data-driven approach, such as sparse coding, predictive coding or CNN, may converge
to, if exposed to a sufficiently rich variety of visual stimuli, and as
acquired from a sufficiently rich variety of objects/events in the
world and geometric viewing conditions.

Contrary to the sparse
coding, predictive coding and CNN approaches, the derivation of
idealized models for visual receptive fields according to the
normative theory does not need to make use of {\em any\/} training data.
Instead, the derivation of idealized receptive field shapes is solely based
on structural properties of the environment in combination with
internal consistency requirements to guarantee theoretically
well-founded treatment of image structures over multiple spatial and
temporal scales (Lindeberg \citeyear{Lin21-Heliyon}).

Thereby, the proposed framework can be applied in
situations where one does simply not have access to sufficient amounts of
training data.
From the presented theory, it is also possible to state very
specific hypotheses and predictions, which could then be answered
by rather few actual biological experiments, and not require
significantly more complex experiments that one may not afford to
carry out.

Thus, the underlying normative theory for visual receptive field
should not be seen as in conflict with or as competing with the sparse
coding, predictive coding or CNN-based modelling approaches.
It can instead {\em explain\/} the visual receptive fields that such learning-based approaches
could be expected to converge to, if exposed to the variabilities in
image structures that arise when a general visual observer observes
sufficiently rich sets of objects in the
environment from a sufficiently rich set of viewing conditions.

Notably, however, using the proposed theory, it is possible to
perform an observation from one viewing condition (for a specific distance,
viewing angle, relative motion and temporal duration) and then perform
recognition with respect to another viewing condition (for other values of
the distance, the viewing direction, the relative motion and the temporal
duration). In this way, the presented theory opens up to {\em generalization to
novel views\/} in a way that is not directly possible for approaches based on
sparse coding, predictive coding or regular (non-covariant) CNNs.

\section{Summary and discussion}
\label{sec-summ-disc}

We have presented a principled theory for the interaction between
geometric image transformations and receptive field responses,
and used results from that theory to address the question about
variabilities in the shapes of the receptive fields of simple cells
in the primary visual cortex.

This theory is based on idealized models of visual receptive fields in
terms of combinations of smoothing with spatial smoothing kernels
of the form (\ref{eq-gauss-fcn-2D}) 
or spatio-temporal smoothing kernels of the form
(\ref{eq-spat-temp-RF-model-again-cov-props-basic})
with
scale-normalized spatial and temporal derivatives according to
Section~\ref{sec-indiv-cov-props-spat-temp-ders}.
In Sections~\ref{sec-joint-cov-props-spat-spattemo-rfs-comp-transf}
and~\ref{sec-expl-ex-cov-props-rfs}, we
have described how the resulting idealized models of spatial or
spatio-temporal receptive fields of the form
(\ref{eq-spat-temp-RF-model-der-norm-caus})
obey provable covariance properties
under compositions of spatial scaling transformations, spatial affine
transformations, Galilean transformations and temporal scaling
transformations.
Specifically, we have in Section~\ref{sec-span-vars}
considered the hypothesis about whether the
receptive fields of simple cells in the primary visual cortex can be
regarded as having their shapes expanded with regard to the degrees of freedom
of the basic types of geometric
image transformations that occur in the image formation process.

By postulating that the responses of idealized models of receptive
fields in terms of scale-normalized spatial and temporal derivative
operators should be possible to match between the image domains before
and after the geometric image transformations, we have predicted a set
of variabilities over (i)~spatial scaling transformations,
(ii)~image rotations, (iii)~the degree of spatial elongation of the
receptive fields, (iv)~a 4th spatial degree of freedom,
(v)~Galilean transformations over joint image space-time, and
(vi)~temporal scaling transformations.

We have presented partial
support for the covariance properties of the either purely spatial or
joint spatio-temporal receptive fields with regard to these
7 degrees of freedom (where the spatial affine transformations
comprise 4~degrees of freedom and the Galilean transformations comprise 2 degrees
of freedom). This has been done in view of neurophysiological evidence and
structural properties regarding how populations of receptive field
responses can be computed based on structural properties of the
families of receptive fields under variabilities over the filter
parameters.
Based on comparisons to existing and publicly available
neurophysiological measurements, we have in
Sections~\ref{sec-variabil-spat} and~\ref{sec-variabil-temp} presented
support for explicit variabilities regarding
(i)~the preferred orientation, (ii)~the degree of elongation and the
(iii)~preferred speed and (iv)~direction in joint space-time of simple cells.
In the absence of sufficient neurophysiological or
psychophysical evidence to perform exhaustive in-depth studies concerning
to what extent the predicted properties would
hold in the primary visual cortex of higher mammals
and what are the actual parameter ranges for the shape parameters for
different types of species, we have
in Section~\ref{sec-outlines-neurophys} pointed
to directions for future neurophysiological investigations to characterize these
properties in more detail.

Concerning a possible expansion of the shapes of the simple cells with
regard to the degrees of freedom of the considered 4 main types of
geometric image transformations in terms of (i)~spatial scaling
transformations, (ii)~affine image transformations, (iii)~Galilean
transformations and (iv)~temporal scaling transformations,
it is interesting to consider the number of receptive fields
in the early layers of the visual hierarchy.
Given that the 1~M output channels from the retina are mapped to 1~M
output channels from the lateral geniculate
nucleus (LGN) to the primary visual cortex (V1) to 190~M
neurons in V1 with 37~M output channels
(see DiCarlo {\em et al.\/} (\citeyear{DiCZocRus12-Neuron}) Figure~3),
the substantial expansion of
the number of receptive fields from the LGN to V1 would indeed be
consistent with an expansion of the shapes of the receptive fields
over shape parameters of the receptive fields.

Furthermore, we can physically interpret the parameters
$(S_x, A, v, S_t)$ of the primitive geometric image transformations
in the composed geometric image transformations according to
(\ref{eq-x-transf}), (\ref{eq-t-transf}) and
(\ref{eq-sc-aff-vel-transf-alt-obs-model}) 
as follows (Lindeberg \citeyear{Lin25-JMIV} Section~9):
\begin{itemize}
  \item
  the spatial scaling factor $S_x$ corresponds to the inverse
  depth $1/Z$, if the affine transformation matrix $A$ in the composed
  monocular image transformation (\ref{eq-x-transf}) is normalized
  to the maximum eigenvalue $\lambda_{\max} = 1$, such that
  the affine transformation matrix $A$ then
  reflects the pure orthographic projection component in
  a scaled orthographic projection model,
\item
  knowledge about the affine transformation matrix $A$ in the composed
  monocular transformation model (\ref{eq-x-transf}) provides direct
  information about the local surface orientation of the viewed local
  surface patch, according to the theoretical analysis in G{\aa}rding and Lindeberg
  (\citeyear{GL94-IJCV}) Section~5.2,
\item
  knowledge about the affine transformation $B$ in the binocular
  transformation model (\ref{eq-sc-aff-vel-transf-alt-obs-model})
  provides direct information about the local
  surface orientation of the viewed local surface patch,
  according to the theoretical analysis in G{\aa}rding and Lindeberg
  (\citeyear{GL94-IJCV}) Section~6.1,
\item
  knowledge about image velocity $u$ in the monocular projection
  model (\ref{eq-x-transf}), in combination with an estimate of the local depth
  $Z$ according to above, reveals the projection of the 3-D motion vector
  $U$ of the viewed object onto the image plane.
\end{itemize}
Hence, a vision system, that is able to extract these parameters of the
geometric image transformations based on processing and comparing
populations of receptive field responses, should in principle have the
ability to compute direct cues to the structure of environment,
directly from established matching relations over
the receptive field responses between the different views,
or in relation to a learned memory of receptive field responses from
previous views.
Thereby, certain functionalities of the vision system could be
formulated directly in terms of the parameters of the primitive
geometric image transformations between different views of the same
scene or the same spatio-temporal event.

Thus, in summary, we have presented a coherent theory for how the
structure of the geometric image transformations, that determine the
formation of the image data that reach the visual sensors in the
perception system, influence the desired structure of the receptive
field responses in an idealized vision system.
Specifically, we have argued how we, based on the requirement of formulating
identity relations between receptive field responses acquired under
different viewing conditions for the same scene, are lead to the
requirement of covariance properties of the receptive fields, which in
turn imply a set of variabilities of the shapes of the receptive
fields, to span the degrees of freedom of the corresponding geometric
image transformations. In this way, we have formulated a set of
theoretical predictions, and also discussed partial support of these
predictions in view of existing neurophysiological evidence and
complementary theoretical analysis in terms of cascade smoothing properties.



\subsection{Perspective and outlook}
\label{sec-persp-outlook}

Let us conclude by relating the presented theory to a set of other related
theories and concepts regarding vision.

Towards his goal of formulating an computational theory of vision,
Marr (\citeyear{Mar82}) proposed that the vision system may be
creating a symbolic representation of intensity changes in terms
of a primal sketch. While Marr primarily developed his notions
around edge, blob and bar primitives using the difference-of-Gaussians
(approximating the Laplacian-of-the-Gaussian) as a spatial filter, and which mimics the spatial component
in many receptive fields in the retina and the lateral geniculate
nucleus (LGN), with the more developed knowledge of the receptive fields
in the primary visual cortex that we have today, we can instead regard
the idealized models of spatial and spatial receptive fields used
in this treatment as providing a much richer set of basis functions for
registering the intensity variations over space and time as
the input for later stage visual processes in the visual pathway.
Specifically, the theory presented in this treatment goes
substantially further, by also addressing intimate connections between
the receptive fields and 3-D properties of the environment,
such as surface orientation, variations in depth and object motion, and notably
incorporating the effects of the perspective and the projective image transformations
directly into the filter shapes. Furthermore, although many of these receptive
fields will often give rise to strong responses near edges, we may not regard
their primary purpose as being edge detectors, but instead as members of a
richer family of spatial and spatial receptive fields applicable for a
multiple purposes in visual perception.

According to the sparse coding model for visual receptive fields
(Olshausen and Field \citeyear{OlsFie96-Nature,OlsFie97-VR}),
the receptive fields are suggested to be shaped like oriented edges,
partly for statistical efficiency, in order to compress natural
images according to the redundancy reduction principle.
This idea does also bear very close relationship to
the efficient coding hypothesis for sensory neurons
proposed by Barlow (\citeyear{Bar61-SensComm}).
According to the here considered normative theory for visual receptive fields,
the spatial primitives ought to, however, be rather shaped as affine Gaussians
derivatives, to obey geometric covariance properties to handle
variations in the viewing direction relative to 3-D objects.

According to the underlying normative theory for visual receptive
fields, there is a set of idealized receptive field models to handle
the influence of geometric image transformations. The underlying
biological substrate in terms of simple cells may, on the other hand,
have drifting properties over time. One may therefore ask if a
biological vision system exposed to variabilities in image data, as
generated from substantial variabilities in the underlying geometric
image transformations, would converge to these idealized receptive
field models by plasticity. In fact, for learning-based algorithms for
computer vision, it has been demonstrated that idealized receptive
field models structurally closely related to Gaussian derivatives have been
demonstrated to well model the receptive fields learned by
state-of-the-art convolutional networks when exposed the ImageNet dataset
(Lindeberg {\em et al.\/} \citeyear{LinBabKia26-JMIV}), however, then with much
lower variabilities in that dataset compared to the type of visual data seen by a possibly
moving visual observer in a dynamic world.

Given the structurally very close
connections between the predictions from the underlying normative
theory for visual receptive fields, derived from structural (physical)
properties of the environment, and properties of visual neurons in the primary
visual cortex, one may hence consider if the vision systems of higher
mammals have either implicitly or explicitly internalized structural/physical
properties of the environment, either by evolution or as obtained by
learning from the visual data that the vision system is exposed to
from birth. Specifically, one may ask if we from this conceptual
background can regard the influence of geometric image transformations
as a primary factor for the development of the vision systems for
higher mammals.

More generally, researchers have for decades worked on finding geometric
invariants for visual perception, see {\em e.g.\/},
Mundy and Zisserman (\citeyear{MunZis92-book}),
Rolls (\citeyear{Rol94-BehavProc}),
Booth and Rolls (\citeyear{BooRol98-CerebrCort}),
Lowe (\citeyear{Low04-IJCV}),
Quiroga {\em et al.\/} (\citeyear{QuiRedKreKocFri05-Nature}),
DiCarlo and Cox (\citeyear{DiCCox07-TICS}),
Tuytelaars and Mikolajczyk (\citeyear{TuyMik08-Book}),
Goris and Op de~Beeck (\citeyear{GorBee09-FCNS}),
Lindeberg (\citeyear{Lin13-PONE}),
Isik {\em et al.\/} (\citeyear{IsiMeyLeiPog13-JNPhys}),
Poggio and Anselmi (\citeyear{PogAns16-book}) and
Rodr{\'\i}guez  {\em et al.\/} (\citeyear{RodDelMor18-SIAM}).
From such a background, the covariant receptive field families
considered in this work behave in a predictable way under the basic classes of
geometric image transformations in vision, and do in this respect
constitute a precursor for computing corresponding invariant
visual representations at higher levels in the visual hierarchy.

\section*{Acknowledgements}

I would like to thank Benjamin Auffarth for valuable suggestions
regarding the conceptual description in Section~\ref{sec-persp-outlook}.

\appendix

\section{Propagating covariance properties of the simple cells to
  covariance properties at higher layers in the visual hierarchy and
  then instead using such derived visual representations for explicit
  matching of receptive field responses}
\label{app-prop-cov-props-high-layers}

  According to the presented theory, the primary
  visual cortex ought to be designed so as to compute receptive field
  responses for a rich variety of shapes of the spatial and
  spatio-temporal receptive fields, in the idealized theory
  corresponding to having different values of the shape parameters
  $(s, \Sigma, v, \tau)$ of the receptive fields. Let us therefore think of each
  such receptive field as being labelled with these shape parameters,
  corresponding to the notion of labelled lines (Rose \citeyear{Ros99-Perc}).

  Then, given say two views of the same object or event in the
  environment, with the respective 
  geometric parameters $(S_x, A, u, S_t)$ and $(S'_x, A', u', S'_t)$
  for the image acquisition process,
  we obtain two sets of receptive field responses
  $R(s, \Sigma, v, \tau)$ and $R'(s', \Sigma', v', \tau')$
  for rich varieties of the filter shape parameters
  $(s, \Sigma, v, \tau)$ and $(s', \Sigma', v', \tau')$ from the two
  viewing conditions.

  For conceptual simplicity, let us initially abstractly assume that the current
  real-time estimates are the ones labelled without primes, while the
  representation with primes constitute a memory of a previous
  observation of the same/similar object or event.
  The next step in a matching process, then consists of establishing
  matching relations between between the receptive field responses
  from the current real-time estimates $R(s, \Sigma, v, \tau)$
  to the previously record memory representation $R'(s', \Sigma', v',
  \tau')$
  given the transformation properties according to
  Equations~(\ref{eq-s-transf-result})--(\ref{eq-v-transf-result})
  \begin{align}
    \begin{split}
      s' & = S_x^2 \, s,
    \end{split}\\
    \begin{split}
      \Sigma' & = A \, \Sigma \, A^{T},
    \end{split}\\
    \begin{split}
      \tau' & = S_t^2 \, \tau,
    \end{split}\\
    \begin{split}
      v' & = \frac{S_x}{S_t} (A \, v + u),
    \end{split}
  \end{align}
  for the parameters of the receptive fields. This is a matching task that
  would then have to be performed by higher layers in the visual system,
  by combining the receptive field responses with a memory
  representation of earlier observation of the receptive field
  responses of a similar object or event, as schematically illustrated
  in Figure~\ref{fig-match-rf-resp-pars}.

  As remarked in connection with the reference to
  Figure~\ref{fig-match-rf-resp-pars} 
  in the introduction to Section~\ref{sec-cov-props},
  one could also conceive that the
  matching operation may not be performed directly on the output from
  the simple cells, but from receptive fields at higher layers in the
  visual hierarchy, that combine and integrate the receptive field
  responses computed for different shape parameters of the simple
  cells.
  
  In core computer science, there is an approach of using hashing
  functions for mapping higher-dimensional representations to
  lower-dimensional representations with lower computational and
  memory needs. In covariant deep networks for image information,
  pooling methods, such as max pooling, average pooling or logsumexp
  pooling, are in turn used for combining multiple receptive field
  responses obtained for different parameters of the receptive fields
  into a unified output format,
  see {\em e.g.\/} (Perzanowski and Lindeberg \citeyear{PerLin25-JMIV}).
  A very interesting question concerns if biological
  vision has developed related types of image representations
  for matching current views of objects to memory representations from
  previous views, based on lower-dimensional abstractions from the
  populations of the receptive field responses from the simple cells
  (possibly complemented with also such multi-parameter responses from
  complex cells).

  Indeed, if the visual system would choose to define derived
  receptive fields at higher layers in the visual hierarchy in an
  either covariant or invariant way, that uses the output from the
  assumed covariant receptive fields of the simple cells as input,
  then the covariance properties of the simple cells in the primary
  visual cortex could be propagated to more compact either covariant or invariant
  receptive field responses at higher layers in the visual hierarchy,
  which could in turn be directly mapped to more compact memory representations
  from previous views of the same object or spatio-temporal events.

  In the lack of available neurophysiological measurements to investigate the
  possible neural implementation of such mechanisms, we will not here aim further
  at developing the possible existence of such neural mechanisms for
  different types of animals. A strength of the presented theory, however, is that it
  provides a theoretical framework for posing such questions, based on
  a normative theory for how an idealized vision system could choose to
  handle the influence of geometric image transformations on the
  receptive field responses, and which could then be studied in
  further neurophysiological experiments.

{\footnotesize
\bibliographystyle{abbrvnat}
\bibliography{bib/defs,bib/tlmac}}

\begin{thebibliography}{149}
\providecommand{\natexlab}[1]{#1}
\providecommand{\url}[1]{\texttt{#1}}
\expandafter\ifx\csname urlstyle\endcsname\relax
  \providecommand{\doi}[1]{doi: #1}\else
  \providecommand{\doi}{doi: \begingroup \urlstyle{rm}\Url}\fi

\bibitem[Abballe and Asari(2022)]{AbbAsa22-PONE}
L.~Abballe and H.~Asari.
\newblock Natural image statistics for mouse vision.
\newblock \emph{PLOS ONE}, 17\penalty0 (1):\penalty0 e0262763, 2022.

\bibitem[Albright et~al.(1984)Albright, Desimone, and
  Gross]{AlbDesGro84-JNeuroPhys}
T.~D. Albright, R.~Desimone, and C.~G. Gross.
\newblock Columnar organization of directionally selective cells in visual area
  {MT} of the macaque.
\newblock \emph{Journal of Neurophysiology}, 51\penalty0 (1):\penalty0 16--31,
  1984.

\bibitem[Bae et~al.(2021)Bae, Kim, and Kim]{BaeKimKim21-FrontSystNeuroSci}
H.~Bae, S.~J. Kim, and C.-E. Kim.
\newblock Lessons from deep neural networks for studying the coding principles
  of biological neural networks.
\newblock \emph{Frontiers in Systems Neuroscience}, 14:\penalty0 615129, 2021.

\bibitem[Barbieri et~al.(2014)Barbieri, Citti, Cocci, and
  Sarti]{BarCitCocSar14-JMIV}
D.~Barbieri, G.~Citti, G.~Cocci, and A.~Sarti.
\newblock A cortical-inspired geometry for contour perception and motion
  integration.
\newblock \emph{Journal of Mathematical Imaging and Vision}, 49\penalty0
  (3):\penalty0 511--529, 2014.

\bibitem[Barlow(1961)]{Bar61-SensComm}
H.~B. Barlow.
\newblock Possible principles underlying the transformation of sensory
  messages.
\newblock \emph{Sensory Communication}, 1\penalty0 (1):\penalty0 217--233,
  1961.

\bibitem[Baspinar et~al.(2018)Baspinar, Citti, and Sarti]{BasCitSar18-JMIV}
E.~Baspinar, G.~Citti, and A.~Sarti.
\newblock A geometric model of multi-scale orientation preference maps via
  {G}abor functions.
\newblock \emph{Journal of Mathematical Imaging and Vision}, 60:\penalty0
  900--912, 2018.

\bibitem[Baspinar et~al.(2020)Baspinar, Sarti, and
  Citti]{BasSarCit20-MathNeuroSci}
E.~Baspinar, A.~Sarti, and G.~Citti.
\newblock A sub-{R}iemannian model of the visual cortex with frequency and
  phase.
\newblock \emph{The Journal of Mathematical Neuroscience}, 10\penalty0
  (1):\penalty0 11, 2020.

\bibitem[Bekkers(2020)]{Bek20-ICLR}
E.~J. Bekkers.
\newblock {B}-spline {CNN}s on {L}ie groups.
\newblock \emph{International Conference on Learning Representations (ICLR
  2020)}, 2020.
\newblock {https}://openreview.net/forum?id=H1gBhkBFDH, preprint at
  arXiv:1909.12057.

\bibitem[Biederman and Cooper(1992)]{BieCoo92-ExpPhys}
I.~Biederman and E.~E. Cooper.
\newblock Size invariance in visual object priming.
\newblock \emph{Journal of Experimental Physiology: Human Perception and
  Performance}, 18\penalty0 (1):\penalty0 121--133, 1992.

\bibitem[Blasdel(1992)]{Bla92-JNeuroSci}
G.~G. Blasdel.
\newblock Orientation selectivity, preference and continuity in monkey striate
  cortex.
\newblock \emph{Journal of Neuroscience}, 12\penalty0 (8):\penalty0 3139--3161,
  1992.

\bibitem[Bonhoeffer and Grinvald(1991)]{BonGri91-Nature}
T.~Bonhoeffer and A.~Grinvald.
\newblock Iso-orientation domains in cat visual cortex are arranged in
  pinwheel-like patterns.
\newblock \emph{Nature}, 353:\penalty0 429--431, 1991.

\bibitem[Booth and Rolls(1998)]{BooRol98-CerebrCort}
M.~C.~A. Booth and E.~T. Rolls.
\newblock View-invariant representations of familiar objects by neurons in the
  inferior temporal visual cortex.
\newblock \emph{Cerebral Cortex}, 8:\penalty0 510--523, 1998.

\bibitem[Born and Bradley(2005)]{BorBra05-AnnRevNeurSci}
R.~T. Born and D.~C. Bradley.
\newblock Structure and function of visual area {MT}.
\newblock \emph{Annual Review of Neurosciences}, 28\penalty0 (1):\penalty0
  157--189, 2005.

\bibitem[Bowers et~al.(2022)Bowers, Malhotra, Dujmovi{\'c}, Montero, Tsvetkov,
  Biscione, Puebla, Adolfi, Hummel, Heaton, Evans, Mitchell, and
  Blything]{BowMalDujMonTsvBisPueAdoHumHeaEvaMitBly22-BehBrainSci}
J.~S. Bowers, G.~Malhotra, M.~Dujmovi{\'c}, M.~L. Montero, C.~Tsvetkov,
  V.~Biscione, G.~Puebla, F.~Adolfi, J.~E. Hummel, R.~F. Heaton, B.~D. Evans,
  J.~Mitchell, and R.~Blything.
\newblock Deep problems with neural network models of human vision.
\newblock \emph{Behavioral and Brain Sciences}, pages 1--74, 2022.

\bibitem[Bronstein et~al.(2021)Bronstein, Bruna, Cohen, and
  Veli{\v{c}}kovi{\'c}]{BroBruCohVel21-arXiv}
M.~M. Bronstein, J.~Bruna, T.~Cohen, and P.~Veli{\v{c}}kovi{\'c}.
\newblock Geometric deep learning: {G}rids, groups, graphs, geodesics, and
  gauges.
\newblock \emph{arXiv preprint arXiv:2104.13478}, 2021.

\bibitem[Churchland et~al.(2005)Churchland, Priebe, and
  Lisberger]{ChuPriLis05-JNeuroPhys}
M.~M. Churchland, N.~J. Priebe, and S.~Lisberger.
\newblock Comparison of the spatial limits on direction selectivity in visual
  areas {MT} and {V1}.
\newblock \emph{Journal of Neurophysiology}, 93\penalty0 (3):\penalty0
  1235--1245, 2005.

\bibitem[Citti and Sarti(2006)]{CitSar06-JMIV}
G.~Citti and A.~Sarti.
\newblock A cortical based model of perceptual completion in the
  roto-translation space.
\newblock \emph{Journal of Mathematical Imaging and Vision}, 24\penalty0
  (3):\penalty0 307--326, 2006.

\bibitem[Cocci et~al.(2011)Cocci, Barbieri, and Sarti]{CocBarSar11-JOSA}
G.~Cocci, D.~Barbieri, and A.~Sarti.
\newblock Spatiotemporal receptive fields of cells in {V1} are optimally shaped
  for stimulus velocity estimation.
\newblock \emph{Journal of the Optical Society of America A}, 29\penalty0
  (1):\penalty0 130--138, 2011.

\bibitem[Conway and Livingstone(2006)]{ConLiv06-JNeurSci}
B.~R. Conway and M.~S. Livingstone.
\newblock Spatial and temporal properties of cone signals in alert macaque
  primary visual cortex.
\newblock \emph{Journal of Neuroscience}, 26\penalty0 (42):\penalty0
  10826--10846, 2006.

\bibitem[De and Horwitz(2021)]{DeHor21-JNPhys}
A.~De and G.~D. Horwitz.
\newblock Spatial receptive field structure of double-opponent cells in macaque
  {V1}.
\newblock \emph{Journal of Neurophysiology}, 125\penalty0 (3):\penalty0
  843--857, 2021.

\bibitem[de~Valois et~al.(2000)de~Valois, Cottaris, Mahon, Elfar, and
  Wilson]{ValCotMahElfWil00-VisRes}
R.~L. de~Valois, N.~P. Cottaris, L.~Mahon, S.~D. Elfar, and J.~A. Wilson.
\newblock Spatial and temporal receptive fields of geniculate and cortical
  cells and directional selectivity.
\newblock \emph{Vision Research}, 40\penalty0 (27):\penalty0 3685--3702, 2000.

\bibitem[DeAngelis and Anzai(2004)]{deAngAnz04-VisNeuroSci}
G.~C. DeAngelis and A.~Anzai.
\newblock A modern view of the classical receptive field: Linear and non-linear
  spatio-temporal processing by {V1} neurons.
\newblock In L.~M. Chalupa and J.~S. Werner, editors, \emph{The Visual
  Neurosciences}, volume~1, pages 704--719. MIT Press, 2004.

\bibitem[DeAngelis et~al.(1995)DeAngelis, Ohzawa, and
  Freeman]{DeAngOhzFre95-TINS}
G.~C. DeAngelis, I.~Ohzawa, and R.~D. Freeman.
\newblock Receptive field dynamics in the central visual pathways.
\newblock \emph{Trends in Neuroscience}, 18\penalty0 (10):\penalty0 451--457,
  1995.

\bibitem[DiCarlo and Cox(2007)]{DiCCox07-TICS}
J.~J. DiCarlo and D.~D. Cox.
\newblock Untangling invariant object recognition.
\newblock \emph{Trends in Cognitive Science}, 11\penalty0 (8):\penalty0
  333--341, 2007.

\bibitem[DiCarlo and Maunsell(2000)]{DiCMau00-Nature}
J.~J. DiCarlo and J.~H.~R. Maunsell.
\newblock Form representation in monkey inferotemporal cortex is virtually
  unaltered by free viewing.
\newblock \emph{Nature Neuroscience}, 3\penalty0 (8):\penalty0 814--821, 2000.

\bibitem[DiCarlo et~al.(2012)DiCarlo, Zoccolan, and Rust]{DiCZocRus12-Neuron}
J.~J. DiCarlo, D.~Zoccolan, and N.~C. Rust.
\newblock How does the brain solve visual object recognition?
\newblock \emph{Neuron}, 73\penalty0 (3):\penalty0 415--434, 2012.

\bibitem[Elstrott and Feller(2009)]{ElsFel09-CurrOpNeurBiol}
J.~Elstrott and M.~B. Feller.
\newblock Vision and the establishment of direction-selectivity: a tale of two
  circuits.
\newblock \emph{Current Opinion in Neurobiology}, 19\penalty0 (3):\penalty0
  293--297, 2009.

\bibitem[Fu et~al.(2024)Fu, A, Willeke, Bashiri, Muhammad, Diamantaki,
  Froudarakis, Restivo, Ponder, Denfield, Sinz, Tolias, and
  Franke]{FuPieWilBasMuhDiaFroResPonDenSinTolFra24-CellRep}
J.~Fu, P.~P. A, K.~F. Willeke, M.~Bashiri, T.~Muhammad, M.~Diamantaki,
  E.~Froudarakis, K.~Restivo, K.~Ponder, G.~H. Denfield, F.~Sinz, A.~S. Tolias,
  and K.~Franke.
\newblock Heterogeneous orientation tuning in the primary visual cortex of mice
  diverges from {G}abor-like receptive fields in primates.
\newblock \emph{Cell Reports}, 43\penalty0 (8), 2024.

\bibitem[Furmanski and Engel(2000)]{FurEng00-VisRes}
C.~S. Furmanski and S.~A. Engel.
\newblock Perceptual learning in object recognition: {O}bject specificity and
  size invariance.
\newblock \emph{Vision Research}, 40:\penalty0 473--484, 2000.

\bibitem[G{\aa}rding and Lindeberg(1996)]{GL94-IJCV}
J.~G{\aa}rding and T.~Lindeberg.
\newblock Direct computation of shape cues using scale-adapted spatial
  derivative operators.
\newblock \emph{International Journal of Computer Vision}, 17\penalty0
  (2):\penalty0 163--191, 1996.

\bibitem[Geisler(2008)]{Wil08-AnnRevPsychol}
W.~S. Geisler.
\newblock Visual perception and the statistical properties of natural scenes.
\newblock \emph{Annual Review of Psychology}, 59:\penalty0 10.1--10.26, 2008.

\bibitem[Georgeson et~al.(2007)Georgeson, May, Freeman, and
  Hesse]{GeoMayFreHes07-JVis}
M.~A. Georgeson, K.~A. May, T.~C.~A. Freeman, and G.~S. Hesse.
\newblock From filters to features: {S}cale-space analysis of edge and blur
  coding in human vision.
\newblock \emph{Journal of Vision}, 7\penalty0 (13):\penalty0 7.1--21, 2007.

\bibitem[Gerken et~al.(2023)Gerken, Aronsson, Carlsson, Linander, Ohlsson,
  Petersson, and Persson]{GerAroCarLinOhlPetPer23-AIRev}
J.~E. Gerken, J.~Aronsson, O.~Carlsson, H.~Linander, F.~Ohlsson, C.~Petersson,
  and D.~Persson.
\newblock Geometric deep learning and equivariant neural networks.
\newblock \emph{Artificial Intelligence Review}, 56\penalty0 (12):\penalty0
  14605--14662, 2023.

\bibitem[Ghodrati et~al.(2017)Ghodrati, Khaligh-Razavi, and
  Lehky]{GhoKhaLeh17-ProNeurobiol}
M.~Ghodrati, S.-M. Khaligh-Razavi, and S.~R. Lehky.
\newblock Towards building a more complex view of the lateral geniculate
  nucleus: {R}ecent advances in understanding its role.
\newblock \emph{Progress in Neurobiology}, 156:\penalty0 214--255, 2017.

\bibitem[Goris and de~Beeck(2009)]{GorBee09-FCNS}
R.~L.~T. Goris and H.~P.~O. de~Beeck.
\newblock Neural representations that support invariant object recognition.
\newblock \emph{Frontiers in Computational Neuroscience}, 3\penalty0
  (Article~3):\penalty0 1--16, 2009.

\bibitem[Goris et~al.(2015)Goris, Simoncelli, and Movshon]{GorSimMov15-Neuron}
R.~L.~T. Goris, E.~P. Simoncelli, and J.~A. Movshon.
\newblock Origin and function of tuning diversity in {M}acaque visual cortex.
\newblock \emph{Neuron}, 88\penalty0 (4):\penalty0 819--831, 2015.

\bibitem[Grimes and Rao(2005)]{GriRao05-NeurComp}
D.~B. Grimes and R.~P.~N. Rao.
\newblock Bilinear sparse coding for invariant vision.
\newblock \emph{Nature Neuroscience}, 3\penalty0 (8):\penalty0 814--821, 2005.

\bibitem[Hansen and Neumann(2008)]{HanNeu09-JVis}
T.~Hansen and H.~Neumann.
\newblock A recurrent model of contour integration in primary visual cortex.
\newblock \emph{Journal of Vision}, 8\penalty0 (8):\penalty0 8.1--25, 2008.

\bibitem[Heinke et~al.(2022)Heinke, Leonardis, and Leek]{HeiLeoLee22-VisRes}
D.~Heinke, A.~Leonardis, and C.~E. Leek.
\newblock What do deep neural networks tell us about biological vision?
\newblock \emph{Vision Research}, 198:\penalty0 108069, 2022.

\bibitem[Hesse and Georgeson(2005)]{HesGeo05-VisRes}
G.~S. Hesse and M.~A. Georgeson.
\newblock Edges and bars: where do people see features in 1-{D} images?
\newblock \emph{Vision Research}, 45\penalty0 (4):\penalty0 507--525, 2005.

\bibitem[Hubel(1959)]{Hub59-JPhys}
D.~H. Hubel.
\newblock Single unit activity in striate cortex of unrestrained cats.
\newblock \emph{The Journal of Physiology}, 147\penalty0 (2):\penalty0 226,
  1959.

\bibitem[Hubel and Wiesel(1959)]{HubWie59-Phys}
D.~H. Hubel and T.~N. Wiesel.
\newblock Receptive fields of single neurones in the cat's striate cortex.
\newblock \emph{J Physiol}, 147:\penalty0 226--238, 1959.

\bibitem[Hubel and Wiesel(1962)]{HubWie62-Phys}
D.~H. Hubel and T.~N. Wiesel.
\newblock Receptive fields, binocular interaction and functional architecture
  in the cat's visual cortex.
\newblock \emph{J Physiol}, 160:\penalty0 106--154, 1962.

\bibitem[Hubel and Wiesel(1968)]{HubWie68-JPhys}
D.~H. Hubel and T.~N. Wiesel.
\newblock Receptive fields and functional architecture of monkey striate
  cortex.
\newblock \emph{The Journal of Physiology}, 195\penalty0 (1):\penalty0
  215--243, 1968.

\bibitem[Hubel and Wiesel(2005)]{HubWie05-book}
D.~H. Hubel and T.~N. Wiesel.
\newblock \emph{Brain and Visual Perception: {T}he Story of a 25-Year
  Collaboration}.
\newblock Oxford University Press, 2005.

\bibitem[Huberman and Niell(2011)]{HubNie11-TINS}
A.~D. Huberman and C.~M. Niell.
\newblock What can mice tell us about how vision works?
\newblock \emph{Trends in Neurosciences}, 34\penalty0 (9):\penalty0 464--473,
  2011.

\bibitem[Hung et~al.(2005)Hung, Kreiman, Poggio, and
  DiCarlo]{HunKrePogDiC05-Science}
C.~P. Hung, G.~Kreiman, T.~Poggio, and J.~J. DiCarlo.
\newblock Fast readout of object indentity from macaque inferior temporal
  cortex.
\newblock \emph{Science}, 310:\penalty0 863--866, 2005.

\bibitem[Hyv{\"a}rinen et~al.(2009)Hyv{\"a}rinen, Hurri, and
  Hoyer]{HyvHurHoy09-NatImgStat}
A.~Hyv{\"a}rinen, J.~Hurri, and P.~O. Hoyer.
\newblock \emph{Natural Image Statistics: {A} Probabilistic Approach to Early
  Computational Vision}.
\newblock Computational Imaging and Vision. Springer, 2009.

\bibitem[Isik et~al.(2013)Isik, Meyers, Leibo, and
  Poggio]{IsiMeyLeiPog13-JNPhys}
L.~Isik, E.~M. Meyers, J.~Z. Leibo, and T.~Poggio.
\newblock The dynamics of invariant object recognition in the human visual
  system.
\newblock \emph{Journal of Neurophysiology}, 111\penalty0 (1):\penalty0
  91--102, 2013.

\bibitem[Ito et~al.(1995)Ito, Tamura, Fujita, and
  Tanaka]{ItoTamFujTan95-JNeuroPhys}
M.~Ito, H.~Tamura, I.~Fujita, and K.~Tanaka.
\newblock Size and position invariance of neuronal responses in monkey
  inferotemporal cortex.
\newblock \emph{Journal of Neurophysiology}, 73\penalty0 (1):\penalty0
  218--226, 1995.

\bibitem[Jacques et~al.(2022)Jacques, Tiganj, Sarkar, Howard, and
  Sederberg]{JacTigSarHowSed22-ICML}
G.~B. Jacques, Z.~Tiganj, A.~Sarkar, M.~Howard, and P.~Sederberg.
\newblock A deep convolutional neural network that is invariant to time
  rescaling.
\newblock In \emph{International Conference on Machine Learning (ICML 2022)},
  pages 9729--9738, 2022.

\bibitem[Jansson and Lindeberg(2022)]{JanLin22-JMIV}
Y.~Jansson and T.~Lindeberg.
\newblock Scale-invariant scale-channel networks: Deep networks that generalise
  to previously unseen scales.
\newblock \emph{Journal of Mathematical Imaging and Vision}, 64\penalty0
  (5):\penalty0 506--536, 2022.

\bibitem[Johnson et~al.(2008)Johnson, Hawken, and
  Shapley]{JohHawSha08-JNeuroSci}
E.~N. Johnson, M.~J. Hawken, and R.~Shapley.
\newblock The orientation selectivity of color-responsive neurons in {M}acaque
  {V1}.
\newblock \emph{The Journal of Neuroscience}, 28\penalty0 (32):\penalty0
  8096--8106, 2008.

\bibitem[Jones and Palmer(1987{\natexlab{a}})]{JonPal87a}
J.~Jones and L.~Palmer.
\newblock The two-dimensional spatial structure of simple receptive fields in
  cat striate cortex.
\newblock \emph{J. of Neurophysiology}, 58:\penalty0 1187--1211,
  1987{\natexlab{a}}.

\bibitem[Jones and Palmer(1987{\natexlab{b}})]{JonPal87b}
J.~Jones and L.~Palmer.
\newblock An evaluation of the two-dimensional {G}abor filter model of simple
  receptive fields in cat striate cortex.
\newblock \emph{J. of Neurophysiology}, 58:\penalty0 1233--1258,
  1987{\natexlab{b}}.

\bibitem[Keller(2025)]{Kel25-NeurIPS}
T.~A. Keller.
\newblock Flow equivariant recurrent neural networks.
\newblock In \emph{Proc.\ Neural Information Processing Systems (NeurIPS
  2025)}, 2025.
\newblock preprint at arXiv:2507.14793.

\bibitem[Keshishian et~al.(2020)Keshishian, Akbari, Khalighinejad, Herrero,
  Mehta, and Mesgarani]{KesAkbKhaHerMehMes20-Elife}
M.~Keshishian, H.~Akbari, B.~Khalighinejad, J.~L. Herrero, A.~D. Mehta, and
  N.~Mesgarani.
\newblock Estimating and interpreting nonlinear receptive field of sensory
  neural responses with deep neural network models.
\newblock \emph{eLife}, 9:\penalty0 e53445, 2020.

\bibitem[King et~al.(2013)King, Zylberberg, and DeWeese]{KinZylDeW13-JNeuroSci}
P.~D. King, J.~Zylberberg, and M.~R. DeWeese.
\newblock Inhibitory interneurons decorrelate excitatory cells to drive sparse
  code formation in a spiking model of {V1}.
\newblock \emph{Journal of Neuroscience}, 33\penalty0 (13):\penalty0
  5475--5485, 2013.

\bibitem[Koch et~al.(2016)Koch, Jin, Alonso, and Zaidi]{KocJinAloZai16-NatComm}
E.~Koch, J.~Jin, J.~M. Alonso, and Q.~Zaidi.
\newblock Functional implications of orientation maps in primary visual cortex.
\newblock \emph{Nature Communications}, 7\penalty0 (1):\penalty0 13529, 2016.

\bibitem[Koenderink(1984)]{Koe84}
J.~J. Koenderink.
\newblock The structure of images.
\newblock \emph{Biological Cybernetics}, 50\penalty0 (5):\penalty0 363--370,
  1984.

\bibitem[Koenderink and {van Doorn}(1978)]{KoeDoo78-BC}
J.~J. Koenderink and A.~J. {van Doorn}.
\newblock Visual detection of spatial contrast; influence of location in the
  visual field, target extent and illuminance level.
\newblock \emph{Biological Cybernetics}, 30:\penalty0 157--167, 1978.

\bibitem[Koenderink and {van Doorn}(1987)]{KoeDoo87-BC}
J.~J. Koenderink and A.~J. {van Doorn}.
\newblock Representation of local geometry in the visual system.
\newblock \emph{Biological Cybernetics}, 55\penalty0 (6):\penalty0 367--375,
  1987.

\bibitem[Koenderink and {van Doorn}(1992)]{KoeDoo92-PAMI}
J.~J. Koenderink and A.~J. {van Doorn}.
\newblock Generic neighborhood operators.
\newblock \emph{IEEE Transactions on Pattern Analysis and Machine
  Intelligence}, 14\penalty0 (6):\penalty0 597--605, Jun. 1992.

\bibitem[Kristensen and Sandberg(2021)]{KriSan21-SciRep}
D.~G. Kristensen and K.~Sandberg.
\newblock Population receptive fields of human primary visual cortex organised
  as {DC}-balanced bandpass filters.
\newblock \emph{Scientific Reports}, 11\penalty0 (1):\penalty0 22423, 2021.

\bibitem[Kwon and Park(2019)]{KwoPar19-ICCV}
Y.-H. Kwon and M.-G. Park.
\newblock Predicting future frames using retrospective cycle {GAN}.
\newblock In \emph{Proc.\ Computer Vision and Pattern Recognition (CVPR 2019)},
  pages 1811--1820, 2019.

\bibitem[Land(1974)]{Lan74-RoyInst}
E.~H. Land.
\newblock The retinex theory of colour vision.
\newblock \emph{Proc.\ Royal Institution of Great Britain}, 57:\penalty0
  23--58, 1974.

\bibitem[Land(1986)]{Lan86-VR}
E.~H. Land.
\newblock Recent advances in retinex theory.
\newblock \emph{Vision Research}, 26\penalty0 (1):\penalty0 7--21, 1986.

\bibitem[Li et~al.(2024)Li, Qiu, Chen, He, and Lin]{LiQiuCheHeLin24-CVPR}
Y.~Li, Y.~Qiu, Y.~Chen, L.~He, and Z.~Lin.
\newblock Affine equivariant networks based on differential invariants.
\newblock In \emph{Proc.\ Computer Vision and Pattern Recognition (CVPR 2024)},
  pages 5546--5556, 2024.

\bibitem[Lindeberg(1993)]{Lin93-Dis}
T.~Lindeberg.
\newblock \emph{Scale-Space Theory in Computer Vision}.
\newblock Springer, 1993.

\bibitem[Lindeberg(1998)]{Lin97-IJCV}
T.~Lindeberg.
\newblock Feature detection with automatic scale selection.
\newblock \emph{International Journal of Computer Vision}, 30\penalty0
  (2):\penalty0 77--116, 1998.

\bibitem[Lindeberg(2011)]{Lin10-JMIV}
T.~Lindeberg.
\newblock Generalized {G}aussian scale-space axiomatics comprising linear
  scale-space, affine scale-space and spatio-temporal scale-space.
\newblock \emph{Journal of Mathematical Imaging and Vision}, 40\penalty0
  (1):\penalty0 36--81, 2011.

\bibitem[Lindeberg(2013{\natexlab{a}})]{Lin13-BICY}
T.~Lindeberg.
\newblock A computational theory of visual receptive fields.
\newblock \emph{Biological Cybernetics}, 107\penalty0 (6):\penalty0 589--635,
  2013{\natexlab{a}}.

\bibitem[Lindeberg(2013{\natexlab{b}})]{Lin13-PONE}
T.~Lindeberg.
\newblock Invariance of visual operations at the level of receptive fields.
\newblock \emph{{PLOS One}}, 8\penalty0 (7):\penalty0 e66990,
  2013{\natexlab{b}}.

\bibitem[Lindeberg(2016)]{Lin16-JMIV}
T.~Lindeberg.
\newblock Time-causal and time-recursive spatio-temporal receptive fields.
\newblock \emph{Journal of Mathematical Imaging and Vision}, 55\penalty0
  (1):\penalty0 50--88, 2016.

\bibitem[Lindeberg(2017)]{Lin17-JMIV}
T.~Lindeberg.
\newblock Temporal scale selection in time-causal scale space.
\newblock \emph{Journal of Mathematical Imaging and Vision}, 58\penalty0
  (1):\penalty0 57--101, 2017.

\bibitem[Lindeberg(2021{\natexlab{a}})]{Lin21-EncCompVis}
T.~Lindeberg.
\newblock Scale selection.
\newblock In K.~Ikeuchi, editor, \emph{Computer Vision}, pages 1110--1123.
  Springer, 2021{\natexlab{a}}.
\newblock {https}://doi.org/10.1007/978-3-030-03243-2\_242-1.

\bibitem[Lindeberg(2021{\natexlab{b}})]{Lin21-Heliyon}
T.~Lindeberg.
\newblock Normative theory of visual receptive fields.
\newblock \emph{Heliyon}, 7\penalty0 (1):\penalty0 e05897:1--20,
  2021{\natexlab{b}}.
\newblock \doi{10.1016/j.heliyon.2021.e05897}.

\bibitem[Lindeberg(2022)]{Lin22-JMIV}
T.~Lindeberg.
\newblock Scale-covariant and scale-invariant {G}aussian derivative networks.
\newblock \emph{Journal of Mathematical Imaging and Vision}, 64\penalty0
  (3):\penalty0 223--242, 2022.

\bibitem[Lindeberg(2023{\natexlab{a}})]{Lin23-BICY}
T.~Lindeberg.
\newblock A time-causal and time-recursive scale-covariant scale-space
  representation of temporal signals and past time.
\newblock \emph{Biological Cybernetics}, 117\penalty0 (1--2):\penalty0 21--59,
  2023{\natexlab{a}}.

\bibitem[Lindeberg(2023{\natexlab{b}})]{Lin23-FrontCompNeuroSci}
T.~Lindeberg.
\newblock Covariance properties under natural image transformations for the
  generalized {G}aussian derivative model for visual receptive fields.
\newblock \emph{Frontiers in Computational Neuroscience}, 17:\penalty0
  1189949:1--23, 2023{\natexlab{b}}.

\bibitem[Lindeberg(2024)]{Lin24-arXiv-UnifiedJointCovProps}
T.~Lindeberg.
\newblock Unified theory for joint covariance properties under geometric image
  transformations for spatio-temporal receptive fields according to the
  generalized {G}aussian derivative model for visual receptive fields.
\newblock \emph{arXiv preprint arXiv:2311.10543}, 2024.

\bibitem[Lindeberg(2025{\natexlab{a}})]{Lin25-BICY}
T.~Lindeberg.
\newblock Relationships between the degrees of freedom in the affine {G}aussian
  derivative model for visual receptive fields and {2-D} affine image
  transformations, with application to covariance properties of simple cells in
  the primary visual cortex.
\newblock \emph{Biological Cybernetics}, 119\penalty0 (2--3):\penalty0
  15:1--25, 2025{\natexlab{a}}.

\bibitem[Lindeberg(2025{\natexlab{b}})]{Lin25-JCompNeurSci-orisel}
T.~Lindeberg.
\newblock Orientation selectivity properties for the affine {G}aussian
  derivative and the affine {G}abor models for visual receptive fields.
\newblock \emph{Journal of Computational Neuroscience}, 53\penalty0
  (1):\penalty0 61--98, 2025{\natexlab{b}}.

\bibitem[Lindeberg(2025{\natexlab{c}})]{Lin25-JCompNeurSci-spanelong}
T.~Lindeberg.
\newblock Do the receptive fields in the primary visual cortex span a
  variability over the degree of elongation of the receptive fields?
\newblock \emph{Journal of Computational Neuroscience}, 53\penalty0
  (3):\penalty0 397--417, 2025{\natexlab{c}}.

\bibitem[Lindeberg(2025{\natexlab{d}})]{Lin25-JMIV}
T.~Lindeberg.
\newblock Unified theory for joint covariance properties under geometric image
  transformations for spatio-temporal receptive fields according to the
  generalized {G}aussian derivative model for visual receptive fields.
\newblock \emph{Journal of Mathematical Imaging and Vision}, 67\penalty0
  (4):\penalty0 44:1--49, 2025{\natexlab{d}}.

\bibitem[Lindeberg(2025{\natexlab{e}})]{Lin25-PONE}
T.~Lindeberg.
\newblock Orientation selectivity properties for integrated affine quasi
  quadrature models of complex cells.
\newblock \emph{PLOS One}, 20\penalty0 (9):\penalty0 e0332139:1--25,
  2025{\natexlab{e}}.

\bibitem[Lindeberg(2025{\natexlab{f}})]{Lin25-arXiv-dirsel}
T.~Lindeberg.
\newblock Direction and speed selectivity properties for spatio-temporal
  receptive fields according to the generalized {G}aussian derivative model for
  visual receptive fields.
\newblock \emph{arXiv preprint ar{X}iv:2511.08101}, 2025{\natexlab{f}}.

\bibitem[Lindeberg(2026)]{Lin26-JMIV}
T.~Lindeberg.
\newblock Hybrid {L}ie semi-group and cascade structures for the generalized
  {G}aussian derivative model for visual receptive fields.
\newblock \emph{Journal of Mathematical Imaging and Vision}, 68\penalty0
  (3):\penalty0 19:1--29, 2026.

\bibitem[Lindeberg and Florack(1994)]{CVAP166}
T.~Lindeberg and L.~Florack.
\newblock Foveal scale-space and linear increase of receptive field size as a
  function of eccentricity.
\newblock report, ISRN KTH/NA/P-{}-94/27-{}-SE, Dept. of Numerical Analysis and
  Computer Science, KTH, Aug. 1994.
\newblock Available from
  http://www.csc.kth.se/$\sim$tony/abstracts/CVAP166.html.

\bibitem[Lindeberg and G{\aa}rding(1997)]{LG96-IVC}
T.~Lindeberg and J.~G{\aa}rding.
\newblock Shape-adapted smoothing in estimation of 3-{D} shape cues from affine
  distortions of local 2-{D} structure.
\newblock \emph{Image and Vision Computing}, 15\penalty0 (6):\penalty0
  415--434, 1997.

\bibitem[Lindeberg et~al.(2026)Lindeberg, Babaiee, and
  Kiasari]{LinBabKia26-JMIV}
T.~Lindeberg, Z.~Babaiee, and P.~M. Kiasari.
\newblock Modelling and analysis of the 8 filters from the 'master key filters
  hypothesis' for depthwise-separable deep networks in relation to idealized
  receptive fields based on scale-space theory.
\newblock \emph{Journal of Mathematical Imaging and Vision}, 68\penalty0
  (3):\penalty0 22:1--26, 2026.

\bibitem[Livingstone and Conway(2003)]{LivCon03-JNeuroPhys}
M.~S. Livingstone and B.~R. Conway.
\newblock Substructure of direction-selective receptive fields in macaque {V1}.
\newblock \emph{Journal of Neurophysiology}, 89\penalty0 (5):\penalty0
  2743--2759, 2003.

\bibitem[Logothetis et~al.(1995)Logothetis, Pauls, and
  Poggio]{LogPauPog95-CurrBiol}
N.~K. Logothetis, J.~Pauls, and T.~Poggio.
\newblock Shape representation in the inferior temporal cortex of monkeys.
\newblock \emph{Current Biology}, 5\penalty0 (2):\penalty0 552--563, 1995.

\bibitem[L{\"o}rincz et~al.(2012)L{\"o}rincz, Palotal, and
  Szirtes]{LoePalSzi12-PLOS-CB}
A.~L{\"o}rincz, Z.~Palotal, and G.~Szirtes.
\newblock Efficient sparse coding in early sensory processing: {L}essons from
  signal recovery.
\newblock \emph{PLOS Computational Biology}, 8\penalty0 (3):\penalty0 e1002372,
  2012.

\bibitem[Lotter et~al.(2020)Lotter, Kreiman, and
  Cox]{LotKreCox20-NatMachIntell}
W.~Lotter, G.~Kreiman, and D.~Cox.
\newblock A neural network trained to predict future video frames mimics
  critical properties of biological neuronal responses and perception.
\newblock \emph{Nature Machine Intelligence}, 4\penalty0 (2):\penalty0
  210--219, 2020.

\bibitem[Lowe(2000)]{Low00-BIO}
D.~G. Lowe.
\newblock Towards a computational model for object recognition in {IT} cortex.
\newblock In \emph{Biologically Motivated Computer Vision}, volume 1811 of
  \emph{Springer LNCS}, pages 20--31. Springer, 2000.

\bibitem[Lowe(2004)]{Low04-IJCV}
D.~G. Lowe.
\newblock Distinctive image features from scale-invariant keypoints.
\newblock \emph{International Journal of Computer Vision}, 60\penalty0
  (2):\penalty0 91--110, 2004.

\bibitem[Marcelja(1980)]{Mar80-JOSA}
S.~Marcelja.
\newblock Mathematical description of the responses of simple cortical cells.
\newblock \emph{Journal of Optical Society of America}, 70\penalty0
  (11):\penalty0 1297--1300, 1980.

\bibitem[Marr(1982)]{Mar82}
D.~Marr.
\newblock \emph{Vision: {A} Computational Investigation into the Human
  Representation and Processing of Visual Information}.
\newblock W.H. Freeman, New York, 1982.

\bibitem[May and Georgeson(2007)]{MayGeo05-VisRes}
K.~A. May and M.~A. Georgeson.
\newblock Blurred edges look faint, and faint edges look sharp: {T}he effect of
  a gradient threshold in a multi-scale edge coding model.
\newblock \emph{Vision Research}, 47\penalty0 (13):\penalty0 1705--1720, 2007.

\bibitem[Meshcherina et~al.(2026)Meshcherina, Auffarth, and
  Lindeberg]{MesAufLin20-bioRxiv}
D.~Meshcherina, B.~Auffarth, and T.~Lindeberg.
\newblock Variabilities in the shapes of the receptive fields of simple cells
  in mice.
\newblock \emph{bioRxiv preprint 2026.10.07.757489}, 2026.

\bibitem[M{\l}ynarski(2025)]{Mly25-VisRes}
W.~F. M{\l}ynarski.
\newblock Origins and objectives of computational diversity in sensory
  populations.
\newblock \emph{Vision Research}, 237:\penalty0 108683, 2025.

\bibitem[Movshon and Newsome(1996)]{MovNew98-JNeuroSci}
J.~A. Movshon and W.~T. Newsome.
\newblock Visual response properties of striate cortical neurons projecting to
  area {MT} in macaque monkeys.
\newblock \emph{Journal of Neuroscience}, 16\penalty0 (23):\penalty0
  7733--7741, 1996.

\bibitem[Mundy and Zisserman(1992)]{MunZis92-book}
J.~L. Mundy and A.~Zisserman, editors.
\newblock \emph{Geometric Invariance in Computer Vision}.
\newblock MIT Press, Cambridge, Massachusetts, 1992.

\bibitem[Nauhaus et~al.(2008)Nauhaus, Benucci, Carandini, and
  Ringach]{NauBenCarRin09-Neuron}
I.~Nauhaus, A.~Benucci, M.~Carandini, and D.~L. Ringach.
\newblock Neuronal selectivity and local map structure in visual cortex.
\newblock \emph{Neuron}, 57\penalty0 (5):\penalty0 673--679, 2008.

\bibitem[Olshausen and Field(1996)]{OlsFie96-Nature}
B.~A. Olshausen and D.~J. Field.
\newblock Emergence of simple-cell receptive field properties by learning a
  sparse code for natural images.
\newblock \emph{Journal of Optical Society of America}, 381:\penalty0 607--609,
  1996.

\bibitem[Olshausen and Field(1997)]{OlsFie97-VR}
B.~A. Olshausen and D.~J. Field.
\newblock Sparse coding with an overcomplete basis set: {A} strategy employed
  by {V}1?
\newblock \emph{Vision Research}, 37\penalty0 (23):\penalty0 3311--3325, 1997.

\bibitem[O'Neill(1966)]{Nei66}
B.~O'Neill.
\newblock \emph{Elementary Differential Geometry}.
\newblock Academic Press, Orlando, Florida, 1966.

\bibitem[Orban(1997)]{Orb97-ExtrStriCortPrim}
G.~A. Orban.
\newblock Visual processing in macaque area {MT/V5} and its satellites ({MSTd}
  and {MSTv}).
\newblock In K.~S. Rockland, J.~H. Kaas, and A.~Peters, editors,
  \emph{Extrastriate Cortex in Primates}, pages 359--434. Springer, 1997.

\bibitem[Orban et~al.(1986)Orban, Kennedy, and Bullier]{OrbKenNul86-JNeurPhys}
G.~A. Orban, H.~Kennedy, and J.~Bullier.
\newblock Velocity sensitivity and direction selectivity of neurons in areas
  {V1} and {V2} of the monkey: influence of eccentricity.
\newblock \emph{Journal of Neurophysiology}, 56\penalty0 (2):\penalty0
  462--480, 1986.

\bibitem[Palmer(1999)]{Pal99-Book}
S.~E. Palmer.
\newblock \emph{Vision Science: Photons to Phenomenology}.
\newblock MIT Press, 1999.
\newblock First Edition.

\bibitem[Pei et~al.(2016)Pei, Gao, Hao, Qiao, and
  Ai]{PeiGaoHaoQiaAi16-NeurRegen}
Z.-J. Pei, G.-X. Gao, B.~Hao, Q.-L. Qiao, and H.-J. Ai.
\newblock A cascade model of information processing and encoding for retinal
  prosthesis.
\newblock \emph{Neural Regeneration Research}, 11\penalty0 (4):\penalty0 646,
  2016.

\bibitem[Peli(1990)]{Pel90-JOSA}
E.~Peli.
\newblock Contrast in complex images.
\newblock \emph{Journal of the Optical Society of America (JOSA A)}, 7\penalty0
  (10):\penalty0 2032--2040, 1990.

\bibitem[Perzanowski and Lindeberg(2025)]{PerLin25-JMIV}
A.~Perzanowski and T.~Lindeberg.
\newblock Scale generalisation properties of extended scale-covariant and
  scale-invariant {G}aussian derivative networks on image datasets with spatial
  scaling variations.
\newblock \emph{Journal of Mathematical Imaging and Vision}, 67\penalty0
  (3):\penalty0 29:1--39, 2025.

\bibitem[Poggio and Anselmi(2016)]{PogAns16-book}
T.~A. Poggio and F.~Anselmi.
\newblock \emph{Visual Cortex and Deep Networks: Learning Invariant
  Representations}.
\newblock MIT Press, 2016.

\bibitem[Porat and Zeevi(1988)]{PorZee88-PAMI}
M.~Porat and Y.~Y. Zeevi.
\newblock The generalized {G}abor scheme of image representation in biological
  and machine vision.
\newblock \emph{IEEE Transactions on Pattern Analysis and Machine
  Intelligence}, 10\penalty0 (4):\penalty0 452--468, 1988.

\bibitem[Quiroga et~al.(2005)Quiroga, Reddy, Kreiman, Koch, and
  Fried]{QuiRedKreKocFri05-Nature}
R.~Q. Quiroga, L.~Reddy, G.~Kreiman, C.~Koch, and I.~Fried.
\newblock Invariant visual representations by single neurons in the human
  brain.
\newblock \emph{Nature}, 435:\penalty0 1102--1107, 2005.

\bibitem[Rao and Ballard(1998)]{RaoBal98-CompNeurSyst}
R.~P.~N. Rao and D.~H. Ballard.
\newblock Development of localized oriented receptive fields by learning a
  translation-invariant code for natural images.
\newblock \emph{Computation in Neural Systems}, 9\penalty0 (2):\penalty0
  219--234, 1998.

\bibitem[Rehn and Sommer(2007)]{RehSom07-JCompNeuroSci}
M.~Rehn and F.~T. Sommer.
\newblock A network that uses few active neurones to code visual input predicts
  the diverse shapes of cortical receptive fields.
\newblock \emph{Journal of Computational Neuroscience}, 22:\penalty0 135--146,
  2007.

\bibitem[Ringach(2002)]{Rin01-JNeuroPhys}
D.~L. Ringach.
\newblock Spatial structure and symmetry of simple-cell receptive fields in
  macaque primary visual cortex.
\newblock \emph{Journal of Neurophysiology}, 88:\penalty0 455--463, 2002.

\bibitem[Ringach(2004)]{Rin04-JPhys}
D.~L. Ringach.
\newblock Mapping receptive fields in primary visual cortex.
\newblock \emph{Journal of Physiology}, 558\penalty0 (3):\penalty0 717--728,
  2004.

\bibitem[Rodr{\'\i}guez et~al.(2018)Rodr{\'\i}guez, Delon, and
  Morel]{RodDelMor18-SIAM}
M.~Rodr{\'\i}guez, J.~Delon, and J.-M. Morel.
\newblock Covering the space of tilts: Application to affine invariant image
  comparison.
\newblock \emph{SIAM Journal on Imaging Sciences}, 11\penalty0 (2):\penalty0
  1230--1267, 2018.

\bibitem[Rolls(1994)]{Rol94-BehavProc}
E.~T. Rolls.
\newblock Brain mechanisms for invariant visual recognition and learning.
\newblock \emph{Behavioural Processes}, 33:\penalty0 113--138, 1994.

\bibitem[Rose(1999)]{Ros99-Perc}
D.~Rose.
\newblock The historical roots of the theories of local signs and labelled
  lines.
\newblock \emph{Perception}, 28\penalty0 (6):\penalty0 675--685, 1999.

\bibitem[Ruslim et~al.(2023)Ruslim, Burkitt, and Lian]{RusBurLia23-bioRxiv}
M.~A. Ruslim, A.~N. Burkitt, and Y.~Lian.
\newblock Learning spatio-temporal {V1} cells from diverse {LGN} inputs.
\newblock \emph{bioRxiv preprint 2023.11.30.569354}, 2023.

\bibitem[Sarti et~al.(2008)Sarti, Citti, and Petitot]{SarCitPet08-BICY}
A.~Sarti, G.~Citti, and J.~Petitot.
\newblock The symplectic structure of the primary visual cortex.
\newblock \emph{Biological Cybernetics}, 98\penalty0 (1):\penalty0 33--48,
  2008.

\bibitem[Schoenberg(1950)]{Sch50}
I.~J. Schoenberg.
\newblock On {P}\`olya frequency functions. ii. {V}ariation-diminishing
  integral operators of the convolution type.
\newblock \emph{Acta Sci. Math. (Szeged)}, 12:\penalty0 97--106, 1950.

\bibitem[Serre et~al.(2007)Serre, Wolf, Bileschi, Riesenhuber, and
  Poggio]{SerWolBilRiePog07-PAMI}
T.~Serre, L.~Wolf, S.~Bileschi, M.~Riesenhuber, and T.~Poggio.
\newblock Robust object recognition with cortex-like mechanisms.
\newblock \emph{IEEE Transactions on Pattern Analysis and Machine
  Intelligence}, 29\penalty0 (3):\penalty0 411--426, 2007.

\bibitem[Simoncelli and Olshausen(2001)]{SimOls01-AnnRevNeurSci}
E.~P. Simoncelli and B.~A. Olshausen.
\newblock Natural image statistics and neural representations.
\newblock \emph{Annual Review of Neuroscience}, 24:\penalty0 1193--1216, 2001.

\bibitem[Singer et~al.(2018)Singer, Teramoto, Willmore, Schnupp, King, and
  Harper]{SinTerWilSchKinHar18-Elife}
Y.~Singer, Y.~Teramoto, B.~D.~B. Willmore, J.~W.~H. Schnupp, A.~J. King, and
  N.~S. Harper.
\newblock Sensory cortex is optimized for prediction of future input.
\newblock \emph{Elife}, 7:\penalty0 e31557, 2018.

\bibitem[Sosnovik et~al.(2020)Sosnovik, Szmaja, and
  Smeulders]{SosSzmSme20-ICLR}
I.~Sosnovik, M.~Szmaja, and A.~Smeulders.
\newblock Scale-equivariant steerable networks.
\newblock \emph{International Conference on Learning Representations (ICLR
  2020)}, 2020.
\newblock preprint at {arXiv}:1910.11093.

\bibitem[Sosnovik et~al.(2021)Sosnovik, Moskalev, and
  Smeulders]{SosMosSme21-BMVC}
I.~Sosnovik, A.~Moskalev, and A.~Smeulders.
\newblock {DISCO}: {A}ccurate discrete scale convolutions.
\newblock \emph{British Machine Vision Conference (BMVC 2021)}, 2021.
\newblock preprint at {arXiv}:2106.02733.

\bibitem[ter Haar~Romeny(1994)]{Haa94-GDDbook}
B.~ter Haar~Romeny, editor.
\newblock \emph{Geometry-Driven Diffusion in Computer Vision}.
\newblock Series in Mathematical Imaging and Vision. Springer, 1994.

\bibitem[ter Haar~Romeny(2003)]{Haa04-book}
B.~ter Haar~Romeny.
\newblock \emph{Front-End Vision and Multi-Scale Image Analysis}.
\newblock Springer, 2003.

\bibitem[Tuytelaars and Mikolajczyk(2008)]{TuyMik08-Book}
T.~Tuytelaars and K.~Mikolajczyk.
\newblock \emph{A Survey on Local Invariant Features}, volume 3(3) of
  \emph{Foundations and Trends in Computer Graphics and Vision}.
\newblock Now Publishers, 2008.

\bibitem[Walker et~al.(2019)Walker, Sinz, Cobos, Muhammad, Froudarakis, Fahey,
  Ecker, Reimer, Pitkow, and
  Tolias]{WalSinCobMuhFroFahEckReiPitTol19-NatNeurSci}
E.~Y. Walker, F.~H. Sinz, E.~Cobos, T.~Muhammad, E.~Froudarakis, P.~G. Fahey,
  A.~S. Ecker, J.~Reimer, X.~Pitkow, and A.~S. Tolias.
\newblock Inception loops discover what excites neurons most using deep
  predictive models.
\newblock \emph{Nature Neuroscience}, 22\penalty0 (12):\penalty0 2060--2065,
  2019.

\bibitem[Wallis and Georgeson(2009)]{WalGeo09-VisRes}
S.~A. Wallis and M.~A. Georgeson.
\newblock Mach edges: Local features predicted by 3rd derivative spatial
  filtering.
\newblock \emph{Vision Research}, 49\penalty0 (14):\penalty0 1886--1893, 2009.

\bibitem[Wang and Spratling(2016)]{WanSpra16-CognComp}
Q.~Wang and M.~W. Spratling.
\newblock Contour detection in colour images using a neurophysiologically
  inspired model.
\newblock \emph{Cognitive Computation}, 8\penalty0 (6):\penalty0 1027--1035,
  2016.

\bibitem[Wendt and Faul(2024)]{WenFay24-JVis}
G.~Wendt and F.~Faul.
\newblock Binocular luster elicited by isoluminant chromatic stimuli relies on
  mechanisms similar to those in the achromatic case.
\newblock \emph{Journal of Vision}, 24\penalty0 (3):\penalty0 7--7, 2024.

\bibitem[Wichmann and Geirhos(2023)]{WichGei23-AnnRevVisSci}
F.~A. Wichmann and R.~Geirhos.
\newblock Are deep neural networks adequate behavioral models of human visual
  perception?
\newblock \emph{Annual Review of Vision Science}, 9, 2023.

\bibitem[Wimmer et~al.(2023)Wimmer, Golkov, Dang, Zaiss, Maier, and
  Cremers]{WimGolDa23-arXiv}
T.~Wimmer, V.~Golkov, H.~N. Dang, M.~Zaiss, A.~Maier, and D.~Cremers.
\newblock Scale-equivariant deep learning for 3{D} data.
\newblock \emph{arXiv preprint ar{X}iv:2304.05864}, 2023.

\bibitem[Worrall and Welling(2019)]{WorWel19-NeuroIPS}
D.~Worrall and M.~Welling.
\newblock Deep scale-spaces: {E}quivariance over scale.
\newblock In \emph{Advances in Neural Information Processing Systems (NeurIPS
  2019)}, pages 7366--7378, 2019.

\bibitem[Yazdanbakhsh and Livingstone(2006)]{YazLiv06-NatNeuroSci}
A.~Yazdanbakhsh and M.~S. Livingstone.
\newblock End stopping in {V1} is sensitive to contrast.
\newblock \emph{Nature Neuroscience}, 9\penalty0 (5):\penalty0 697--702, 2006.

\bibitem[Young(1987)]{You87-SV}
R.~A. Young.
\newblock The {G}aussian derivative model for spatial vision: {I}. {R}etinal
  mechanisms.
\newblock \emph{Spatial Vision}, 2\penalty0 (4):\penalty0 273--293, 1987.

\bibitem[Young and Lesperance(2001)]{YouLes01-SV}
R.~A. Young and R.~M. Lesperance.
\newblock The {G}aussian derivative model for spatio-temporal vision: {II}.
  {C}ortical data.
\newblock \emph{Spatial Vision}, 14\penalty0 (3, 4):\penalty0 321--389, 2001.

\bibitem[Young et~al.(2001)Young, Lesperance, and Meyer]{YouLesMey01-SV}
R.~A. Young, R.~M. Lesperance, and W.~W. Meyer.
\newblock The {G}aussian derivative model for spatio-temporal vision: {I}.
  {C}ortical model.
\newblock \emph{Spatial Vision}, 14\penalty0 (3, 4):\penalty0 261--319, 2001.

\bibitem[Zhan et~al.(2022)Zhan, Sun, and Li]{ZhaSunLi22-ICCRE}
W.~Zhan, G.~Sun, and Y.~Li.
\newblock Scale-equivariant steerable networks for crowd counting.
\newblock In \emph{Proc.\ International Conference on Control and Robotics
  Engineering (ICCRE 2022)}, pages 174--179, 2022.

\bibitem[Zhu et~al.(2022)Zhu, Qiu, Calderbank, Sapiro, and
  Cheng]{ZhuQiuCalSapChe22-JMLR}
W.~Zhu, Q.~Qiu, R.~Calderbank, G.~Sapiro, and X.~Cheng.
\newblock Scale-translation-equivariant neural networks with decomposed
  convolutional filters.
\newblock \emph{Journal of Machine Learning Research}, 23\penalty0
  (68):\penalty0 1--45, 2022.

\bibitem[Zylberberg et~al.(2011)Zylberberg, Murphy, and
  DeWeese]{ZylMurDeW11-PlosCompBio}
J.~Zylberberg, J.~T. Murphy, and M.~R. DeWeese.
\newblock A sparse coding model with synaptically local plasticity and spiking
  neurons can account for the diverse shapes of v1 simple cell receptive
  fields.
\newblock \emph{PLOS Computational Biology}, 7\penalty0 (10):\penalty0
  e1002250, 2011.

\end{thebibliography}

\end{document}